\documentclass[aps,reprint,twocolumn,10pt,superscriptaddress,english,showpacs,longbibliography,nofootinbib,floatfix]{revtex4-1}

\usepackage{amsmath}
\usepackage{amssymb}
\usepackage{amsthm}
\usepackage{amsfonts}
\usepackage{listings}
\usepackage{longtable}
\usepackage{enumerate}
\usepackage{latexsym}
\usepackage{color}
\usepackage{multirow}
\usepackage[figuresright]{rotating}
\usepackage[table]{xcolor}
\usepackage{setspace} 
\usepackage{blindtext}
\usepackage{dcolumn}
\usepackage{braket}
\usepackage{xr}
\usepackage{xr-hyper}
\usepackage{changepage}

\usepackage{array,etoolbox}
\preto\tabular{\setcounter{magicrownumbers}{0}}
\newcounter{magicrownumbers}

\usepackage{array,etoolbox}

\usepackage{bm}
\usepackage{hyperref}
\hypersetup{
 pdfnewwindow=true, colorlinks=true,
 linkcolor=blue, anchorcolor=blue,
 citecolor=blue, filecolor=blue,
 menucolor=blue, urlcolor=blue}
\definecolor{red}{rgb}{1,0,0}
\definecolor{blue}{rgb}{0,0,1}
\definecolor{black}{rgb}{0,0,0}

\usepackage{graphicx}
\usepackage{subfigure}
\usepackage{appendix}
\usepackage{array}
\usepackage{booktabs}
\newcommand{\webBCSMAG}{\href{http://www.cryst.ehu.es/magndata/}{MAGNDATA}}

\newcommand{\webMTQC}{\href{https://www.topologicalquantumchemistry.fr/magnetic}{Topological Magnetic Materials}}

\newcommand{\bcsidwebformula}[2]{\href{https://www.topologicalquantumchemistry.fr/magnetic/index.html\#?BCSID=#1}{#2}}

\newcommand{\bcsidmagndatashort}[1]{\href{http://www.cryst.ehu.es/magndata/index.php?index=#1}{#1}}

\newcommand{\bcsidmagndata}[1]{\href{http://www.cryst.ehu.es/magndata/index.php?index=#1}{BCSID #1}}

\newcommand{\webTQC}{\href{https://www.topologicalquantumchemistry.org/}{Topological Quantum Chemistry website}}

\newcommand{\MTQCDBNbrBCSIDsWithPhaseDiagram}{894}

\newcommand{\MTQCDBNbrBCSIDsNonTrivialNonAtomicPercent}{43.29\%}

\newcommand{\MTQCDBRunOneNbrBCSIDsWithPhaseDiagram}{372}

\newcommand{\MTQCDBRunOneNbrBCSIDsNonTrivialNonAtomic}{137}

\newcommand{\MTQCDBRunTwoNbrBCSIDsWithPhaseDiagram}{522}

\newcommand{\MTQCDBRunTwoNbrBCSIDsTI}{98}

\newcommand{\MTQCDBRunTwoNbrBCSIDsSM}{182}

\newcommand{\MTQCDBRunTwoNbrBCSIDsNonTrivialNonAtomic}{250}
\newcommand{\MTQCDBRunTwoNbrBCSIDsNonTrivialNonAtomicPercent}{47.89\%}

\newcommand{\MTQCDBRunTwoNbrBCSIDsMOAI}{50}

\newcommand{\MTQCDBRunTwoNbrBCSIDsAXI}{84}

\newcommand{\MTQCDBRunTwoNbrBCSIDsThreeDQAH}{34}

\newcommand{\MTQCDBRunTwoNbrBCSIDsMTCI}{3}

\newcommand{\MTQCDBRunTwoNbrBCSIDsSISM}{13}

\newcommand{\msgsymb}[2]{\ifnum#1=1
\ifnum#2=1
$P1$\else
\ifnum#2=2
$P11'$\else
\ifnum#2=3
$P_{S}1$\else
{\color{red}{Invalid MSG number}}
\fi
\fi
\fi
\else
\ifnum#1=2
\ifnum#2=4
$P\bar{1}$\else
\ifnum#2=5
$P\bar{1}1'$\else
\ifnum#2=6
$P\bar{1}'$\else
\ifnum#2=7
$P_{S}\bar{1}$\else
{\color{red}{Invalid MSG number}}
\fi
\fi
\fi
\fi
\else
\ifnum#1=3
\ifnum#2=1
$P2$\else
\ifnum#2=2
$P21'$\else
\ifnum#2=3
$P2'$\else
\ifnum#2=4
$P_{a}2$\else
\ifnum#2=5
$P_{b}2$\else
\ifnum#2=6
$P_{C}2$\else
{\color{red}{Invalid MSG number}}
\fi
\fi
\fi
\fi
\fi
\fi
\else
\ifnum#1=4
\ifnum#2=7
$P2_{1}$\else
\ifnum#2=8
$P2_{1}1'$\else
\ifnum#2=9
$P2_{1}'$\else
\ifnum#2=10
$P_{a}2_{1}$\else
\ifnum#2=11
$P_{b}2_{1}$\else
\ifnum#2=12
$P_{C}2_{1}$\else
{\color{red}{Invalid MSG number}}
\fi
\fi
\fi
\fi
\fi
\fi
\else
\ifnum#1=5
\ifnum#2=13
$C2$\else
\ifnum#2=14
$C21'$\else
\ifnum#2=15
$C2'$\else
\ifnum#2=16
$C_{c}2$\else
\ifnum#2=17
$C_{a}2$\else
{\color{red}{Invalid MSG number}}
\fi
\fi
\fi
\fi
\fi
\else
\ifnum#1=6
\ifnum#2=18
$Pm$\else
\ifnum#2=19
$Pm1'$\else
\ifnum#2=20
$Pm'$\else
\ifnum#2=21
$P_{a}m$\else
\ifnum#2=22
$P_{b}m$\else
\ifnum#2=23
$P_{C}m$\else
{\color{red}{Invalid MSG number}}
\fi
\fi
\fi
\fi
\fi
\fi
\else
\ifnum#1=7
\ifnum#2=24
$Pc$\else
\ifnum#2=25
$Pc1'$\else
\ifnum#2=26
$Pc'$\else
\ifnum#2=27
$P_{a}c$\else
\ifnum#2=28
$P_{c}c$\else
\ifnum#2=29
$P_{b}c$\else
\ifnum#2=30
$P_{C}c$\else
\ifnum#2=31
$P_{A}c$\else
{\color{red}{Invalid MSG number}}
\fi
\fi
\fi
\fi
\fi
\fi
\fi
\fi
\else
\ifnum#1=8
\ifnum#2=32
$Cm$\else
\ifnum#2=33
$Cm1'$\else
\ifnum#2=34
$Cm'$\else
\ifnum#2=35
$C_{c}m$\else
\ifnum#2=36
$C_{a}m$\else
{\color{red}{Invalid MSG number}}
\fi
\fi
\fi
\fi
\fi
\else
\ifnum#1=9
\ifnum#2=37
$Cc$\else
\ifnum#2=38
$Cc1'$\else
\ifnum#2=39
$Cc'$\else
\ifnum#2=40
$C_{c}c$\else
\ifnum#2=41
$C_{a}c$\else
{\color{red}{Invalid MSG number}}
\fi
\fi
\fi
\fi
\fi
\else
\ifnum#1=10
\ifnum#2=42
$P2/m$\else
\ifnum#2=43
$P2/m1'$\else
\ifnum#2=44
$P2'/m$\else
\ifnum#2=45
$P2/m'$\else
\ifnum#2=46
$P2'/m'$\else
\ifnum#2=47
$P_{a}2/m$\else
\ifnum#2=48
$P_{b}2/m$\else
\ifnum#2=49
$P_{C}2/m$\else
{\color{red}{Invalid MSG number}}
\fi
\fi
\fi
\fi
\fi
\fi
\fi
\fi
\else
\ifnum#1=11
\ifnum#2=50
$P2_{1}/m$\else
\ifnum#2=51
$P2_{1}/m1'$\else
\ifnum#2=52
$P2_{1}'/m$\else
\ifnum#2=53
$P2_{1}/m'$\else
\ifnum#2=54
$P2_{1}'/m'$\else
\ifnum#2=55
$P_{a}2_{1}/m$\else
\ifnum#2=56
$P_{b}2_{1}/m$\else
\ifnum#2=57
$P_{C}2_{1}/m$\else
{\color{red}{Invalid MSG number}}
\fi
\fi
\fi
\fi
\fi
\fi
\fi
\fi
\else
\ifnum#1=12
\ifnum#2=58
$C2/m$\else
\ifnum#2=59
$C2/m1'$\else
\ifnum#2=60
$C2'/m$\else
\ifnum#2=61
$C2/m'$\else
\ifnum#2=62
$C2'/m'$\else
\ifnum#2=63
$C_{c}2/m$\else
\ifnum#2=64
$C_{a}2/m$\else
{\color{red}{Invalid MSG number}}
\fi
\fi
\fi
\fi
\fi
\fi
\fi
\else
\ifnum#1=13
\ifnum#2=65
$P2/c$\else
\ifnum#2=66
$P2/c1'$\else
\ifnum#2=67
$P2'/c$\else
\ifnum#2=68
$P2/c'$\else
\ifnum#2=69
$P2'/c'$\else
\ifnum#2=70
$P_{a}2/c$\else
\ifnum#2=71
$P_{b}2/c$\else
\ifnum#2=72
$P_{c}2/c$\else
\ifnum#2=73
$P_{A}2/c$\else
\ifnum#2=74
$P_{C}2/c$\else
{\color{red}{Invalid MSG number}}
\fi
\fi
\fi
\fi
\fi
\fi
\fi
\fi
\fi
\fi
\else
\ifnum#1=14
\ifnum#2=75
$P2_{1}/c$\else
\ifnum#2=76
$P2_{1}/c1'$\else
\ifnum#2=77
$P2_{1}'/c$\else
\ifnum#2=78
$P2_{1}/c'$\else
\ifnum#2=79
$P2_{1}'/c'$\else
\ifnum#2=80
$P_{a}2_{1}/c$\else
\ifnum#2=81
$P_{b}2_{1}/c$\else
\ifnum#2=82
$P_{c}2_{1}/c$\else
\ifnum#2=83
$P_{A}2_{1}/c$\else
\ifnum#2=84
$P_{C}2_{1}/c$\else
{\color{red}{Invalid MSG number}}
\fi
\fi
\fi
\fi
\fi
\fi
\fi
\fi
\fi
\fi
\else
\ifnum#1=15
\ifnum#2=85
$C2/c$\else
\ifnum#2=86
$C2/c1'$\else
\ifnum#2=87
$C2'/c$\else
\ifnum#2=88
$C2/c'$\else
\ifnum#2=89
$C2'/c'$\else
\ifnum#2=90
$C_{c}2/c$\else
\ifnum#2=91
$C_{a}2/c$\else
{\color{red}{Invalid MSG number}}
\fi
\fi
\fi
\fi
\fi
\fi
\fi
\else
\ifnum#1=16
\ifnum#2=1
$P222$\else
\ifnum#2=2
$P2221'$\else
\ifnum#2=3
$P2'2'2$\else
\ifnum#2=4
$P_{a}222$\else
\ifnum#2=5
$P_{C}222$\else
\ifnum#2=6
$P_{I}222$\else
{\color{red}{Invalid MSG number}}
\fi
\fi
\fi
\fi
\fi
\fi
\else
\ifnum#1=17
\ifnum#2=7
$P222_{1}$\else
\ifnum#2=8
$P222_{1}1'$\else
\ifnum#2=9
$P2'2'2_{1}$\else
\ifnum#2=10
$P22'2_{1}'$\else
\ifnum#2=11
$P_{a}222_{1}$\else
\ifnum#2=12
$P_{c}222_{1}$\else
\ifnum#2=13
$P_{B}222_{1}$\else
\ifnum#2=14
$P_{C}222_{1}$\else
\ifnum#2=15
$P_{I}222_{1}$\else
{\color{red}{Invalid MSG number}}
\fi
\fi
\fi
\fi
\fi
\fi
\fi
\fi
\fi
\else
\ifnum#1=18
\ifnum#2=16
$P2_{1}2_{1}2$\else
\ifnum#2=17
$P2_{1}2_{1}21'$\else
\ifnum#2=18
$P2_{1}'2_{1}'2$\else
\ifnum#2=19
$P2_{1}2_{1}'2'$\else
\ifnum#2=20
$P_{b}2_{1}2_{1}2$\else
\ifnum#2=21
$P_{c}2_{1}2_{1}2$\else
\ifnum#2=22
$P_{B}2_{1}2_{1}2$\else
\ifnum#2=23
$P_{C}2_{1}2_{1}2$\else
\ifnum#2=24
$P_{I}2_{1}2_{1}2$\else
{\color{red}{Invalid MSG number}}
\fi
\fi
\fi
\fi
\fi
\fi
\fi
\fi
\fi
\else
\ifnum#1=19
\ifnum#2=25
$P2_{1}2_{1}2_{1}$\else
\ifnum#2=26
$P2_{1}2_{1}2_{1}1'$\else
\ifnum#2=27
$P2_{1}'2_{1}'2_{1}$\else
\ifnum#2=28
$P_{c}2_{1}2_{1}2_{1}$\else
\ifnum#2=29
$P_{C}2_{1}2_{1}2_{1}$\else
\ifnum#2=30
$P_{I}2_{1}2_{1}2_{1}$\else
{\color{red}{Invalid MSG number}}
\fi
\fi
\fi
\fi
\fi
\fi
\else
\ifnum#1=20
\ifnum#2=31
$C222_{1}$\else
\ifnum#2=32
$C222_{1}1'$\else
\ifnum#2=33
$C2'2'2_{1}$\else
\ifnum#2=34
$C22'2_{1}'$\else
\ifnum#2=35
$C_{c}222_{1}$\else
\ifnum#2=36
$C_{a}222_{1}$\else
\ifnum#2=37
$C_{A}222_{1}$\else
{\color{red}{Invalid MSG number}}
\fi
\fi
\fi
\fi
\fi
\fi
\fi
\else
\ifnum#1=21
\ifnum#2=38
$C222$\else
\ifnum#2=39
$C2221'$\else
\ifnum#2=40
$C2'2'2$\else
\ifnum#2=41
$C22'2'$\else
\ifnum#2=42
$C_{c}222$\else
\ifnum#2=43
$C_{a}222$\else
\ifnum#2=44
$C_{A}222$\else
{\color{red}{Invalid MSG number}}
\fi
\fi
\fi
\fi
\fi
\fi
\fi
\else
\ifnum#1=22
\ifnum#2=45
$F222$\else
\ifnum#2=46
$F2221'$\else
\ifnum#2=47
$F2'2'2$\else
\ifnum#2=48
$F_{S}222$\else
{\color{red}{Invalid MSG number}}
\fi
\fi
\fi
\fi
\else
\ifnum#1=23
\ifnum#2=49
$I222$\else
\ifnum#2=50
$I2221'$\else
\ifnum#2=51
$I2'2'2$\else
\ifnum#2=52
$I_{c}222$\else
{\color{red}{Invalid MSG number}}
\fi
\fi
\fi
\fi
\else
\ifnum#1=24
\ifnum#2=53
$I2_{1}2_{1}2_{1}$\else
\ifnum#2=54
$I2_{1}2_{1}2_{1}1'$\else
\ifnum#2=55
$I2_{1}'2_{1}'2_{1}$\else
\ifnum#2=56
$I_{c}2_{1}2_{1}2_{1}$\else
{\color{red}{Invalid MSG number}}
\fi
\fi
\fi
\fi
\else
\ifnum#1=25
\ifnum#2=57
$Pmm2$\else
\ifnum#2=58
$Pmm21'$\else
\ifnum#2=59
$Pm'm2'$\else
\ifnum#2=60
$Pm'm'2$\else
\ifnum#2=61
$P_{c}mm2$\else
\ifnum#2=62
$P_{a}mm2$\else
\ifnum#2=63
$P_{C}mm2$\else
\ifnum#2=64
$P_{A}mm2$\else
\ifnum#2=65
$P_{I}mm2$\else
{\color{red}{Invalid MSG number}}
\fi
\fi
\fi
\fi
\fi
\fi
\fi
\fi
\fi
\else
\ifnum#1=26
\ifnum#2=66
$Pmc2_{1}$\else
\ifnum#2=67
$Pmc2_{1}1'$\else
\ifnum#2=68
$Pm'c2_{1}'$\else
\ifnum#2=69
$Pmc'2_{1}'$\else
\ifnum#2=70
$Pm'c'2_{1}$\else
\ifnum#2=71
$P_{a}mc2_{1}$\else
\ifnum#2=72
$P_{b}mc2_{1}$\else
\ifnum#2=73
$P_{c}mc2_{1}$\else
\ifnum#2=74
$P_{A}mc2_{1}$\else
\ifnum#2=75
$P_{B}mc2_{1}$\else
\ifnum#2=76
$P_{C}mc2_{1}$\else
\ifnum#2=77
$P_{I}mc2_{1}$\else
{\color{red}{Invalid MSG number}}
\fi
\fi
\fi
\fi
\fi
\fi
\fi
\fi
\fi
\fi
\fi
\fi
\else
\ifnum#1=27
\ifnum#2=78
$Pcc2$\else
\ifnum#2=79
$Pcc21'$\else
\ifnum#2=80
$Pc'c2'$\else
\ifnum#2=81
$Pc'c'2$\else
\ifnum#2=82
$P_{c}cc2$\else
\ifnum#2=83
$P_{a}cc2$\else
\ifnum#2=84
$P_{C}cc2$\else
\ifnum#2=85
$P_{A}cc2$\else
\ifnum#2=86
$P_{I}cc2$\else
{\color{red}{Invalid MSG number}}
\fi
\fi
\fi
\fi
\fi
\fi
\fi
\fi
\fi
\else
\ifnum#1=28
\ifnum#2=87
$Pma2$\else
\ifnum#2=88
$Pma21'$\else
\ifnum#2=89
$Pm'a2'$\else
\ifnum#2=90
$Pma'2'$\else
\ifnum#2=91
$Pm'a'2$\else
\ifnum#2=92
$P_{a}ma2$\else
\ifnum#2=93
$P_{b}ma2$\else
\ifnum#2=94
$P_{c}ma2$\else
\ifnum#2=95
$P_{A}ma2$\else
\ifnum#2=96
$P_{B}ma2$\else
\ifnum#2=97
$P_{C}ma2$\else
\ifnum#2=98
$P_{I}ma2$\else
{\color{red}{Invalid MSG number}}
\fi
\fi
\fi
\fi
\fi
\fi
\fi
\fi
\fi
\fi
\fi
\fi
\else
\ifnum#1=29
\ifnum#2=99
$Pca2_{1}$\else
\ifnum#2=100
$Pca2_{1}1'$\else
\ifnum#2=101
$Pc'a2_{1}'$\else
\ifnum#2=102
$Pca'2_{1}'$\else
\ifnum#2=103
$Pc'a'2_{1}$\else
\ifnum#2=104
$P_{a}ca2_{1}$\else
\ifnum#2=105
$P_{b}ca2_{1}$\else
\ifnum#2=106
$P_{c}ca2_{1}$\else
\ifnum#2=107
$P_{A}ca2_{1}$\else
\ifnum#2=108
$P_{B}ca2_{1}$\else
\ifnum#2=109
$P_{C}ca2_{1}$\else
\ifnum#2=110
$P_{I}ca2_{1}$\else
{\color{red}{Invalid MSG number}}
\fi
\fi
\fi
\fi
\fi
\fi
\fi
\fi
\fi
\fi
\fi
\fi
\else
\ifnum#1=30
\ifnum#2=111
$Pnc2$\else
\ifnum#2=112
$Pnc21'$\else
\ifnum#2=113
$Pn'c2'$\else
\ifnum#2=114
$Pnc'2'$\else
\ifnum#2=115
$Pn'c'2$\else
\ifnum#2=116
$P_{a}nc2$\else
\ifnum#2=117
$P_{b}nc2$\else
\ifnum#2=118
$P_{c}nc2$\else
\ifnum#2=119
$P_{A}nc2$\else
\ifnum#2=120
$P_{B}nc2$\else
\ifnum#2=121
$P_{C}nc2$\else
\ifnum#2=122
$P_{I}nc2$\else
{\color{red}{Invalid MSG number}}
\fi
\fi
\fi
\fi
\fi
\fi
\fi
\fi
\fi
\fi
\fi
\fi
\else
\ifnum#1=31
\ifnum#2=123
$Pmn2_{1}$\else
\ifnum#2=124
$Pmn2_{1}1'$\else
\ifnum#2=125
$Pm'n2_{1}'$\else
\ifnum#2=126
$Pmn'2_{1}'$\else
\ifnum#2=127
$Pm'n'2_{1}$\else
\ifnum#2=128
$P_{a}mn2_{1}$\else
\ifnum#2=129
$P_{b}mn2_{1}$\else
\ifnum#2=130
$P_{c}mn2_{1}$\else
\ifnum#2=131
$P_{A}mn2_{1}$\else
\ifnum#2=132
$P_{B}mn2_{1}$\else
\ifnum#2=133
$P_{C}mn2_{1}$\else
\ifnum#2=134
$P_{I}mn2_{1}$\else
{\color{red}{Invalid MSG number}}
\fi
\fi
\fi
\fi
\fi
\fi
\fi
\fi
\fi
\fi
\fi
\fi
\else
\ifnum#1=32
\ifnum#2=135
$Pba2$\else
\ifnum#2=136
$Pba21'$\else
\ifnum#2=137
$Pb'a2'$\else
\ifnum#2=138
$Pb'a'2$\else
\ifnum#2=139
$P_{c}ba2$\else
\ifnum#2=140
$P_{b}ba2$\else
\ifnum#2=141
$P_{C}ba2$\else
\ifnum#2=142
$P_{A}ba2$\else
\ifnum#2=143
$P_{I}ba2$\else
{\color{red}{Invalid MSG number}}
\fi
\fi
\fi
\fi
\fi
\fi
\fi
\fi
\fi
\else
\ifnum#1=33
\ifnum#2=144
$Pna2_{1}$\else
\ifnum#2=145
$Pna2_{1}1'$\else
\ifnum#2=146
$Pn'a2_{1}'$\else
\ifnum#2=147
$Pna'2_{1}'$\else
\ifnum#2=148
$Pn'a'2_{1}$\else
\ifnum#2=149
$P_{a}na2_{1}$\else
\ifnum#2=150
$P_{b}na2_{1}$\else
\ifnum#2=151
$P_{c}na2_{1}$\else
\ifnum#2=152
$P_{A}na2_{1}$\else
\ifnum#2=153
$P_{B}na2_{1}$\else
\ifnum#2=154
$P_{C}na2_{1}$\else
\ifnum#2=155
$P_{I}na2_{1}$\else
{\color{red}{Invalid MSG number}}
\fi
\fi
\fi
\fi
\fi
\fi
\fi
\fi
\fi
\fi
\fi
\fi
\else
\ifnum#1=34
\ifnum#2=156
$Pnn2$\else
\ifnum#2=157
$Pnn21'$\else
\ifnum#2=158
$Pn'n2'$\else
\ifnum#2=159
$Pn'n'2$\else
\ifnum#2=160
$P_{a}nn2$\else
\ifnum#2=161
$P_{c}nn2$\else
\ifnum#2=162
$P_{A}nn2$\else
\ifnum#2=163
$P_{C}nn2$\else
\ifnum#2=164
$P_{I}nn2$\else
{\color{red}{Invalid MSG number}}
\fi
\fi
\fi
\fi
\fi
\fi
\fi
\fi
\fi
\else
\ifnum#1=35
\ifnum#2=165
$Cmm2$\else
\ifnum#2=166
$Cmm21'$\else
\ifnum#2=167
$Cm'm2'$\else
\ifnum#2=168
$Cm'm'2$\else
\ifnum#2=169
$C_{c}mm2$\else
\ifnum#2=170
$C_{a}mm2$\else
\ifnum#2=171
$C_{A}mm2$\else
{\color{red}{Invalid MSG number}}
\fi
\fi
\fi
\fi
\fi
\fi
\fi
\else
\ifnum#1=36
\ifnum#2=172
$Cmc2_{1}$\else
\ifnum#2=173
$Cmc2_{1}1'$\else
\ifnum#2=174
$Cm'c2_{1}'$\else
\ifnum#2=175
$Cmc'2_{1}'$\else
\ifnum#2=176
$Cm'c'2_{1}$\else
\ifnum#2=177
$C_{c}mc2_{1}$\else
\ifnum#2=178
$C_{a}mc2_{1}$\else
\ifnum#2=179
$C_{A}mc2_{1}$\else
{\color{red}{Invalid MSG number}}
\fi
\fi
\fi
\fi
\fi
\fi
\fi
\fi
\else
\ifnum#1=37
\ifnum#2=180
$Ccc2$\else
\ifnum#2=181
$Ccc21'$\else
\ifnum#2=182
$Cc'c2'$\else
\ifnum#2=183
$Cc'c'2$\else
\ifnum#2=184
$C_{c}cc2$\else
\ifnum#2=185
$C_{a}cc2$\else
\ifnum#2=186
$C_{A}cc2$\else
{\color{red}{Invalid MSG number}}
\fi
\fi
\fi
\fi
\fi
\fi
\fi
\else
\ifnum#1=38
\ifnum#2=187
$Amm2$\else
\ifnum#2=188
$Amm21'$\else
\ifnum#2=189
$Am'm2'$\else
\ifnum#2=190
$Amm'2'$\else
\ifnum#2=191
$Am'm'2$\else
\ifnum#2=192
$A_{a}mm2$\else
\ifnum#2=193
$A_{b}mm2$\else
\ifnum#2=194
$A_{B}mm2$\else
{\color{red}{Invalid MSG number}}
\fi
\fi
\fi
\fi
\fi
\fi
\fi
\fi
\else
\ifnum#1=39
\ifnum#2=195
$Abm2$\else
\ifnum#2=196
$Abm21'$\else
\ifnum#2=197
$Ab'm2'$\else
\ifnum#2=198
$Abm'2'$\else
\ifnum#2=199
$Ab'm'2$\else
\ifnum#2=200
$A_{a}bm2$\else
\ifnum#2=201
$A_{b}bm2$\else
\ifnum#2=202
$A_{B}bm2$\else
{\color{red}{Invalid MSG number}}
\fi
\fi
\fi
\fi
\fi
\fi
\fi
\fi
\else
\ifnum#1=40
\ifnum#2=203
$Ama2$\else
\ifnum#2=204
$Ama21'$\else
\ifnum#2=205
$Am'a2'$\else
\ifnum#2=206
$Ama'2'$\else
\ifnum#2=207
$Am'a'2$\else
\ifnum#2=208
$A_{a}ma2$\else
\ifnum#2=209
$A_{b}ma2$\else
\ifnum#2=210
$A_{B}ma2$\else
{\color{red}{Invalid MSG number}}
\fi
\fi
\fi
\fi
\fi
\fi
\fi
\fi
\else
\ifnum#1=41
\ifnum#2=211
$Aba2$\else
\ifnum#2=212
$Aba21'$\else
\ifnum#2=213
$Ab'a2'$\else
\ifnum#2=214
$Aba'2'$\else
\ifnum#2=215
$Ab'a'2$\else
\ifnum#2=216
$A_{a}ba2$\else
\ifnum#2=217
$A_{b}ba2$\else
\ifnum#2=218
$A_{B}ba2$\else
{\color{red}{Invalid MSG number}}
\fi
\fi
\fi
\fi
\fi
\fi
\fi
\fi
\else
\ifnum#1=42
\ifnum#2=219
$Fmm2$\else
\ifnum#2=220
$Fmm21'$\else
\ifnum#2=221
$Fm'm2'$\else
\ifnum#2=222
$Fm'm'2$\else
\ifnum#2=223
$F_{S}mm2$\else
{\color{red}{Invalid MSG number}}
\fi
\fi
\fi
\fi
\fi
\else
\ifnum#1=43
\ifnum#2=224
$Fdd2$\else
\ifnum#2=225
$Fdd21'$\else
\ifnum#2=226
$Fd'd2'$\else
\ifnum#2=227
$Fd'd'2$\else
\ifnum#2=228
$F_{S}dd2$\else
{\color{red}{Invalid MSG number}}
\fi
\fi
\fi
\fi
\fi
\else
\ifnum#1=44
\ifnum#2=229
$Imm2$\else
\ifnum#2=230
$Imm21'$\else
\ifnum#2=231
$Im'm2'$\else
\ifnum#2=232
$Im'm'2$\else
\ifnum#2=233
$I_{c}mm2$\else
\ifnum#2=234
$I_{a}mm2$\else
{\color{red}{Invalid MSG number}}
\fi
\fi
\fi
\fi
\fi
\fi
\else
\ifnum#1=45
\ifnum#2=235
$Iba2$\else
\ifnum#2=236
$Iba21'$\else
\ifnum#2=237
$Ib'a2'$\else
\ifnum#2=238
$Ib'a'2$\else
\ifnum#2=239
$I_{c}ba2$\else
\ifnum#2=240
$I_{a}ba2$\else
{\color{red}{Invalid MSG number}}
\fi
\fi
\fi
\fi
\fi
\fi
\else
\ifnum#1=46
\ifnum#2=241
$Ima2$\else
\ifnum#2=242
$Ima21'$\else
\ifnum#2=243
$Im'a2'$\else
\ifnum#2=244
$Ima'2'$\else
\ifnum#2=245
$Im'a'2$\else
\ifnum#2=246
$I_{c}ma2$\else
\ifnum#2=247
$I_{a}ma2$\else
\ifnum#2=248
$I_{b}ma2$\else
{\color{red}{Invalid MSG number}}
\fi
\fi
\fi
\fi
\fi
\fi
\fi
\fi
\else
\ifnum#1=47
\ifnum#2=249
$Pmmm$\else
\ifnum#2=250
$Pmmm1'$\else
\ifnum#2=251
$Pm'mm$\else
\ifnum#2=252
$Pm'm'm$\else
\ifnum#2=253
$Pm'm'm'$\else
\ifnum#2=254
$P_{a}mmm$\else
\ifnum#2=255
$P_{C}mmm$\else
\ifnum#2=256
$P_{I}mmm$\else
{\color{red}{Invalid MSG number}}
\fi
\fi
\fi
\fi
\fi
\fi
\fi
\fi
\else
\ifnum#1=48
\ifnum#2=257
$Pnnn$\else
\ifnum#2=258
$Pnnn1'$\else
\ifnum#2=259
$Pn'nn$\else
\ifnum#2=260
$Pn'n'n$\else
\ifnum#2=261
$Pn'n'n'$\else
\ifnum#2=262
$P_{c}nnn$\else
\ifnum#2=263
$P_{C}nnn$\else
\ifnum#2=264
$P_{I}nnn$\else
{\color{red}{Invalid MSG number}}
\fi
\fi
\fi
\fi
\fi
\fi
\fi
\fi
\else
\ifnum#1=49
\ifnum#2=265
$Pccm$\else
\ifnum#2=266
$Pccm1'$\else
\ifnum#2=267
$Pc'cm$\else
\ifnum#2=268
$Pccm'$\else
\ifnum#2=269
$Pc'c'm$\else
\ifnum#2=270
$Pc'cm'$\else
\ifnum#2=271
$Pc'c'm'$\else
\ifnum#2=272
$P_{a}ccm$\else
\ifnum#2=273
$P_{c}ccm$\else
\ifnum#2=274
$P_{B}ccm$\else
\ifnum#2=275
$P_{C}ccm$\else
\ifnum#2=276
$P_{I}ccm$\else
{\color{red}{Invalid MSG number}}
\fi
\fi
\fi
\fi
\fi
\fi
\fi
\fi
\fi
\fi
\fi
\fi
\else
\ifnum#1=50
\ifnum#2=277
$Pban$\else
\ifnum#2=278
$Pban1'$\else
\ifnum#2=279
$Pb'an$\else
\ifnum#2=280
$Pban'$\else
\ifnum#2=281
$Pb'a'n$\else
\ifnum#2=282
$Pb'an'$\else
\ifnum#2=283
$Pb'a'n'$\else
\ifnum#2=284
$P_{a}ban$\else
\ifnum#2=285
$P_{c}ban$\else
\ifnum#2=286
$P_{A}ban$\else
\ifnum#2=287
$P_{C}ban$\else
\ifnum#2=288
$P_{I}ban$\else
{\color{red}{Invalid MSG number}}
\fi
\fi
\fi
\fi
\fi
\fi
\fi
\fi
\fi
\fi
\fi
\fi
\else
\ifnum#1=51
\ifnum#2=289
$Pmma$\else
\ifnum#2=290
$Pmma1'$\else
\ifnum#2=291
$Pm'ma$\else
\ifnum#2=292
$Pmm'a$\else
\ifnum#2=293
$Pmma'$\else
\ifnum#2=294
$Pm'm'a$\else
\ifnum#2=295
$Pmm'a'$\else
\ifnum#2=296
$Pm'ma'$\else
\ifnum#2=297
$Pm'm'a'$\else
\ifnum#2=298
$P_{a}mma$\else
\ifnum#2=299
$P_{b}mma$\else
\ifnum#2=300
$P_{c}mma$\else
\ifnum#2=301
$P_{A}mma$\else
\ifnum#2=302
$P_{B}mma$\else
\ifnum#2=303
$P_{C}mma$\else
\ifnum#2=304
$P_{I}mma$\else
{\color{red}{Invalid MSG number}}
\fi
\fi
\fi
\fi
\fi
\fi
\fi
\fi
\fi
\fi
\fi
\fi
\fi
\fi
\fi
\fi
\else
\ifnum#1=52
\ifnum#2=305
$Pnna$\else
\ifnum#2=306
$Pnna1'$\else
\ifnum#2=307
$Pn'na$\else
\ifnum#2=308
$Pnn'a$\else
\ifnum#2=309
$Pnna'$\else
\ifnum#2=310
$Pn'n'a$\else
\ifnum#2=311
$Pnn'a'$\else
\ifnum#2=312
$Pn'na'$\else
\ifnum#2=313
$Pn'n'a'$\else
\ifnum#2=314
$P_{a}nna$\else
\ifnum#2=315
$P_{b}nna$\else
\ifnum#2=316
$P_{c}nna$\else
\ifnum#2=317
$P_{A}nna$\else
\ifnum#2=318
$P_{B}nna$\else
\ifnum#2=319
$P_{C}nna$\else
\ifnum#2=320
$P_{I}nna$\else
{\color{red}{Invalid MSG number}}
\fi
\fi
\fi
\fi
\fi
\fi
\fi
\fi
\fi
\fi
\fi
\fi
\fi
\fi
\fi
\fi
\else
\ifnum#1=53
\ifnum#2=321
$Pmna$\else
\ifnum#2=322
$Pmna1'$\else
\ifnum#2=323
$Pm'na$\else
\ifnum#2=324
$Pmn'a$\else
\ifnum#2=325
$Pmna'$\else
\ifnum#2=326
$Pm'n'a$\else
\ifnum#2=327
$Pmn'a'$\else
\ifnum#2=328
$Pm'na'$\else
\ifnum#2=329
$Pm'n'a'$\else
\ifnum#2=330
$P_{a}mna$\else
\ifnum#2=331
$P_{b}mna$\else
\ifnum#2=332
$P_{c}mna$\else
\ifnum#2=333
$P_{A}mna$\else
\ifnum#2=334
$P_{B}mna$\else
\ifnum#2=335
$P_{C}mna$\else
\ifnum#2=336
$P_{I}mna$\else
{\color{red}{Invalid MSG number}}
\fi
\fi
\fi
\fi
\fi
\fi
\fi
\fi
\fi
\fi
\fi
\fi
\fi
\fi
\fi
\fi
\else
\ifnum#1=54
\ifnum#2=337
$Pcca$\else
\ifnum#2=338
$Pcca1'$\else
\ifnum#2=339
$Pc'ca$\else
\ifnum#2=340
$Pcc'a$\else
\ifnum#2=341
$Pcca'$\else
\ifnum#2=342
$Pc'c'a$\else
\ifnum#2=343
$Pcc'a'$\else
\ifnum#2=344
$Pc'ca'$\else
\ifnum#2=345
$Pc'c'a'$\else
\ifnum#2=346
$P_{a}cca$\else
\ifnum#2=347
$P_{b}cca$\else
\ifnum#2=348
$P_{c}cca$\else
\ifnum#2=349
$P_{A}cca$\else
\ifnum#2=350
$P_{B}cca$\else
\ifnum#2=351
$P_{C}cca$\else
\ifnum#2=352
$P_{I}cca$\else
{\color{red}{Invalid MSG number}}
\fi
\fi
\fi
\fi
\fi
\fi
\fi
\fi
\fi
\fi
\fi
\fi
\fi
\fi
\fi
\fi
\else
\ifnum#1=55
\ifnum#2=353
$Pbam$\else
\ifnum#2=354
$Pbam1'$\else
\ifnum#2=355
$Pb'am$\else
\ifnum#2=356
$Pbam'$\else
\ifnum#2=357
$Pb'a'm$\else
\ifnum#2=358
$Pb'am'$\else
\ifnum#2=359
$Pb'a'm'$\else
\ifnum#2=360
$P_{a}bam$\else
\ifnum#2=361
$P_{c}bam$\else
\ifnum#2=362
$P_{A}bam$\else
\ifnum#2=363
$P_{C}bam$\else
\ifnum#2=364
$P_{I}bam$\else
{\color{red}{Invalid MSG number}}
\fi
\fi
\fi
\fi
\fi
\fi
\fi
\fi
\fi
\fi
\fi
\fi
\else
\ifnum#1=56
\ifnum#2=365
$Pccn$\else
\ifnum#2=366
$Pccn1'$\else
\ifnum#2=367
$Pc'cn$\else
\ifnum#2=368
$Pccn'$\else
\ifnum#2=369
$Pc'c'n$\else
\ifnum#2=370
$Pc'cn'$\else
\ifnum#2=371
$Pc'c'n'$\else
\ifnum#2=372
$P_{b}ccn$\else
\ifnum#2=373
$P_{c}ccn$\else
\ifnum#2=374
$P_{A}ccn$\else
\ifnum#2=375
$P_{C}ccn$\else
\ifnum#2=376
$P_{I}ccn$\else
{\color{red}{Invalid MSG number}}
\fi
\fi
\fi
\fi
\fi
\fi
\fi
\fi
\fi
\fi
\fi
\fi
\else
\ifnum#1=57
\ifnum#2=377
$Pbcm$\else
\ifnum#2=378
$Pbcm1'$\else
\ifnum#2=379
$Pb'cm$\else
\ifnum#2=380
$Pbc'm$\else
\ifnum#2=381
$Pbcm'$\else
\ifnum#2=382
$Pb'c'm$\else
\ifnum#2=383
$Pbc'm'$\else
\ifnum#2=384
$Pb'cm'$\else
\ifnum#2=385
$Pb'c'm'$\else
\ifnum#2=386
$P_{a}bcm$\else
\ifnum#2=387
$P_{b}bcm$\else
\ifnum#2=388
$P_{c}bcm$\else
\ifnum#2=389
$P_{A}bcm$\else
\ifnum#2=390
$P_{B}bcm$\else
\ifnum#2=391
$P_{C}bcm$\else
\ifnum#2=392
$P_{I}bcm$\else
{\color{red}{Invalid MSG number}}
\fi
\fi
\fi
\fi
\fi
\fi
\fi
\fi
\fi
\fi
\fi
\fi
\fi
\fi
\fi
\fi
\else
\ifnum#1=58
\ifnum#2=393
$Pnnm$\else
\ifnum#2=394
$Pnnm1'$\else
\ifnum#2=395
$Pn'nm$\else
\ifnum#2=396
$Pnnm'$\else
\ifnum#2=397
$Pn'n'm$\else
\ifnum#2=398
$Pnn'm'$\else
\ifnum#2=399
$Pn'n'm'$\else
\ifnum#2=400
$P_{a}nnm$\else
\ifnum#2=401
$P_{c}nnm$\else
\ifnum#2=402
$P_{B}nnm$\else
\ifnum#2=403
$P_{C}nnm$\else
\ifnum#2=404
$P_{I}nnm$\else
{\color{red}{Invalid MSG number}}
\fi
\fi
\fi
\fi
\fi
\fi
\fi
\fi
\fi
\fi
\fi
\fi
\else
\ifnum#1=59
\ifnum#2=405
$Pmmn$\else
\ifnum#2=406
$Pmmn1'$\else
\ifnum#2=407
$Pm'mn$\else
\ifnum#2=408
$Pmmn'$\else
\ifnum#2=409
$Pm'm'n$\else
\ifnum#2=410
$Pmm'n'$\else
\ifnum#2=411
$Pm'm'n'$\else
\ifnum#2=412
$P_{b}mmn$\else
\ifnum#2=413
$P_{c}mmn$\else
\ifnum#2=414
$P_{B}mmn$\else
\ifnum#2=415
$P_{C}mmn$\else
\ifnum#2=416
$P_{I}mmn$\else
{\color{red}{Invalid MSG number}}
\fi
\fi
\fi
\fi
\fi
\fi
\fi
\fi
\fi
\fi
\fi
\fi
\else
\ifnum#1=60
\ifnum#2=417
$Pbcn$\else
\ifnum#2=418
$Pbcn1'$\else
\ifnum#2=419
$Pb'cn$\else
\ifnum#2=420
$Pbc'n$\else
\ifnum#2=421
$Pbcn'$\else
\ifnum#2=422
$Pb'c'n$\else
\ifnum#2=423
$Pbc'n'$\else
\ifnum#2=424
$Pb'cn'$\else
\ifnum#2=425
$Pb'c'n'$\else
\ifnum#2=426
$P_{a}bcn$\else
\ifnum#2=427
$P_{b}bcn$\else
\ifnum#2=428
$P_{c}bcn$\else
\ifnum#2=429
$P_{A}bcn$\else
\ifnum#2=430
$P_{B}bcn$\else
\ifnum#2=431
$P_{C}bcn$\else
\ifnum#2=432
$P_{I}bcn$\else
{\color{red}{Invalid MSG number}}
\fi
\fi
\fi
\fi
\fi
\fi
\fi
\fi
\fi
\fi
\fi
\fi
\fi
\fi
\fi
\fi
\else
\ifnum#1=61
\ifnum#2=433
$Pbca$\else
\ifnum#2=434
$Pbca1'$\else
\ifnum#2=435
$Pb'ca$\else
\ifnum#2=436
$Pb'c'a$\else
\ifnum#2=437
$Pb'c'a'$\else
\ifnum#2=438
$P_{a}bca$\else
\ifnum#2=439
$P_{C}bca$\else
\ifnum#2=440
$P_{I}bca$\else
{\color{red}{Invalid MSG number}}
\fi
\fi
\fi
\fi
\fi
\fi
\fi
\fi
\else
\ifnum#1=62
\ifnum#2=441
$Pnma$\else
\ifnum#2=442
$Pnma1'$\else
\ifnum#2=443
$Pn'ma$\else
\ifnum#2=444
$Pnm'a$\else
\ifnum#2=445
$Pnma'$\else
\ifnum#2=446
$Pn'm'a$\else
\ifnum#2=447
$Pnm'a'$\else
\ifnum#2=448
$Pn'ma'$\else
\ifnum#2=449
$Pn'm'a'$\else
\ifnum#2=450
$P_{a}nma$\else
\ifnum#2=451
$P_{b}nma$\else
\ifnum#2=452
$P_{c}nma$\else
\ifnum#2=453
$P_{A}nma$\else
\ifnum#2=454
$P_{B}nma$\else
\ifnum#2=455
$P_{C}nma$\else
\ifnum#2=456
$P_{I}nma$\else
{\color{red}{Invalid MSG number}}
\fi
\fi
\fi
\fi
\fi
\fi
\fi
\fi
\fi
\fi
\fi
\fi
\fi
\fi
\fi
\fi
\else
\ifnum#1=63
\ifnum#2=457
$Cmcm$\else
\ifnum#2=458
$Cmcm1'$\else
\ifnum#2=459
$Cm'cm$\else
\ifnum#2=460
$Cmc'm$\else
\ifnum#2=461
$Cmcm'$\else
\ifnum#2=462
$Cm'c'm$\else
\ifnum#2=463
$Cmc'm'$\else
\ifnum#2=464
$Cm'cm'$\else
\ifnum#2=465
$Cm'c'm'$\else
\ifnum#2=466
$C_{c}mcm$\else
\ifnum#2=467
$C_{a}mcm$\else
\ifnum#2=468
$C_{A}mcm$\else
{\color{red}{Invalid MSG number}}
\fi
\fi
\fi
\fi
\fi
\fi
\fi
\fi
\fi
\fi
\fi
\fi
\else
\ifnum#1=64
\ifnum#2=469
$Cmca$\else
\ifnum#2=470
$Cmca1'$\else
\ifnum#2=471
$Cm'ca$\else
\ifnum#2=472
$Cmc'a$\else
\ifnum#2=473
$Cmca'$\else
\ifnum#2=474
$Cm'c'a$\else
\ifnum#2=475
$Cmc'a'$\else
\ifnum#2=476
$Cm'ca'$\else
\ifnum#2=477
$Cm'c'a'$\else
\ifnum#2=478
$C_{c}mca$\else
\ifnum#2=479
$C_{a}mca$\else
\ifnum#2=480
$C_{A}mca$\else
{\color{red}{Invalid MSG number}}
\fi
\fi
\fi
\fi
\fi
\fi
\fi
\fi
\fi
\fi
\fi
\fi
\else
\ifnum#1=65
\ifnum#2=481
$Cmmm$\else
\ifnum#2=482
$Cmmm1'$\else
\ifnum#2=483
$Cm'mm$\else
\ifnum#2=484
$Cmmm'$\else
\ifnum#2=485
$Cm'm'm$\else
\ifnum#2=486
$Cmm'm'$\else
\ifnum#2=487
$Cm'm'm'$\else
\ifnum#2=488
$C_{c}mmm$\else
\ifnum#2=489
$C_{a}mmm$\else
\ifnum#2=490
$C_{A}mmm$\else
{\color{red}{Invalid MSG number}}
\fi
\fi
\fi
\fi
\fi
\fi
\fi
\fi
\fi
\fi
\else
\ifnum#1=66
\ifnum#2=491
$Cccm$\else
\ifnum#2=492
$Cccm1'$\else
\ifnum#2=493
$Cc'cm$\else
\ifnum#2=494
$Cccm'$\else
\ifnum#2=495
$Cc'c'm$\else
\ifnum#2=496
$Ccc'm'$\else
\ifnum#2=497
$Cc'c'm'$\else
\ifnum#2=498
$C_{c}ccm$\else
\ifnum#2=499
$C_{a}ccm$\else
\ifnum#2=500
$C_{A}ccm$\else
{\color{red}{Invalid MSG number}}
\fi
\fi
\fi
\fi
\fi
\fi
\fi
\fi
\fi
\fi
\else
\ifnum#1=67
\ifnum#2=501
$Cmma$\else
\ifnum#2=502
$Cmma1'$\else
\ifnum#2=503
$Cm'ma$\else
\ifnum#2=504
$Cmma'$\else
\ifnum#2=505
$Cm'm'a$\else
\ifnum#2=506
$Cmm'a'$\else
\ifnum#2=507
$Cm'm'a'$\else
\ifnum#2=508
$C_{c}mma$\else
\ifnum#2=509
$C_{a}mma$\else
\ifnum#2=510
$C_{A}mma$\else
{\color{red}{Invalid MSG number}}
\fi
\fi
\fi
\fi
\fi
\fi
\fi
\fi
\fi
\fi
\else
\ifnum#1=68
\ifnum#2=511
$Ccca$\else
\ifnum#2=512
$Ccca1'$\else
\ifnum#2=513
$Cc'ca$\else
\ifnum#2=514
$Ccca'$\else
\ifnum#2=515
$Cc'c'a$\else
\ifnum#2=516
$Ccc'a'$\else
\ifnum#2=517
$Cc'c'a'$\else
\ifnum#2=518
$C_{c}cca$\else
\ifnum#2=519
$C_{a}cca$\else
\ifnum#2=520
$C_{A}cca$\else
{\color{red}{Invalid MSG number}}
\fi
\fi
\fi
\fi
\fi
\fi
\fi
\fi
\fi
\fi
\else
\ifnum#1=69
\ifnum#2=521
$Fmmm$\else
\ifnum#2=522
$Fmmm1'$\else
\ifnum#2=523
$Fm'mm$\else
\ifnum#2=524
$Fm'm'm$\else
\ifnum#2=525
$Fm'm'm'$\else
\ifnum#2=526
$F_{S}mmm$\else
{\color{red}{Invalid MSG number}}
\fi
\fi
\fi
\fi
\fi
\fi
\else
\ifnum#1=70
\ifnum#2=527
$Fddd$\else
\ifnum#2=528
$Fddd1'$\else
\ifnum#2=529
$Fd'dd$\else
\ifnum#2=530
$Fd'd'd$\else
\ifnum#2=531
$Fd'd'd'$\else
\ifnum#2=532
$F_{S}ddd$\else
{\color{red}{Invalid MSG number}}
\fi
\fi
\fi
\fi
\fi
\fi
\else
\ifnum#1=71
\ifnum#2=533
$Immm$\else
\ifnum#2=534
$Immm1'$\else
\ifnum#2=535
$Im'mm$\else
\ifnum#2=536
$Im'm'm$\else
\ifnum#2=537
$Im'm'm'$\else
\ifnum#2=538
$I_{c}mmm$\else
{\color{red}{Invalid MSG number}}
\fi
\fi
\fi
\fi
\fi
\fi
\else
\ifnum#1=72
\ifnum#2=539
$Ibam$\else
\ifnum#2=540
$Ibam1'$\else
\ifnum#2=541
$Ib'am$\else
\ifnum#2=542
$Ibam'$\else
\ifnum#2=543
$Ib'a'm$\else
\ifnum#2=544
$Iba'm'$\else
\ifnum#2=545
$Ib'a'm'$\else
\ifnum#2=546
$I_{c}bam$\else
\ifnum#2=547
$I_{b}bam$\else
{\color{red}{Invalid MSG number}}
\fi
\fi
\fi
\fi
\fi
\fi
\fi
\fi
\fi
\else
\ifnum#1=73
\ifnum#2=548
$Ibca$\else
\ifnum#2=549
$Ibca1'$\else
\ifnum#2=550
$Ib'ca$\else
\ifnum#2=551
$Ib'c'a$\else
\ifnum#2=552
$Ib'c'a'$\else
\ifnum#2=553
$I_{c}bca$\else
{\color{red}{Invalid MSG number}}
\fi
\fi
\fi
\fi
\fi
\fi
\else
\ifnum#1=74
\ifnum#2=554
$Imma$\else
\ifnum#2=555
$Imma1'$\else
\ifnum#2=556
$Im'ma$\else
\ifnum#2=557
$Imma'$\else
\ifnum#2=558
$Im'm'a$\else
\ifnum#2=559
$Imm'a'$\else
\ifnum#2=560
$Im'm'a'$\else
\ifnum#2=561
$I_{c}mma$\else
\ifnum#2=562
$I_{b}mma$\else
{\color{red}{Invalid MSG number}}
\fi
\fi
\fi
\fi
\fi
\fi
\fi
\fi
\fi
\else
\ifnum#1=75
\ifnum#2=1
$P4$\else
\ifnum#2=2
$P41'$\else
\ifnum#2=3
$P4'$\else
\ifnum#2=4
$P_{c}4$\else
\ifnum#2=5
$P_{C}4$\else
\ifnum#2=6
$P_{I}4$\else
{\color{red}{Invalid MSG number}}
\fi
\fi
\fi
\fi
\fi
\fi
\else
\ifnum#1=76
\ifnum#2=7
$P4_{1}$\else
\ifnum#2=8
$P4_{1}1'$\else
\ifnum#2=9
$P4_{1}'$\else
\ifnum#2=10
$P_{c}4_{1}$\else
\ifnum#2=11
$P_{C}4_{1}$\else
\ifnum#2=12
$P_{I}4_{1}$\else
{\color{red}{Invalid MSG number}}
\fi
\fi
\fi
\fi
\fi
\fi
\else
\ifnum#1=77
\ifnum#2=13
$P4_{2}$\else
\ifnum#2=14
$P4_{2}1'$\else
\ifnum#2=15
$P4_{2}'$\else
\ifnum#2=16
$P_{c}4_{2}$\else
\ifnum#2=17
$P_{C}4_{2}$\else
\ifnum#2=18
$P_{I}4_{2}$\else
{\color{red}{Invalid MSG number}}
\fi
\fi
\fi
\fi
\fi
\fi
\else
\ifnum#1=78
\ifnum#2=19
$P4_{3}$\else
\ifnum#2=20
$P4_{3}1'$\else
\ifnum#2=21
$P4_{3}'$\else
\ifnum#2=22
$P_{c}4_{3}$\else
\ifnum#2=23
$P_{C}4_{3}$\else
\ifnum#2=24
$P_{I}4_{3}$\else
{\color{red}{Invalid MSG number}}
\fi
\fi
\fi
\fi
\fi
\fi
\else
\ifnum#1=79
\ifnum#2=25
$I4$\else
\ifnum#2=26
$I41'$\else
\ifnum#2=27
$I4'$\else
\ifnum#2=28
$I_{c}4$\else
{\color{red}{Invalid MSG number}}
\fi
\fi
\fi
\fi
\else
\ifnum#1=80
\ifnum#2=29
$I4_{1}$\else
\ifnum#2=30
$I4_{1}1'$\else
\ifnum#2=31
$I4_{1}'$\else
\ifnum#2=32
$I_{c}4_{1}$\else
{\color{red}{Invalid MSG number}}
\fi
\fi
\fi
\fi
\else
\ifnum#1=81
\ifnum#2=33
$P\bar{4}$\else
\ifnum#2=34
$P\bar{4}1'$\else
\ifnum#2=35
$P\bar{4}'$\else
\ifnum#2=36
$P_{c}\bar{4}$\else
\ifnum#2=37
$P_{C}\bar{4}$\else
\ifnum#2=38
$P_{I}\bar{4}$\else
{\color{red}{Invalid MSG number}}
\fi
\fi
\fi
\fi
\fi
\fi
\else
\ifnum#1=82
\ifnum#2=39
$I\bar{4}$\else
\ifnum#2=40
$I\bar{4}1'$\else
\ifnum#2=41
$I\bar{4}'$\else
\ifnum#2=42
$I_{c}\bar{4}$\else
{\color{red}{Invalid MSG number}}
\fi
\fi
\fi
\fi
\else
\ifnum#1=83
\ifnum#2=43
$P4/m$\else
\ifnum#2=44
$P4/m1'$\else
\ifnum#2=45
$P4'/m$\else
\ifnum#2=46
$P4/m'$\else
\ifnum#2=47
$P4'/m'$\else
\ifnum#2=48
$P_{c}4/m$\else
\ifnum#2=49
$P_{C}4/m$\else
\ifnum#2=50
$P_{I}4/m$\else
{\color{red}{Invalid MSG number}}
\fi
\fi
\fi
\fi
\fi
\fi
\fi
\fi
\else
\ifnum#1=84
\ifnum#2=51
$P4_{2}/m$\else
\ifnum#2=52
$P4_{2}/m1'$\else
\ifnum#2=53
$P4_{2}'/m$\else
\ifnum#2=54
$P4_{2}/m'$\else
\ifnum#2=55
$P4_{2}'/m'$\else
\ifnum#2=56
$P_{c}4_{2}/m$\else
\ifnum#2=57
$P_{C}4_{2}/m$\else
\ifnum#2=58
$P_{I}4_{2}/m$\else
{\color{red}{Invalid MSG number}}
\fi
\fi
\fi
\fi
\fi
\fi
\fi
\fi
\else
\ifnum#1=85
\ifnum#2=59
$P4/n$\else
\ifnum#2=60
$P4/n1'$\else
\ifnum#2=61
$P4'/n$\else
\ifnum#2=62
$P4/n'$\else
\ifnum#2=63
$P4'/n'$\else
\ifnum#2=64
$P_{c}4/n$\else
\ifnum#2=65
$P_{C}4/n$\else
\ifnum#2=66
$P_{I}4/n$\else
{\color{red}{Invalid MSG number}}
\fi
\fi
\fi
\fi
\fi
\fi
\fi
\fi
\else
\ifnum#1=86
\ifnum#2=67
$P4_{2}/n$\else
\ifnum#2=68
$P4_{2}/n1'$\else
\ifnum#2=69
$P4_{2}'/n$\else
\ifnum#2=70
$P4_{2}/n'$\else
\ifnum#2=71
$P4_{2}'/n'$\else
\ifnum#2=72
$P_{c}4_{2}/n$\else
\ifnum#2=73
$P_{C}4_{2}/n$\else
\ifnum#2=74
$P_{I}4_{2}/n$\else
{\color{red}{Invalid MSG number}}
\fi
\fi
\fi
\fi
\fi
\fi
\fi
\fi
\else
\ifnum#1=87
\ifnum#2=75
$I4/m$\else
\ifnum#2=76
$I4/m1'$\else
\ifnum#2=77
$I4'/m$\else
\ifnum#2=78
$I4/m'$\else
\ifnum#2=79
$I4'/m'$\else
\ifnum#2=80
$I_{c}4/m$\else
{\color{red}{Invalid MSG number}}
\fi
\fi
\fi
\fi
\fi
\fi
\else
\ifnum#1=88
\ifnum#2=81
$I4_{1}/a$\else
\ifnum#2=82
$I4_{1}/a1'$\else
\ifnum#2=83
$I4_{1}'/a$\else
\ifnum#2=84
$I4_{1}/a'$\else
\ifnum#2=85
$I4_{1}'/a'$\else
\ifnum#2=86
$I_{c}4_{1}/a$\else
{\color{red}{Invalid MSG number}}
\fi
\fi
\fi
\fi
\fi
\fi
\else
\ifnum#1=89
\ifnum#2=87
$P422$\else
\ifnum#2=88
$P4221'$\else
\ifnum#2=89
$P4'22'$\else
\ifnum#2=90
$P42'2'$\else
\ifnum#2=91
$P4'2'2$\else
\ifnum#2=92
$P_{c}422$\else
\ifnum#2=93
$P_{C}422$\else
\ifnum#2=94
$P_{I}422$\else
{\color{red}{Invalid MSG number}}
\fi
\fi
\fi
\fi
\fi
\fi
\fi
\fi
\else
\ifnum#1=90
\ifnum#2=95
$P42_{1}2$\else
\ifnum#2=96
$P42_{1}21'$\else
\ifnum#2=97
$P4'2_{1}2'$\else
\ifnum#2=98
$P42_{1}'2'$\else
\ifnum#2=99
$P4'2_{1}'2$\else
\ifnum#2=100
$P_{c}42_{1}2$\else
\ifnum#2=101
$P_{C}42_{1}2$\else
\ifnum#2=102
$P_{I}42_{1}2$\else
{\color{red}{Invalid MSG number}}
\fi
\fi
\fi
\fi
\fi
\fi
\fi
\fi
\else
\ifnum#1=91
\ifnum#2=103
$P4_{1}22$\else
\ifnum#2=104
$P4_{1}221'$\else
\ifnum#2=105
$P4_{1}'22'$\else
\ifnum#2=106
$P4_{1}2'2'$\else
\ifnum#2=107
$P4_{1}'2'2$\else
\ifnum#2=108
$P_{c}4_{1}22$\else
\ifnum#2=109
$P_{C}4_{1}22$\else
\ifnum#2=110
$P_{I}4_{1}22$\else
{\color{red}{Invalid MSG number}}
\fi
\fi
\fi
\fi
\fi
\fi
\fi
\fi
\else
\ifnum#1=92
\ifnum#2=111
$P4_{1}2_{1}2$\else
\ifnum#2=112
$P4_{1}2_{1}21'$\else
\ifnum#2=113
$P4_{1}'2_{1}2'$\else
\ifnum#2=114
$P4_{1}2_{1}'2'$\else
\ifnum#2=115
$P4_{1}'2_{1}'2$\else
\ifnum#2=116
$P_{c}4_{1}2_{1}2$\else
\ifnum#2=117
$P_{C}4_{1}2_{1}2$\else
\ifnum#2=118
$P_{I}4_{1}2_{1}2$\else
{\color{red}{Invalid MSG number}}
\fi
\fi
\fi
\fi
\fi
\fi
\fi
\fi
\else
\ifnum#1=93
\ifnum#2=119
$P4_{2}22$\else
\ifnum#2=120
$P4_{2}221'$\else
\ifnum#2=121
$P4_{2}'22'$\else
\ifnum#2=122
$P4_{2}2'2'$\else
\ifnum#2=123
$P4_{2}'2'2$\else
\ifnum#2=124
$P_{c}4_{2}22$\else
\ifnum#2=125
$P_{C}4_{2}22$\else
\ifnum#2=126
$P_{I}4_{2}22$\else
{\color{red}{Invalid MSG number}}
\fi
\fi
\fi
\fi
\fi
\fi
\fi
\fi
\else
\ifnum#1=94
\ifnum#2=127
$P4_{2}2_{1}2$\else
\ifnum#2=128
$P4_{2}2_{1}21'$\else
\ifnum#2=129
$P4_{2}'2_{1}2'$\else
\ifnum#2=130
$P4_{2}2_{1}'2'$\else
\ifnum#2=131
$P4_{2}'2_{1}'2$\else
\ifnum#2=132
$P_{c}4_{2}2_{1}2$\else
\ifnum#2=133
$P_{C}4_{2}2_{1}2$\else
\ifnum#2=134
$P_{I}4_{2}2_{1}2$\else
{\color{red}{Invalid MSG number}}
\fi
\fi
\fi
\fi
\fi
\fi
\fi
\fi
\else
\ifnum#1=95
\ifnum#2=135
$P4_{3}22$\else
\ifnum#2=136
$P4_{3}221'$\else
\ifnum#2=137
$P4_{3}'22'$\else
\ifnum#2=138
$P4_{3}2'2'$\else
\ifnum#2=139
$P4_{3}'2'2$\else
\ifnum#2=140
$P_{c}4_{3}22$\else
\ifnum#2=141
$P_{C}4_{3}22$\else
\ifnum#2=142
$P_{I}4_{3}22$\else
{\color{red}{Invalid MSG number}}
\fi
\fi
\fi
\fi
\fi
\fi
\fi
\fi
\else
\ifnum#1=96
\ifnum#2=143
$P4_{3}2_{1}2$\else
\ifnum#2=144
$P4_{3}2_{1}21'$\else
\ifnum#2=145
$P4_{3}'2_{1}2'$\else
\ifnum#2=146
$P4_{3}2_{1}'2'$\else
\ifnum#2=147
$P4_{3}'2_{1}'2$\else
\ifnum#2=148
$P_{c}4_{3}2_{1}2$\else
\ifnum#2=149
$P_{C}4_{3}2_{1}2$\else
\ifnum#2=150
$P_{I}4_{3}2_{1}2$\else
{\color{red}{Invalid MSG number}}
\fi
\fi
\fi
\fi
\fi
\fi
\fi
\fi
\else
\ifnum#1=97
\ifnum#2=151
$I422$\else
\ifnum#2=152
$I4221'$\else
\ifnum#2=153
$I4'22'$\else
\ifnum#2=154
$I42'2'$\else
\ifnum#2=155
$I4'2'2$\else
\ifnum#2=156
$I_{c}422$\else
{\color{red}{Invalid MSG number}}
\fi
\fi
\fi
\fi
\fi
\fi
\else
\ifnum#1=98
\ifnum#2=157
$I4_{1}22$\else
\ifnum#2=158
$I4_{1}221'$\else
\ifnum#2=159
$I4_{1}'22'$\else
\ifnum#2=160
$I4_{1}2'2'$\else
\ifnum#2=161
$I4_{1}'2'2$\else
\ifnum#2=162
$I_{c}4_{1}22$\else
{\color{red}{Invalid MSG number}}
\fi
\fi
\fi
\fi
\fi
\fi
\else
\ifnum#1=99
\ifnum#2=163
$P4mm$\else
\ifnum#2=164
$P4mm1'$\else
\ifnum#2=165
$P4'm'm$\else
\ifnum#2=166
$P4'mm'$\else
\ifnum#2=167
$P4m'm'$\else
\ifnum#2=168
$P_{c}4mm$\else
\ifnum#2=169
$P_{C}4mm$\else
\ifnum#2=170
$P_{I}4mm$\else
{\color{red}{Invalid MSG number}}
\fi
\fi
\fi
\fi
\fi
\fi
\fi
\fi
\else
\ifnum#1=100
\ifnum#2=171
$P4bm$\else
\ifnum#2=172
$P4bm1'$\else
\ifnum#2=173
$P4'b'm$\else
\ifnum#2=174
$P4'bm'$\else
\ifnum#2=175
$P4b'm'$\else
\ifnum#2=176
$P_{c}4bm$\else
\ifnum#2=177
$P_{C}4bm$\else
\ifnum#2=178
$P_{I}4bm$\else
{\color{red}{Invalid MSG number}}
\fi
\fi
\fi
\fi
\fi
\fi
\fi
\fi
\else
\ifnum#1=101
\ifnum#2=179
$P4_{2}cm$\else
\ifnum#2=180
$P4_{2}cm1'$\else
\ifnum#2=181
$P4_{2}'c'm$\else
\ifnum#2=182
$P4_{2}'cm'$\else
\ifnum#2=183
$P4_{2}c'm'$\else
\ifnum#2=184
$P_{c}4_{2}cm$\else
\ifnum#2=185
$P_{C}4_{2}cm$\else
\ifnum#2=186
$P_{I}4_{2}cm$\else
{\color{red}{Invalid MSG number}}
\fi
\fi
\fi
\fi
\fi
\fi
\fi
\fi
\else
\ifnum#1=102
\ifnum#2=187
$P4_{2}nm$\else
\ifnum#2=188
$P4_{2}nm1'$\else
\ifnum#2=189
$P4_{2}'n'm$\else
\ifnum#2=190
$P4_{2}'nm'$\else
\ifnum#2=191
$P4_{2}n'm'$\else
\ifnum#2=192
$P_{c}4_{2}nm$\else
\ifnum#2=193
$P_{C}4_{2}nm$\else
\ifnum#2=194
$P_{I}4_{2}nm$\else
{\color{red}{Invalid MSG number}}
\fi
\fi
\fi
\fi
\fi
\fi
\fi
\fi
\else
\ifnum#1=103
\ifnum#2=195
$P4cc$\else
\ifnum#2=196
$P4cc1'$\else
\ifnum#2=197
$P4'c'c$\else
\ifnum#2=198
$P4'cc'$\else
\ifnum#2=199
$P4c'c'$\else
\ifnum#2=200
$P_{c}4cc$\else
\ifnum#2=201
$P_{C}4cc$\else
\ifnum#2=202
$P_{I}4cc$\else
{\color{red}{Invalid MSG number}}
\fi
\fi
\fi
\fi
\fi
\fi
\fi
\fi
\else
\ifnum#1=104
\ifnum#2=203
$P4nc$\else
\ifnum#2=204
$P4nc1'$\else
\ifnum#2=205
$P4'n'c$\else
\ifnum#2=206
$P4'nc'$\else
\ifnum#2=207
$P4n'c'$\else
\ifnum#2=208
$P_{c}4nc$\else
\ifnum#2=209
$P_{C}4nc$\else
\ifnum#2=210
$P_{I}4nc$\else
{\color{red}{Invalid MSG number}}
\fi
\fi
\fi
\fi
\fi
\fi
\fi
\fi
\else
\ifnum#1=105
\ifnum#2=211
$P4_{2}mc$\else
\ifnum#2=212
$P4_{2}mc1'$\else
\ifnum#2=213
$P4_{2}'m'c$\else
\ifnum#2=214
$P4_{2}'mc'$\else
\ifnum#2=215
$P4_{2}m'c'$\else
\ifnum#2=216
$P_{c}4_{2}mc$\else
\ifnum#2=217
$P_{C}4_{2}mc$\else
\ifnum#2=218
$P_{I}4_{2}mc$\else
{\color{red}{Invalid MSG number}}
\fi
\fi
\fi
\fi
\fi
\fi
\fi
\fi
\else
\ifnum#1=106
\ifnum#2=219
$P4_{2}bc$\else
\ifnum#2=220
$P4_{2}bc1'$\else
\ifnum#2=221
$P4_{2}'b'c$\else
\ifnum#2=222
$P4_{2}'bc'$\else
\ifnum#2=223
$P4_{2}b'c'$\else
\ifnum#2=224
$P_{c}4_{2}bc$\else
\ifnum#2=225
$P_{C}4_{2}bc$\else
\ifnum#2=226
$P_{I}4_{2}bc$\else
{\color{red}{Invalid MSG number}}
\fi
\fi
\fi
\fi
\fi
\fi
\fi
\fi
\else
\ifnum#1=107
\ifnum#2=227
$I4mm$\else
\ifnum#2=228
$I4mm1'$\else
\ifnum#2=229
$I4'm'm$\else
\ifnum#2=230
$I4'mm'$\else
\ifnum#2=231
$I4m'm'$\else
\ifnum#2=232
$I_{c}4mm$\else
{\color{red}{Invalid MSG number}}
\fi
\fi
\fi
\fi
\fi
\fi
\else
\ifnum#1=108
\ifnum#2=233
$I4cm$\else
\ifnum#2=234
$I4cm1'$\else
\ifnum#2=235
$I4'c'm$\else
\ifnum#2=236
$I4'cm'$\else
\ifnum#2=237
$I4c'm'$\else
\ifnum#2=238
$I_{c}4cm$\else
{\color{red}{Invalid MSG number}}
\fi
\fi
\fi
\fi
\fi
\fi
\else
\ifnum#1=109
\ifnum#2=239
$I4_{1}md$\else
\ifnum#2=240
$I4_{1}md1'$\else
\ifnum#2=241
$I4_{1}'m'd$\else
\ifnum#2=242
$I4_{1}'md'$\else
\ifnum#2=243
$I4_{1}m'd'$\else
\ifnum#2=244
$I_{c}4_{1}md$\else
{\color{red}{Invalid MSG number}}
\fi
\fi
\fi
\fi
\fi
\fi
\else
\ifnum#1=110
\ifnum#2=245
$I4_{1}cd$\else
\ifnum#2=246
$I4_{1}cd1'$\else
\ifnum#2=247
$I4_{1}'c'd$\else
\ifnum#2=248
$I4_{1}'cd'$\else
\ifnum#2=249
$I4_{1}c'd'$\else
\ifnum#2=250
$I_{c}4_{1}cd$\else
{\color{red}{Invalid MSG number}}
\fi
\fi
\fi
\fi
\fi
\fi
\else
\ifnum#1=111
\ifnum#2=251
$P\bar{4}2m$\else
\ifnum#2=252
$P\bar{4}2m1'$\else
\ifnum#2=253
$P\bar{4}'2'm$\else
\ifnum#2=254
$P\bar{4}'2m'$\else
\ifnum#2=255
$P\bar{4}2'm'$\else
\ifnum#2=256
$P_{c}\bar{4}2m$\else
\ifnum#2=257
$P_{C}\bar{4}2m$\else
\ifnum#2=258
$P_{I}\bar{4}2m$\else
{\color{red}{Invalid MSG number}}
\fi
\fi
\fi
\fi
\fi
\fi
\fi
\fi
\else
\ifnum#1=112
\ifnum#2=259
$P\bar{4}2c$\else
\ifnum#2=260
$P\bar{4}2c1'$\else
\ifnum#2=261
$P\bar{4}'2'c$\else
\ifnum#2=262
$P\bar{4}'2c'$\else
\ifnum#2=263
$P\bar{4}2'c'$\else
\ifnum#2=264
$P_{c}\bar{4}2c$\else
\ifnum#2=265
$P_{C}\bar{4}2c$\else
\ifnum#2=266
$P_{I}\bar{4}2c$\else
{\color{red}{Invalid MSG number}}
\fi
\fi
\fi
\fi
\fi
\fi
\fi
\fi
\else
\ifnum#1=113
\ifnum#2=267
$P\bar{4}2_{1}m$\else
\ifnum#2=268
$P\bar{4}2_{1}m1'$\else
\ifnum#2=269
$P\bar{4}'2_{1}'m$\else
\ifnum#2=270
$P\bar{4}'2_{1}m'$\else
\ifnum#2=271
$P\bar{4}2_{1}'m'$\else
\ifnum#2=272
$P_{c}\bar{4}2_{1}m$\else
\ifnum#2=273
$P_{C}\bar{4}2_{1}m$\else
\ifnum#2=274
$P_{I}\bar{4}2_{1}m$\else
{\color{red}{Invalid MSG number}}
\fi
\fi
\fi
\fi
\fi
\fi
\fi
\fi
\else
\ifnum#1=114
\ifnum#2=275
$P\bar{4}2_{1}c$\else
\ifnum#2=276
$P\bar{4}2_{1}c1'$\else
\ifnum#2=277
$P\bar{4}'2_{1}'c$\else
\ifnum#2=278
$P\bar{4}'2_{1}c'$\else
\ifnum#2=279
$P\bar{4}2_{1}'c'$\else
\ifnum#2=280
$P_{c}\bar{4}2_{1}c$\else
\ifnum#2=281
$P_{C}\bar{4}2_{1}c$\else
\ifnum#2=282
$P_{I}\bar{4}2_{1}c$\else
{\color{red}{Invalid MSG number}}
\fi
\fi
\fi
\fi
\fi
\fi
\fi
\fi
\else
\ifnum#1=115
\ifnum#2=283
$P\bar{4}m2$\else
\ifnum#2=284
$P\bar{4}m21'$\else
\ifnum#2=285
$P\bar{4}'m'2$\else
\ifnum#2=286
$P\bar{4}'m2'$\else
\ifnum#2=287
$P\bar{4}m'2'$\else
\ifnum#2=288
$P_{c}\bar{4}m2$\else
\ifnum#2=289
$P_{C}\bar{4}m2$\else
\ifnum#2=290
$P_{I}\bar{4}m2$\else
{\color{red}{Invalid MSG number}}
\fi
\fi
\fi
\fi
\fi
\fi
\fi
\fi
\else
\ifnum#1=116
\ifnum#2=291
$P\bar{4}c2$\else
\ifnum#2=292
$P\bar{4}c21'$\else
\ifnum#2=293
$P\bar{4}'c'2$\else
\ifnum#2=294
$P\bar{4}'c2'$\else
\ifnum#2=295
$P\bar{4}c'2'$\else
\ifnum#2=296
$P_{c}\bar{4}c2$\else
\ifnum#2=297
$P_{C}\bar{4}c2$\else
\ifnum#2=298
$P_{I}\bar{4}c2$\else
{\color{red}{Invalid MSG number}}
\fi
\fi
\fi
\fi
\fi
\fi
\fi
\fi
\else
\ifnum#1=117
\ifnum#2=299
$P\bar{4}b2$\else
\ifnum#2=300
$P\bar{4}b21'$\else
\ifnum#2=301
$P\bar{4}'b'2$\else
\ifnum#2=302
$P\bar{4}'b2'$\else
\ifnum#2=303
$P\bar{4}b'2'$\else
\ifnum#2=304
$P_{c}\bar{4}b2$\else
\ifnum#2=305
$P_{C}\bar{4}b2$\else
\ifnum#2=306
$P_{I}\bar{4}b2$\else
{\color{red}{Invalid MSG number}}
\fi
\fi
\fi
\fi
\fi
\fi
\fi
\fi
\else
\ifnum#1=118
\ifnum#2=307
$P\bar{4}n2$\else
\ifnum#2=308
$P\bar{4}n21'$\else
\ifnum#2=309
$P\bar{4}'n'2$\else
\ifnum#2=310
$P\bar{4}'n2'$\else
\ifnum#2=311
$P\bar{4}n'2'$\else
\ifnum#2=312
$P_{c}\bar{4}n2$\else
\ifnum#2=313
$P_{C}\bar{4}n2$\else
\ifnum#2=314
$P_{I}\bar{4}n2$\else
{\color{red}{Invalid MSG number}}
\fi
\fi
\fi
\fi
\fi
\fi
\fi
\fi
\else
\ifnum#1=119
\ifnum#2=315
$I\bar{4}m2$\else
\ifnum#2=316
$I\bar{4}m21'$\else
\ifnum#2=317
$I\bar{4}'m'2$\else
\ifnum#2=318
$I\bar{4}'m2'$\else
\ifnum#2=319
$I\bar{4}m'2'$\else
\ifnum#2=320
$I_{c}\bar{4}m2$\else
{\color{red}{Invalid MSG number}}
\fi
\fi
\fi
\fi
\fi
\fi
\else
\ifnum#1=120
\ifnum#2=321
$I\bar{4}c2$\else
\ifnum#2=322
$I\bar{4}c21'$\else
\ifnum#2=323
$I\bar{4}'c'2$\else
\ifnum#2=324
$I\bar{4}'c2'$\else
\ifnum#2=325
$I\bar{4}c'2'$\else
\ifnum#2=326
$I_{c}\bar{4}c2$\else
{\color{red}{Invalid MSG number}}
\fi
\fi
\fi
\fi
\fi
\fi
\else
\ifnum#1=121
\ifnum#2=327
$I\bar{4}2m$\else
\ifnum#2=328
$I\bar{4}2m1'$\else
\ifnum#2=329
$I\bar{4}'2'm$\else
\ifnum#2=330
$I\bar{4}'2m'$\else
\ifnum#2=331
$I\bar{4}2'm'$\else
\ifnum#2=332
$I_{c}\bar{4}2m$\else
{\color{red}{Invalid MSG number}}
\fi
\fi
\fi
\fi
\fi
\fi
\else
\ifnum#1=122
\ifnum#2=333
$I\bar{4}2d$\else
\ifnum#2=334
$I\bar{4}2d1'$\else
\ifnum#2=335
$I\bar{4}'2'd$\else
\ifnum#2=336
$I\bar{4}'2d'$\else
\ifnum#2=337
$I\bar{4}2'd'$\else
\ifnum#2=338
$I_{c}\bar{4}2d$\else
{\color{red}{Invalid MSG number}}
\fi
\fi
\fi
\fi
\fi
\fi
\else
\ifnum#1=123
\ifnum#2=339
$P4/mmm$\else
\ifnum#2=340
$P4/mmm1'$\else
\ifnum#2=341
$P4/m'mm$\else
\ifnum#2=342
$P4'/mm'm$\else
\ifnum#2=343
$P4'/mmm'$\else
\ifnum#2=344
$P4'/m'm'm$\else
\ifnum#2=345
$P4/mm'm'$\else
\ifnum#2=346
$P4'/m'mm'$\else
\ifnum#2=347
$P4/m'm'm'$\else
\ifnum#2=348
$P_{c}4/mmm$\else
\ifnum#2=349
$P_{C}4/mmm$\else
\ifnum#2=350
$P_{I}4/mmm$\else
{\color{red}{Invalid MSG number}}
\fi
\fi
\fi
\fi
\fi
\fi
\fi
\fi
\fi
\fi
\fi
\fi
\else
\ifnum#1=124
\ifnum#2=351
$P4/mcc$\else
\ifnum#2=352
$P4/mcc1'$\else
\ifnum#2=353
$P4/m'cc$\else
\ifnum#2=354
$P4'/mc'c$\else
\ifnum#2=355
$P4'/mcc'$\else
\ifnum#2=356
$P4'/m'c'c$\else
\ifnum#2=357
$P4/mc'c'$\else
\ifnum#2=358
$P4'/m'cc'$\else
\ifnum#2=359
$P4/m'c'c'$\else
\ifnum#2=360
$P_{c}4/mcc$\else
\ifnum#2=361
$P_{C}4/mcc$\else
\ifnum#2=362
$P_{I}4/mcc$\else
{\color{red}{Invalid MSG number}}
\fi
\fi
\fi
\fi
\fi
\fi
\fi
\fi
\fi
\fi
\fi
\fi
\else
\ifnum#1=125
\ifnum#2=363
$P4/nbm$\else
\ifnum#2=364
$P4/nbm1'$\else
\ifnum#2=365
$P4/n'bm$\else
\ifnum#2=366
$P4'/nb'm$\else
\ifnum#2=367
$P4'/nbm'$\else
\ifnum#2=368
$P4'/n'b'm$\else
\ifnum#2=369
$P4/nb'm'$\else
\ifnum#2=370
$P4'/n'bm'$\else
\ifnum#2=371
$P4/n'b'm'$\else
\ifnum#2=372
$P_{c}4/nbm$\else
\ifnum#2=373
$P_{C}4/nbm$\else
\ifnum#2=374
$P_{I}4/nbm$\else
{\color{red}{Invalid MSG number}}
\fi
\fi
\fi
\fi
\fi
\fi
\fi
\fi
\fi
\fi
\fi
\fi
\else
\ifnum#1=126
\ifnum#2=375
$P4/nnc$\else
\ifnum#2=376
$P4/nnc1'$\else
\ifnum#2=377
$P4/n'nc$\else
\ifnum#2=378
$P4'/nn'c$\else
\ifnum#2=379
$P4'/nnc'$\else
\ifnum#2=380
$P4'/n'n'c$\else
\ifnum#2=381
$P4/nn'c'$\else
\ifnum#2=382
$P4'/n'nc'$\else
\ifnum#2=383
$P4/n'n'c'$\else
\ifnum#2=384
$P_{c}4/nnc$\else
\ifnum#2=385
$P_{C}4/nnc$\else
\ifnum#2=386
$P_{I}4/nnc$\else
{\color{red}{Invalid MSG number}}
\fi
\fi
\fi
\fi
\fi
\fi
\fi
\fi
\fi
\fi
\fi
\fi
\else
\ifnum#1=127
\ifnum#2=387
$P4/mbm$\else
\ifnum#2=388
$P4/mbm1'$\else
\ifnum#2=389
$P4/m'bm$\else
\ifnum#2=390
$P4'/mb'm$\else
\ifnum#2=391
$P4'/mbm'$\else
\ifnum#2=392
$P4'/m'b'm$\else
\ifnum#2=393
$P4/mb'm'$\else
\ifnum#2=394
$P4'/m'bm'$\else
\ifnum#2=395
$P4/m'b'm'$\else
\ifnum#2=396
$P_{c}4/mbm$\else
\ifnum#2=397
$P_{C}4/mbm$\else
\ifnum#2=398
$P_{I}4/mbm$\else
{\color{red}{Invalid MSG number}}
\fi
\fi
\fi
\fi
\fi
\fi
\fi
\fi
\fi
\fi
\fi
\fi
\else
\ifnum#1=128
\ifnum#2=399
$P4/mnc$\else
\ifnum#2=400
$P4/mnc1'$\else
\ifnum#2=401
$P4/m'nc$\else
\ifnum#2=402
$P4'/mn'c$\else
\ifnum#2=403
$P4'/mnc'$\else
\ifnum#2=404
$P4'/m'n'c$\else
\ifnum#2=405
$P4/mn'c'$\else
\ifnum#2=406
$P4'/m'nc'$\else
\ifnum#2=407
$P4/m'n'c'$\else
\ifnum#2=408
$P_{c}4/mnc$\else
\ifnum#2=409
$P_{C}4/mnc$\else
\ifnum#2=410
$P_{I}4/mnc$\else
{\color{red}{Invalid MSG number}}
\fi
\fi
\fi
\fi
\fi
\fi
\fi
\fi
\fi
\fi
\fi
\fi
\else
\ifnum#1=129
\ifnum#2=411
$P4/nmm$\else
\ifnum#2=412
$P4/nmm1'$\else
\ifnum#2=413
$P4/n'mm$\else
\ifnum#2=414
$P4'/nm'm$\else
\ifnum#2=415
$P4'/nmm'$\else
\ifnum#2=416
$P4'/n'm'm$\else
\ifnum#2=417
$P4/nm'm'$\else
\ifnum#2=418
$P4'/n'mm'$\else
\ifnum#2=419
$P4/n'm'm'$\else
\ifnum#2=420
$P_{c}4/nmm$\else
\ifnum#2=421
$P_{C}4/nmm$\else
\ifnum#2=422
$P_{I}4/nmm$\else
{\color{red}{Invalid MSG number}}
\fi
\fi
\fi
\fi
\fi
\fi
\fi
\fi
\fi
\fi
\fi
\fi
\else
\ifnum#1=130
\ifnum#2=423
$P4/ncc$\else
\ifnum#2=424
$P4/ncc1'$\else
\ifnum#2=425
$P4/n'cc$\else
\ifnum#2=426
$P4'/nc'c$\else
\ifnum#2=427
$P4'/ncc'$\else
\ifnum#2=428
$P4'/n'c'c$\else
\ifnum#2=429
$P4/nc'c'$\else
\ifnum#2=430
$P4'/n'cc'$\else
\ifnum#2=431
$P4/n'c'c'$\else
\ifnum#2=432
$P_{c}4/ncc$\else
\ifnum#2=433
$P_{C}4/ncc$\else
\ifnum#2=434
$P_{I}4/ncc$\else
{\color{red}{Invalid MSG number}}
\fi
\fi
\fi
\fi
\fi
\fi
\fi
\fi
\fi
\fi
\fi
\fi
\else
\ifnum#1=131
\ifnum#2=435
$P4_{2}/mmc$\else
\ifnum#2=436
$P4_{2}/mmc1'$\else
\ifnum#2=437
$P4_{2}/m'mc$\else
\ifnum#2=438
$P4_{2}'/mm'c$\else
\ifnum#2=439
$P4_{2}'/mmc'$\else
\ifnum#2=440
$P4_{2}'/m'm'c$\else
\ifnum#2=441
$P4_{2}/mm'c'$\else
\ifnum#2=442
$P4_{2}'/m'mc'$\else
\ifnum#2=443
$P4_{2}/m'm'c'$\else
\ifnum#2=444
$P_{c}4_{2}/mmc$\else
\ifnum#2=445
$P_{C}4_{2}/mmc$\else
\ifnum#2=446
$P_{I}4_{2}/mmc$\else
{\color{red}{Invalid MSG number}}
\fi
\fi
\fi
\fi
\fi
\fi
\fi
\fi
\fi
\fi
\fi
\fi
\else
\ifnum#1=132
\ifnum#2=447
$P4_{2}/mcm$\else
\ifnum#2=448
$P4_{2}/mcm1'$\else
\ifnum#2=449
$P4_{2}/m'cm$\else
\ifnum#2=450
$P4_{2}'/mc'm$\else
\ifnum#2=451
$P4_{2}'/mcm'$\else
\ifnum#2=452
$P4_{2}'/m'c'm$\else
\ifnum#2=453
$P4_{2}/mc'm'$\else
\ifnum#2=454
$P4_{2}'/m'cm'$\else
\ifnum#2=455
$P4_{2}/m'c'm'$\else
\ifnum#2=456
$P_{c}4_{2}/mcm$\else
\ifnum#2=457
$P_{C}4_{2}/mcm$\else
\ifnum#2=458
$P_{I}4_{2}/mcm$\else
{\color{red}{Invalid MSG number}}
\fi
\fi
\fi
\fi
\fi
\fi
\fi
\fi
\fi
\fi
\fi
\fi
\else
\ifnum#1=133
\ifnum#2=459
$P4_{2}/nbc$\else
\ifnum#2=460
$P4_{2}/nbc1'$\else
\ifnum#2=461
$P4_{2}/n'bc$\else
\ifnum#2=462
$P4_{2}'/nb'c$\else
\ifnum#2=463
$P4_{2}'/nbc'$\else
\ifnum#2=464
$P4_{2}'/n'b'c$\else
\ifnum#2=465
$P4_{2}/nb'c'$\else
\ifnum#2=466
$P4_{2}'/n'bc'$\else
\ifnum#2=467
$P4_{2}/n'b'c'$\else
\ifnum#2=468
$P_{c}4_{2}/nbc$\else
\ifnum#2=469
$P_{C}4_{2}/nbc$\else
\ifnum#2=470
$P_{I}4_{2}/nbc$\else
{\color{red}{Invalid MSG number}}
\fi
\fi
\fi
\fi
\fi
\fi
\fi
\fi
\fi
\fi
\fi
\fi
\else
\ifnum#1=134
\ifnum#2=471
$P4_{2}/nnm$\else
\ifnum#2=472
$P4_{2}/nnm1'$\else
\ifnum#2=473
$P4_{2}/n'nm$\else
\ifnum#2=474
$P4_{2}'/nn'm$\else
\ifnum#2=475
$P4_{2}'/nnm'$\else
\ifnum#2=476
$P4_{2}'/n'n'm$\else
\ifnum#2=477
$P4_{2}/nn'm'$\else
\ifnum#2=478
$P4_{2}'/n'nm'$\else
\ifnum#2=479
$P4_{2}/n'n'm'$\else
\ifnum#2=480
$P_{c}4_{2}/nnm$\else
\ifnum#2=481
$P_{C}4_{2}/nnm$\else
\ifnum#2=482
$P_{I}4_{2}/nnm$\else
{\color{red}{Invalid MSG number}}
\fi
\fi
\fi
\fi
\fi
\fi
\fi
\fi
\fi
\fi
\fi
\fi
\else
\ifnum#1=135
\ifnum#2=483
$P4_{2}/mbc$\else
\ifnum#2=484
$P4_{2}/mbc1'$\else
\ifnum#2=485
$P4_{2}/m'bc$\else
\ifnum#2=486
$P4_{2}'/mb'c$\else
\ifnum#2=487
$P4_{2}'/mbc'$\else
\ifnum#2=488
$P4_{2}'/m'b'c$\else
\ifnum#2=489
$P4_{2}/mb'c'$\else
\ifnum#2=490
$P4_{2}'/m'bc'$\else
\ifnum#2=491
$P4_{2}/m'b'c'$\else
\ifnum#2=492
$P_{c}4_{2}/mbc$\else
\ifnum#2=493
$P_{C}4_{2}/mbc$\else
\ifnum#2=494
$P_{I}4_{2}/mbc$\else
{\color{red}{Invalid MSG number}}
\fi
\fi
\fi
\fi
\fi
\fi
\fi
\fi
\fi
\fi
\fi
\fi
\else
\ifnum#1=136
\ifnum#2=495
$P4_{2}/mnm$\else
\ifnum#2=496
$P4_{2}/mnm1'$\else
\ifnum#2=497
$P4_{2}/m'nm$\else
\ifnum#2=498
$P4_{2}'/mn'm$\else
\ifnum#2=499
$P4_{2}'/mnm'$\else
\ifnum#2=500
$P4_{2}'/m'n'm$\else
\ifnum#2=501
$P4_{2}/mn'm'$\else
\ifnum#2=502
$P4_{2}'/m'nm'$\else
\ifnum#2=503
$P4_{2}/m'n'm'$\else
\ifnum#2=504
$P_{c}4_{2}/mnm$\else
\ifnum#2=505
$P_{C}4_{2}/mnm$\else
\ifnum#2=506
$P_{I}4_{2}/mnm$\else
{\color{red}{Invalid MSG number}}
\fi
\fi
\fi
\fi
\fi
\fi
\fi
\fi
\fi
\fi
\fi
\fi
\else
\ifnum#1=137
\ifnum#2=507
$P4_{2}/nmc$\else
\ifnum#2=508
$P4_{2}/nmc1'$\else
\ifnum#2=509
$P4_{2}/n'mc$\else
\ifnum#2=510
$P4_{2}'/nm'c$\else
\ifnum#2=511
$P4_{2}'/nmc'$\else
\ifnum#2=512
$P4_{2}'/n'm'c$\else
\ifnum#2=513
$P4_{2}/nm'c'$\else
\ifnum#2=514
$P4_{2}'/n'mc'$\else
\ifnum#2=515
$P4_{2}/n'm'c'$\else
\ifnum#2=516
$P_{c}4_{2}/nmc$\else
\ifnum#2=517
$P_{C}4_{2}/nmc$\else
\ifnum#2=518
$P_{I}4_{2}/nmc$\else
{\color{red}{Invalid MSG number}}
\fi
\fi
\fi
\fi
\fi
\fi
\fi
\fi
\fi
\fi
\fi
\fi
\else
\ifnum#1=138
\ifnum#2=519
$P4_{2}/ncm$\else
\ifnum#2=520
$P4_{2}/ncm1'$\else
\ifnum#2=521
$P4_{2}/n'cm$\else
\ifnum#2=522
$P4_{2}'/nc'm$\else
\ifnum#2=523
$P4_{2}'/ncm'$\else
\ifnum#2=524
$P4_{2}'/n'c'm$\else
\ifnum#2=525
$P4_{2}/nc'm'$\else
\ifnum#2=526
$P4_{2}'/n'cm'$\else
\ifnum#2=527
$P4_{2}/n'c'm'$\else
\ifnum#2=528
$P_{c}4_{2}/ncm$\else
\ifnum#2=529
$P_{C}4_{2}/ncm$\else
\ifnum#2=530
$P_{I}4_{2}/ncm$\else
{\color{red}{Invalid MSG number}}
\fi
\fi
\fi
\fi
\fi
\fi
\fi
\fi
\fi
\fi
\fi
\fi
\else
\ifnum#1=139
\ifnum#2=531
$I4/mmm$\else
\ifnum#2=532
$I4/mmm1'$\else
\ifnum#2=533
$I4/m'mm$\else
\ifnum#2=534
$I4'/mm'm$\else
\ifnum#2=535
$I4'/mmm'$\else
\ifnum#2=536
$I4'/m'm'm$\else
\ifnum#2=537
$I4/mm'm'$\else
\ifnum#2=538
$I4'/m'mm'$\else
\ifnum#2=539
$I4/m'm'm'$\else
\ifnum#2=540
$I_{c}4/mmm$\else
{\color{red}{Invalid MSG number}}
\fi
\fi
\fi
\fi
\fi
\fi
\fi
\fi
\fi
\fi
\else
\ifnum#1=140
\ifnum#2=541
$I4/mcm$\else
\ifnum#2=542
$I4/mcm1'$\else
\ifnum#2=543
$I4/m'cm$\else
\ifnum#2=544
$I4'/mc'm$\else
\ifnum#2=545
$I4'/mcm'$\else
\ifnum#2=546
$I4'/m'c'm$\else
\ifnum#2=547
$I4/mc'm'$\else
\ifnum#2=548
$I4'/m'cm'$\else
\ifnum#2=549
$I4/m'c'm'$\else
\ifnum#2=550
$I_{c}4/mcm$\else
{\color{red}{Invalid MSG number}}
\fi
\fi
\fi
\fi
\fi
\fi
\fi
\fi
\fi
\fi
\else
\ifnum#1=141
\ifnum#2=551
$I4_{1}/amd$\else
\ifnum#2=552
$I4_{1}/amd1'$\else
\ifnum#2=553
$I4_{1}/a'md$\else
\ifnum#2=554
$I4_{1}'/am'd$\else
\ifnum#2=555
$I4_{1}'/amd'$\else
\ifnum#2=556
$I4_{1}'/a'm'd$\else
\ifnum#2=557
$I4_{1}/am'd'$\else
\ifnum#2=558
$I4_{1}'/a'md'$\else
\ifnum#2=559
$I4_{1}/a'm'd'$\else
\ifnum#2=560
$I_{c}4_{1}/amd$\else
{\color{red}{Invalid MSG number}}
\fi
\fi
\fi
\fi
\fi
\fi
\fi
\fi
\fi
\fi
\else
\ifnum#1=142
\ifnum#2=561
$I4_{1}/acd$\else
\ifnum#2=562
$I4_{1}/acd1'$\else
\ifnum#2=563
$I4_{1}/a'cd$\else
\ifnum#2=564
$I4_{1}'/ac'd$\else
\ifnum#2=565
$I4_{1}'/acd'$\else
\ifnum#2=566
$I4_{1}'/a'c'd$\else
\ifnum#2=567
$I4_{1}/ac'd'$\else
\ifnum#2=568
$I4_{1}'/a'cd'$\else
\ifnum#2=569
$I4_{1}/a'c'd'$\else
\ifnum#2=570
$I_{c}4_{1}/acd$\else
{\color{red}{Invalid MSG number}}
\fi
\fi
\fi
\fi
\fi
\fi
\fi
\fi
\fi
\fi
\else
\ifnum#1=143
\ifnum#2=1
$P3$\else
\ifnum#2=2
$P31'$\else
\ifnum#2=3
$P_{c}3$\else
{\color{red}{Invalid MSG number}}
\fi
\fi
\fi
\else
\ifnum#1=144
\ifnum#2=4
$P3_{1}$\else
\ifnum#2=5
$P3_{1}1'$\else
\ifnum#2=6
$P_{c}3_{1}$\else
{\color{red}{Invalid MSG number}}
\fi
\fi
\fi
\else
\ifnum#1=145
\ifnum#2=7
$P3_{2}$\else
\ifnum#2=8
$P3_{2}1'$\else
\ifnum#2=9
$P_{c}3_{2}$\else
{\color{red}{Invalid MSG number}}
\fi
\fi
\fi
\else
\ifnum#1=146
\ifnum#2=10
$R3$\else
\ifnum#2=11
$R31'$\else
\ifnum#2=12
$R_{I}3$\else
{\color{red}{Invalid MSG number}}
\fi
\fi
\fi
\else
\ifnum#1=147
\ifnum#2=13
$P\bar{3}$\else
\ifnum#2=14
$P\bar{3}1'$\else
\ifnum#2=15
$P\bar{3}'$\else
\ifnum#2=16
$P_{c}\bar{3}$\else
{\color{red}{Invalid MSG number}}
\fi
\fi
\fi
\fi
\else
\ifnum#1=148
\ifnum#2=17
$R\bar{3}$\else
\ifnum#2=18
$R\bar{3}1'$\else
\ifnum#2=19
$R\bar{3}'$\else
\ifnum#2=20
$R_{I}\bar{3}$\else
{\color{red}{Invalid MSG number}}
\fi
\fi
\fi
\fi
\else
\ifnum#1=149
\ifnum#2=21
$P312$\else
\ifnum#2=22
$P3121'$\else
\ifnum#2=23
$P312'$\else
\ifnum#2=24
$P_{c}312$\else
{\color{red}{Invalid MSG number}}
\fi
\fi
\fi
\fi
\else
\ifnum#1=150
\ifnum#2=25
$P321$\else
\ifnum#2=26
$P3211'$\else
\ifnum#2=27
$P32'1$\else
\ifnum#2=28
$P_{c}321$\else
{\color{red}{Invalid MSG number}}
\fi
\fi
\fi
\fi
\else
\ifnum#1=151
\ifnum#2=29
$P3_{1}12$\else
\ifnum#2=30
$P3_{1}121'$\else
\ifnum#2=31
$P3_{1}12'$\else
\ifnum#2=32
$P_{c}3_{1}12$\else
{\color{red}{Invalid MSG number}}
\fi
\fi
\fi
\fi
\else
\ifnum#1=152
\ifnum#2=33
$P3_{1}21$\else
\ifnum#2=34
$P3_{1}211'$\else
\ifnum#2=35
$P3_{1}2'1$\else
\ifnum#2=36
$P_{c}3_{1}21$\else
{\color{red}{Invalid MSG number}}
\fi
\fi
\fi
\fi
\else
\ifnum#1=153
\ifnum#2=37
$P3_{2}12$\else
\ifnum#2=38
$P3_{2}121'$\else
\ifnum#2=39
$P3_{2}12'$\else
\ifnum#2=40
$P_{c}3_{2}12$\else
{\color{red}{Invalid MSG number}}
\fi
\fi
\fi
\fi
\else
\ifnum#1=154
\ifnum#2=41
$P3_{2}21$\else
\ifnum#2=42
$P3_{2}211'$\else
\ifnum#2=43
$P3_{2}2'1$\else
\ifnum#2=44
$P_{c}3_{2}21$\else
{\color{red}{Invalid MSG number}}
\fi
\fi
\fi
\fi
\else
\ifnum#1=155
\ifnum#2=45
$R32$\else
\ifnum#2=46
$R321'$\else
\ifnum#2=47
$R32'$\else
\ifnum#2=48
$R_{I}32$\else
{\color{red}{Invalid MSG number}}
\fi
\fi
\fi
\fi
\else
\ifnum#1=156
\ifnum#2=49
$P3m1$\else
\ifnum#2=50
$P3m11'$\else
\ifnum#2=51
$P3m'1$\else
\ifnum#2=52
$P_{c}3m1$\else
{\color{red}{Invalid MSG number}}
\fi
\fi
\fi
\fi
\else
\ifnum#1=157
\ifnum#2=53
$P31m$\else
\ifnum#2=54
$P31m1'$\else
\ifnum#2=55
$P31m'$\else
\ifnum#2=56
$P_{c}31m$\else
{\color{red}{Invalid MSG number}}
\fi
\fi
\fi
\fi
\else
\ifnum#1=158
\ifnum#2=57
$P3c1$\else
\ifnum#2=58
$P3c11'$\else
\ifnum#2=59
$P3c'1$\else
\ifnum#2=60
$P_{c}3c1$\else
{\color{red}{Invalid MSG number}}
\fi
\fi
\fi
\fi
\else
\ifnum#1=159
\ifnum#2=61
$P31c$\else
\ifnum#2=62
$P31c1'$\else
\ifnum#2=63
$P31c'$\else
\ifnum#2=64
$P_{c}31c$\else
{\color{red}{Invalid MSG number}}
\fi
\fi
\fi
\fi
\else
\ifnum#1=160
\ifnum#2=65
$R3m$\else
\ifnum#2=66
$R3m1'$\else
\ifnum#2=67
$R3m'$\else
\ifnum#2=68
$R_{I}3m$\else
{\color{red}{Invalid MSG number}}
\fi
\fi
\fi
\fi
\else
\ifnum#1=161
\ifnum#2=69
$R3c$\else
\ifnum#2=70
$R3c1'$\else
\ifnum#2=71
$R3c'$\else
\ifnum#2=72
$R_{I}3c$\else
{\color{red}{Invalid MSG number}}
\fi
\fi
\fi
\fi
\else
\ifnum#1=162
\ifnum#2=73
$P\bar{3}1m$\else
\ifnum#2=74
$P\bar{3}1m1'$\else
\ifnum#2=75
$P\bar{3}'1m$\else
\ifnum#2=76
$P\bar{3}'1m'$\else
\ifnum#2=77
$P\bar{3}1m'$\else
\ifnum#2=78
$P_{c}\bar{3}1m$\else
{\color{red}{Invalid MSG number}}
\fi
\fi
\fi
\fi
\fi
\fi
\else
\ifnum#1=163
\ifnum#2=79
$P\bar{3}1c$\else
\ifnum#2=80
$P\bar{3}1c1'$\else
\ifnum#2=81
$P\bar{3}'1c$\else
\ifnum#2=82
$P\bar{3}'1c'$\else
\ifnum#2=83
$P\bar{3}1c'$\else
\ifnum#2=84
$P_{c}\bar{3}1c$\else
{\color{red}{Invalid MSG number}}
\fi
\fi
\fi
\fi
\fi
\fi
\else
\ifnum#1=164
\ifnum#2=85
$P\bar{3}m1$\else
\ifnum#2=86
$P\bar{3}m11'$\else
\ifnum#2=87
$P\bar{3}'m1$\else
\ifnum#2=88
$P\bar{3}'m'1$\else
\ifnum#2=89
$P\bar{3}m'1$\else
\ifnum#2=90
$P_{c}\bar{3}m1$\else
{\color{red}{Invalid MSG number}}
\fi
\fi
\fi
\fi
\fi
\fi
\else
\ifnum#1=165
\ifnum#2=91
$P\bar{3}c1$\else
\ifnum#2=92
$P\bar{3}c11'$\else
\ifnum#2=93
$P\bar{3}'c1$\else
\ifnum#2=94
$P\bar{3}'c'1$\else
\ifnum#2=95
$P\bar{3}c'1$\else
\ifnum#2=96
$P_{c}\bar{3}c1$\else
{\color{red}{Invalid MSG number}}
\fi
\fi
\fi
\fi
\fi
\fi
\else
\ifnum#1=166
\ifnum#2=97
$R\bar{3}m$\else
\ifnum#2=98
$R\bar{3}m1'$\else
\ifnum#2=99
$R\bar{3}'m$\else
\ifnum#2=100
$R\bar{3}'m'$\else
\ifnum#2=101
$R\bar{3}m'$\else
\ifnum#2=102
$R_{I}\bar{3}m$\else
{\color{red}{Invalid MSG number}}
\fi
\fi
\fi
\fi
\fi
\fi
\else
\ifnum#1=167
\ifnum#2=103
$R\bar{3}c$\else
\ifnum#2=104
$R\bar{3}c1'$\else
\ifnum#2=105
$R\bar{3}'c$\else
\ifnum#2=106
$R\bar{3}'c'$\else
\ifnum#2=107
$R\bar{3}c'$\else
\ifnum#2=108
$R_{I}\bar{3}c$\else
{\color{red}{Invalid MSG number}}
\fi
\fi
\fi
\fi
\fi
\fi
\else
\ifnum#1=168
\ifnum#2=109
$P6$\else
\ifnum#2=110
$P61'$\else
\ifnum#2=111
$P6'$\else
\ifnum#2=112
$P_{c}6$\else
{\color{red}{Invalid MSG number}}
\fi
\fi
\fi
\fi
\else
\ifnum#1=169
\ifnum#2=113
$P6_{1}$\else
\ifnum#2=114
$P6_{1}1'$\else
\ifnum#2=115
$P6_{1}'$\else
\ifnum#2=116
$P_{c}6_{1}$\else
{\color{red}{Invalid MSG number}}
\fi
\fi
\fi
\fi
\else
\ifnum#1=170
\ifnum#2=117
$P6_{5}$\else
\ifnum#2=118
$P6_{5}1'$\else
\ifnum#2=119
$P6_{5}'$\else
\ifnum#2=120
$P_{c}6_{5}$\else
{\color{red}{Invalid MSG number}}
\fi
\fi
\fi
\fi
\else
\ifnum#1=171
\ifnum#2=121
$P6_{2}$\else
\ifnum#2=122
$P6_{2}1'$\else
\ifnum#2=123
$P6_{2}'$\else
\ifnum#2=124
$P_{c}6_{2}$\else
{\color{red}{Invalid MSG number}}
\fi
\fi
\fi
\fi
\else
\ifnum#1=172
\ifnum#2=125
$P6_{4}$\else
\ifnum#2=126
$P6_{4}1'$\else
\ifnum#2=127
$P6_{4}'$\else
\ifnum#2=128
$P_{c}6_{4}$\else
{\color{red}{Invalid MSG number}}
\fi
\fi
\fi
\fi
\else
\ifnum#1=173
\ifnum#2=129
$P6_{3}$\else
\ifnum#2=130
$P6_{3}1'$\else
\ifnum#2=131
$P6_{3}'$\else
\ifnum#2=132
$P_{c}6_{3}$\else
{\color{red}{Invalid MSG number}}
\fi
\fi
\fi
\fi
\else
\ifnum#1=174
\ifnum#2=133
$P\bar{6}$\else
\ifnum#2=134
$P\bar{6}1'$\else
\ifnum#2=135
$P\bar{6}'$\else
\ifnum#2=136
$P_{c}\bar{6}$\else
{\color{red}{Invalid MSG number}}
\fi
\fi
\fi
\fi
\else
\ifnum#1=175
\ifnum#2=137
$P6/m$\else
\ifnum#2=138
$P6/m1'$\else
\ifnum#2=139
$P6'/m$\else
\ifnum#2=140
$P6/m'$\else
\ifnum#2=141
$P6'/m'$\else
\ifnum#2=142
$P_{c}6/m$\else
{\color{red}{Invalid MSG number}}
\fi
\fi
\fi
\fi
\fi
\fi
\else
\ifnum#1=176
\ifnum#2=143
$P6_{3}/m$\else
\ifnum#2=144
$P6_{3}/m1'$\else
\ifnum#2=145
$P6_{3}'/m$\else
\ifnum#2=146
$P6_{3}/m'$\else
\ifnum#2=147
$P6_{3}'/m'$\else
\ifnum#2=148
$P_{c}6_{3}/m$\else
{\color{red}{Invalid MSG number}}
\fi
\fi
\fi
\fi
\fi
\fi
\else
\ifnum#1=177
\ifnum#2=149
$P622$\else
\ifnum#2=150
$P6221'$\else
\ifnum#2=151
$P6'2'2$\else
\ifnum#2=152
$P6'22'$\else
\ifnum#2=153
$P62'2'$\else
\ifnum#2=154
$P_{c}622$\else
{\color{red}{Invalid MSG number}}
\fi
\fi
\fi
\fi
\fi
\fi
\else
\ifnum#1=178
\ifnum#2=155
$P6_{1}22$\else
\ifnum#2=156
$P6_{1}221'$\else
\ifnum#2=157
$P6_{1}'2'2$\else
\ifnum#2=158
$P6_{1}'22'$\else
\ifnum#2=159
$P6_{1}2'2'$\else
\ifnum#2=160
$P_{c}6_{1}22$\else
{\color{red}{Invalid MSG number}}
\fi
\fi
\fi
\fi
\fi
\fi
\else
\ifnum#1=179
\ifnum#2=161
$P6_{5}22$\else
\ifnum#2=162
$P6_{5}221'$\else
\ifnum#2=163
$P6_{5}'2'2$\else
\ifnum#2=164
$P6_{5}'22'$\else
\ifnum#2=165
$P6_{5}2'2'$\else
\ifnum#2=166
$P_{c}6_{5}22$\else
{\color{red}{Invalid MSG number}}
\fi
\fi
\fi
\fi
\fi
\fi
\else
\ifnum#1=180
\ifnum#2=167
$P6_{2}22$\else
\ifnum#2=168
$P6_{2}221'$\else
\ifnum#2=169
$P6_{2}'2'2$\else
\ifnum#2=170
$P6_{2}'22'$\else
\ifnum#2=171
$P6_{2}2'2'$\else
\ifnum#2=172
$P_{c}6_{2}22$\else
{\color{red}{Invalid MSG number}}
\fi
\fi
\fi
\fi
\fi
\fi
\else
\ifnum#1=181
\ifnum#2=173
$P6_{4}22$\else
\ifnum#2=174
$P6_{4}221'$\else
\ifnum#2=175
$P6_{4}'2'2$\else
\ifnum#2=176
$P6_{4}'22'$\else
\ifnum#2=177
$P6_{4}2'2'$\else
\ifnum#2=178
$P_{c}6_{4}22$\else
{\color{red}{Invalid MSG number}}
\fi
\fi
\fi
\fi
\fi
\fi
\else
\ifnum#1=182
\ifnum#2=179
$P6_{3}22$\else
\ifnum#2=180
$P6_{3}221'$\else
\ifnum#2=181
$P6_{3}'2'2$\else
\ifnum#2=182
$P6_{3}'22'$\else
\ifnum#2=183
$P6_{3}2'2'$\else
\ifnum#2=184
$P_{c}6_{3}22$\else
{\color{red}{Invalid MSG number}}
\fi
\fi
\fi
\fi
\fi
\fi
\else
\ifnum#1=183
\ifnum#2=185
$P6mm$\else
\ifnum#2=186
$P6mm1'$\else
\ifnum#2=187
$P6'm'm$\else
\ifnum#2=188
$P6'mm'$\else
\ifnum#2=189
$P6m'm'$\else
\ifnum#2=190
$P_{c}6mm$\else
{\color{red}{Invalid MSG number}}
\fi
\fi
\fi
\fi
\fi
\fi
\else
\ifnum#1=184
\ifnum#2=191
$P6cc$\else
\ifnum#2=192
$P6cc1'$\else
\ifnum#2=193
$P6'c'c$\else
\ifnum#2=194
$P6'cc'$\else
\ifnum#2=195
$P6c'c'$\else
\ifnum#2=196
$P_{c}6cc$\else
{\color{red}{Invalid MSG number}}
\fi
\fi
\fi
\fi
\fi
\fi
\else
\ifnum#1=185
\ifnum#2=197
$P6_{3}cm$\else
\ifnum#2=198
$P6_{3}cm1'$\else
\ifnum#2=199
$P6_{3}'c'm$\else
\ifnum#2=200
$P6_{3}'cm'$\else
\ifnum#2=201
$P6_{3}c'm'$\else
\ifnum#2=202
$P_{c}6_{3}cm$\else
{\color{red}{Invalid MSG number}}
\fi
\fi
\fi
\fi
\fi
\fi
\else
\ifnum#1=186
\ifnum#2=203
$P6_{3}mc$\else
\ifnum#2=204
$P6_{3}mc1'$\else
\ifnum#2=205
$P6_{3}'m'c$\else
\ifnum#2=206
$P6_{3}'mc'$\else
\ifnum#2=207
$P6_{3}m'c'$\else
\ifnum#2=208
$P_{c}6_{3}mc$\else
{\color{red}{Invalid MSG number}}
\fi
\fi
\fi
\fi
\fi
\fi
\else
\ifnum#1=187
\ifnum#2=209
$P\bar{6}m2$\else
\ifnum#2=210
$P\bar{6}m21'$\else
\ifnum#2=211
$P\bar{6}'m'2$\else
\ifnum#2=212
$P\bar{6}'m2'$\else
\ifnum#2=213
$P\bar{6}m'2'$\else
\ifnum#2=214
$P_{c}\bar{6}m2$\else
{\color{red}{Invalid MSG number}}
\fi
\fi
\fi
\fi
\fi
\fi
\else
\ifnum#1=188
\ifnum#2=215
$P\bar{6}c2$\else
\ifnum#2=216
$P\bar{6}c21'$\else
\ifnum#2=217
$P\bar{6}'c'2$\else
\ifnum#2=218
$P\bar{6}'c2'$\else
\ifnum#2=219
$P\bar{6}c'2'$\else
\ifnum#2=220
$P_{c}\bar{6}c2$\else
{\color{red}{Invalid MSG number}}
\fi
\fi
\fi
\fi
\fi
\fi
\else
\ifnum#1=189
\ifnum#2=221
$P\bar{6}2m$\else
\ifnum#2=222
$P\bar{6}2m1'$\else
\ifnum#2=223
$P\bar{6}'2'm$\else
\ifnum#2=224
$P\bar{6}'2m'$\else
\ifnum#2=225
$P\bar{6}2'm'$\else
\ifnum#2=226
$P_{c}\bar{6}2m$\else
{\color{red}{Invalid MSG number}}
\fi
\fi
\fi
\fi
\fi
\fi
\else
\ifnum#1=190
\ifnum#2=227
$P\bar{6}2c$\else
\ifnum#2=228
$P\bar{6}2c1'$\else
\ifnum#2=229
$P\bar{6}'2'c$\else
\ifnum#2=230
$P\bar{6}'2c'$\else
\ifnum#2=231
$P\bar{6}2'c'$\else
\ifnum#2=232
$P_{c}\bar{6}2c$\else
{\color{red}{Invalid MSG number}}
\fi
\fi
\fi
\fi
\fi
\fi
\else
\ifnum#1=191
\ifnum#2=233
$P6/mmm$\else
\ifnum#2=234
$P6/mmm1'$\else
\ifnum#2=235
$P6/m'mm$\else
\ifnum#2=236
$P6'/mm'm$\else
\ifnum#2=237
$P6'/mmm'$\else
\ifnum#2=238
$P6'/m'm'm$\else
\ifnum#2=239
$P6'/m'mm'$\else
\ifnum#2=240
$P6/mm'm'$\else
\ifnum#2=241
$P6/m'm'm'$\else
\ifnum#2=242
$P_{c}6/mmm$\else
{\color{red}{Invalid MSG number}}
\fi
\fi
\fi
\fi
\fi
\fi
\fi
\fi
\fi
\fi
\else
\ifnum#1=192
\ifnum#2=243
$P6/mcc$\else
\ifnum#2=244
$P6/mcc1'$\else
\ifnum#2=245
$P6/m'cc$\else
\ifnum#2=246
$P6'/mc'c$\else
\ifnum#2=247
$P6'/mcc'$\else
\ifnum#2=248
$P6'/m'c'c$\else
\ifnum#2=249
$P6'/m'cc'$\else
\ifnum#2=250
$P6/mc'c'$\else
\ifnum#2=251
$P6/m'c'c'$\else
\ifnum#2=252
$P_{c}6/mcc$\else
{\color{red}{Invalid MSG number}}
\fi
\fi
\fi
\fi
\fi
\fi
\fi
\fi
\fi
\fi
\else
\ifnum#1=193
\ifnum#2=253
$P6_{3}/mcm$\else
\ifnum#2=254
$P6_{3}/mcm1'$\else
\ifnum#2=255
$P6_{3}/m'cm$\else
\ifnum#2=256
$P6_{3}'/mc'm$\else
\ifnum#2=257
$P6_{3}'/mcm'$\else
\ifnum#2=258
$P6_{3}'/m'c'm$\else
\ifnum#2=259
$P6_{3}'/m'cm'$\else
\ifnum#2=260
$P6_{3}/mc'm'$\else
\ifnum#2=261
$P6_{3}/m'c'm'$\else
\ifnum#2=262
$P_{c}6_{3}/mcm$\else
{\color{red}{Invalid MSG number}}
\fi
\fi
\fi
\fi
\fi
\fi
\fi
\fi
\fi
\fi
\else
\ifnum#1=194
\ifnum#2=263
$P6_{3}/mmc$\else
\ifnum#2=264
$P6_{3}/mmc1'$\else
\ifnum#2=265
$P6_{3}/m'mc$\else
\ifnum#2=266
$P6_{3}'/mm'c$\else
\ifnum#2=267
$P6_{3}'/mmc'$\else
\ifnum#2=268
$P6_{3}'/m'm'c$\else
\ifnum#2=269
$P6_{3}'/m'mc'$\else
\ifnum#2=270
$P6_{3}/mm'c'$\else
\ifnum#2=271
$P6_{3}/m'm'c'$\else
\ifnum#2=272
$P_{c}6_{3}/mmc$\else
{\color{red}{Invalid MSG number}}
\fi
\fi
\fi
\fi
\fi
\fi
\fi
\fi
\fi
\fi
\else
\ifnum#1=195
\ifnum#2=1
$P23$\else
\ifnum#2=2
$P231'$\else
\ifnum#2=3
$P_{I}23$\else
{\color{red}{Invalid MSG number}}
\fi
\fi
\fi
\else
\ifnum#1=196
\ifnum#2=4
$F23$\else
\ifnum#2=5
$F231'$\else
\ifnum#2=6
$F_{S}23$\else
{\color{red}{Invalid MSG number}}
\fi
\fi
\fi
\else
\ifnum#1=197
\ifnum#2=7
$I23$\else
\ifnum#2=8
$I231'$\else
{\color{red}{Invalid MSG number}}
\fi
\fi
\else
\ifnum#1=198
\ifnum#2=9
$P2_{1}3$\else
\ifnum#2=10
$P2_{1}31'$\else
\ifnum#2=11
$P_{I}2_{1}3$\else
{\color{red}{Invalid MSG number}}
\fi
\fi
\fi
\else
\ifnum#1=199
\ifnum#2=12
$I2_{1}3$\else
\ifnum#2=13
$I2_{1}31'$\else
{\color{red}{Invalid MSG number}}
\fi
\fi
\else
\ifnum#1=200
\ifnum#2=14
$Pm\bar{3}$\else
\ifnum#2=15
$Pm\bar{3}1'$\else
\ifnum#2=16
$Pm'\bar{3}'$\else
\ifnum#2=17
$P_{I}m\bar{3}$\else
{\color{red}{Invalid MSG number}}
\fi
\fi
\fi
\fi
\else
\ifnum#1=201
\ifnum#2=18
$Pn\bar{3}$\else
\ifnum#2=19
$Pn\bar{3}1'$\else
\ifnum#2=20
$Pn'\bar{3}'$\else
\ifnum#2=21
$P_{I}n\bar{3}$\else
{\color{red}{Invalid MSG number}}
\fi
\fi
\fi
\fi
\else
\ifnum#1=202
\ifnum#2=22
$Fm\bar{3}$\else
\ifnum#2=23
$Fm\bar{3}1'$\else
\ifnum#2=24
$Fm'\bar{3}'$\else
\ifnum#2=25
$F_{S}m\bar{3}$\else
{\color{red}{Invalid MSG number}}
\fi
\fi
\fi
\fi
\else
\ifnum#1=203
\ifnum#2=26
$Fd\bar{3}$\else
\ifnum#2=27
$Fd\bar{3}1'$\else
\ifnum#2=28
$Fd'\bar{3}'$\else
\ifnum#2=29
$F_{S}d\bar{3}$\else
{\color{red}{Invalid MSG number}}
\fi
\fi
\fi
\fi
\else
\ifnum#1=204
\ifnum#2=30
$Im\bar{3}$\else
\ifnum#2=31
$Im\bar{3}1'$\else
\ifnum#2=32
$Im'\bar{3}'$\else
{\color{red}{Invalid MSG number}}
\fi
\fi
\fi
\else
\ifnum#1=205
\ifnum#2=33
$Pa\bar{3}$\else
\ifnum#2=34
$Pa\bar{3}1'$\else
\ifnum#2=35
$Pa'\bar{3}'$\else
\ifnum#2=36
$P_{I}a\bar{3}$\else
{\color{red}{Invalid MSG number}}
\fi
\fi
\fi
\fi
\else
\ifnum#1=206
\ifnum#2=37
$Ia\bar{3}$\else
\ifnum#2=38
$Ia\bar{3}1'$\else
\ifnum#2=39
$Ia'\bar{3}'$\else
{\color{red}{Invalid MSG number}}
\fi
\fi
\fi
\else
\ifnum#1=207
\ifnum#2=40
$P432$\else
\ifnum#2=41
$P4321'$\else
\ifnum#2=42
$P4'32'$\else
\ifnum#2=43
$P_{I}432$\else
{\color{red}{Invalid MSG number}}
\fi
\fi
\fi
\fi
\else
\ifnum#1=208
\ifnum#2=44
$P4_{2}32$\else
\ifnum#2=45
$P4_{2}321'$\else
\ifnum#2=46
$P4_{2}'32'$\else
\ifnum#2=47
$P_{I}4_{2}32$\else
{\color{red}{Invalid MSG number}}
\fi
\fi
\fi
\fi
\else
\ifnum#1=209
\ifnum#2=48
$F432$\else
\ifnum#2=49
$F4321'$\else
\ifnum#2=50
$F4'32'$\else
\ifnum#2=51
$F_{S}432$\else
{\color{red}{Invalid MSG number}}
\fi
\fi
\fi
\fi
\else
\ifnum#1=210
\ifnum#2=52
$F4_{1}32$\else
\ifnum#2=53
$F4_{1}321'$\else
\ifnum#2=54
$F4_{1}'32'$\else
\ifnum#2=55
$F_{S}4_{1}32$\else
{\color{red}{Invalid MSG number}}
\fi
\fi
\fi
\fi
\else
\ifnum#1=211
\ifnum#2=56
$I432$\else
\ifnum#2=57
$I4321'$\else
\ifnum#2=58
$I4'32'$\else
{\color{red}{Invalid MSG number}}
\fi
\fi
\fi
\else
\ifnum#1=212
\ifnum#2=59
$P4_{3}32$\else
\ifnum#2=60
$P4_{3}321'$\else
\ifnum#2=61
$P4_{3}'32'$\else
\ifnum#2=62
$P_{I}4_{3}32$\else
{\color{red}{Invalid MSG number}}
\fi
\fi
\fi
\fi
\else
\ifnum#1=213
\ifnum#2=63
$P4_{1}32$\else
\ifnum#2=64
$P4_{1}321'$\else
\ifnum#2=65
$P4_{1}'32'$\else
\ifnum#2=66
$P_{I}4_{1}32$\else
{\color{red}{Invalid MSG number}}
\fi
\fi
\fi
\fi
\else
\ifnum#1=214
\ifnum#2=67
$I4_{1}32$\else
\ifnum#2=68
$I4_{1}321'$\else
\ifnum#2=69
$I4_{1}'32'$\else
{\color{red}{Invalid MSG number}}
\fi
\fi
\fi
\else
\ifnum#1=215
\ifnum#2=70
$P\bar{4}3m$\else
\ifnum#2=71
$P\bar{4}3m1'$\else
\ifnum#2=72
$P\bar{4}'3m'$\else
\ifnum#2=73
$P_{I}\bar{4}3m$\else
{\color{red}{Invalid MSG number}}
\fi
\fi
\fi
\fi
\else
\ifnum#1=216
\ifnum#2=74
$F\bar{4}3m$\else
\ifnum#2=75
$F\bar{4}3m1'$\else
\ifnum#2=76
$F\bar{4}'3m'$\else
\ifnum#2=77
$F_{S}\bar{4}3m$\else
{\color{red}{Invalid MSG number}}
\fi
\fi
\fi
\fi
\else
\ifnum#1=217
\ifnum#2=78
$I\bar{4}3m$\else
\ifnum#2=79
$I\bar{4}3m1'$\else
\ifnum#2=80
$I\bar{4}'3m'$\else
{\color{red}{Invalid MSG number}}
\fi
\fi
\fi
\else
\ifnum#1=218
\ifnum#2=81
$P\bar{4}3n$\else
\ifnum#2=82
$P\bar{4}3n1'$\else
\ifnum#2=83
$P\bar{4}'3n'$\else
\ifnum#2=84
$P_{I}\bar{4}3n$\else
{\color{red}{Invalid MSG number}}
\fi
\fi
\fi
\fi
\else
\ifnum#1=219
\ifnum#2=85
$F\bar{4}3c$\else
\ifnum#2=86
$F\bar{4}3c1'$\else
\ifnum#2=87
$F\bar{4}'3c'$\else
\ifnum#2=88
$F_{S}\bar{4}3c$\else
{\color{red}{Invalid MSG number}}
\fi
\fi
\fi
\fi
\else
\ifnum#1=220
\ifnum#2=89
$I\bar{4}3d$\else
\ifnum#2=90
$I\bar{4}3d1'$\else
\ifnum#2=91
$I\bar{4}'3d'$\else
{\color{red}{Invalid MSG number}}
\fi
\fi
\fi
\else
\ifnum#1=221
\ifnum#2=92
$Pm\bar{3}m$\else
\ifnum#2=93
$Pm\bar{3}m1'$\else
\ifnum#2=94
$Pm'\bar{3}'m$\else
\ifnum#2=95
$Pm\bar{3}m'$\else
\ifnum#2=96
$Pm'\bar{3}'m'$\else
\ifnum#2=97
$P_{I}m\bar{3}m$\else
{\color{red}{Invalid MSG number}}
\fi
\fi
\fi
\fi
\fi
\fi
\else
\ifnum#1=222
\ifnum#2=98
$Pn\bar{3}n$\else
\ifnum#2=99
$Pn\bar{3}n1'$\else
\ifnum#2=100
$Pn'\bar{3}'n$\else
\ifnum#2=101
$Pn\bar{3}n'$\else
\ifnum#2=102
$Pn'\bar{3}'n'$\else
\ifnum#2=103
$P_{I}n\bar{3}n$\else
{\color{red}{Invalid MSG number}}
\fi
\fi
\fi
\fi
\fi
\fi
\else
\ifnum#1=223
\ifnum#2=104
$Pm\bar{3}n$\else
\ifnum#2=105
$Pm\bar{3}n1'$\else
\ifnum#2=106
$Pm'\bar{3}'n$\else
\ifnum#2=107
$Pm\bar{3}n'$\else
\ifnum#2=108
$Pm'\bar{3}'n'$\else
\ifnum#2=109
$P_{I}m\bar{3}n$\else
{\color{red}{Invalid MSG number}}
\fi
\fi
\fi
\fi
\fi
\fi
\else
\ifnum#1=224
\ifnum#2=110
$Pn\bar{3}m$\else
\ifnum#2=111
$Pn\bar{3}m1'$\else
\ifnum#2=112
$Pn'\bar{3}'m$\else
\ifnum#2=113
$Pn\bar{3}m'$\else
\ifnum#2=114
$Pn'\bar{3}'m'$\else
\ifnum#2=115
$P_{I}n\bar{3}m$\else
{\color{red}{Invalid MSG number}}
\fi
\fi
\fi
\fi
\fi
\fi
\else
\ifnum#1=225
\ifnum#2=116
$Fm\bar{3}m$\else
\ifnum#2=117
$Fm\bar{3}m1'$\else
\ifnum#2=118
$Fm'\bar{3}'m$\else
\ifnum#2=119
$Fm\bar{3}m'$\else
\ifnum#2=120
$Fm'\bar{3}'m'$\else
\ifnum#2=121
$F_{S}m\bar{3}m$\else
{\color{red}{Invalid MSG number}}
\fi
\fi
\fi
\fi
\fi
\fi
\else
\ifnum#1=226
\ifnum#2=122
$Fm\bar{3}c$\else
\ifnum#2=123
$Fm\bar{3}c1'$\else
\ifnum#2=124
$Fm'\bar{3}'c$\else
\ifnum#2=125
$Fm\bar{3}c'$\else
\ifnum#2=126
$Fm'\bar{3}'c'$\else
\ifnum#2=127
$F_{S}m\bar{3}c$\else
{\color{red}{Invalid MSG number}}
\fi
\fi
\fi
\fi
\fi
\fi
\else
\ifnum#1=227
\ifnum#2=128
$Fd\bar{3}m$\else
\ifnum#2=129
$Fd\bar{3}m1'$\else
\ifnum#2=130
$Fd'\bar{3}'m$\else
\ifnum#2=131
$Fd\bar{3}m'$\else
\ifnum#2=132
$Fd'\bar{3}'m'$\else
\ifnum#2=133
$F_{S}d\bar{3}m$\else
{\color{red}{Invalid MSG number}}
\fi
\fi
\fi
\fi
\fi
\fi
\else
\ifnum#1=228
\ifnum#2=134
$Fd\bar{3}c$\else
\ifnum#2=135
$Fd\bar{3}c1'$\else
\ifnum#2=136
$Fd'\bar{3}'c$\else
\ifnum#2=137
$Fd\bar{3}c'$\else
\ifnum#2=138
$Fd'\bar{3}'c'$\else
\ifnum#2=139
$F_{S}d\bar{3}c$\else
{\color{red}{Invalid MSG number}}
\fi
\fi
\fi
\fi
\fi
\fi
\else
\ifnum#1=229
\ifnum#2=140
$Im\bar{3}m$\else
\ifnum#2=141
$Im\bar{3}m1'$\else
\ifnum#2=142
$Im'\bar{3}'m$\else
\ifnum#2=143
$Im\bar{3}m'$\else
\ifnum#2=144
$Im'\bar{3}'m'$\else
{\color{red}{Invalid MSG number}}
\fi
\fi
\fi
\fi
\fi
\else
\ifnum#1=230
\ifnum#2=145
$Ia\bar{3}d$\else
\ifnum#2=146
$Ia\bar{3}d1'$\else
\ifnum#2=147
$Ia'\bar{3}'d$\else
\ifnum#2=148
$Ia\bar{3}d'$\else
\ifnum#2=149
$Ia'\bar{3}'d'$\else
{\color{red}{Invalid MSG number}}
\fi
\fi
\fi
\fi
\fi
\fi
\fi
\fi
\fi
\fi
\fi
\fi
\fi
\fi
\fi
\fi
\fi
\fi
\fi
\fi
\fi
\fi
\fi
\fi
\fi
\fi
\fi
\fi
\fi
\fi
\fi
\fi
\fi
\fi
\fi
\fi
\fi
\fi
\fi
\fi
\fi
\fi
\fi
\fi
\fi
\fi
\fi
\fi
\fi
\fi
\fi
\fi
\fi
\fi
\fi
\fi
\fi
\fi
\fi
\fi
\fi
\fi
\fi
\fi
\fi
\fi
\fi
\fi
\fi
\fi
\fi
\fi
\fi
\fi
\fi
\fi
\fi
\fi
\fi
\fi
\fi
\fi
\fi
\fi
\fi
\fi
\fi
\fi
\fi
\fi
\fi
\fi
\fi
\fi
\fi
\fi
\fi
\fi
\fi
\fi
\fi
\fi
\fi
\fi
\fi
\fi
\fi
\fi
\fi
\fi
\fi
\fi
\fi
\fi
\fi
\fi
\fi
\fi
\fi
\fi
\fi
\fi
\fi
\fi
\fi
\fi
\fi
\fi
\fi
\fi
\fi
\fi
\fi
\fi
\fi
\fi
\fi
\fi
\fi
\fi
\fi
\fi
\fi
\fi
\fi
\fi
\fi
\fi
\fi
\fi
\fi
\fi
\fi
\fi
\fi
\fi
\fi
\fi
\fi
\fi
\fi
\fi
\fi
\fi
\fi
\fi
\fi
\fi
\fi
\fi
\fi
\fi
\fi
\fi
\fi
\fi
\fi
\fi
\fi
\fi
\fi
\fi
\fi
\fi
\fi
\fi
\fi
\fi
\fi
\fi
\fi
\fi
\fi
\fi
\fi
\fi
\fi
\fi
\fi
\fi
\fi
\fi
\fi
\fi
\fi
\fi
\fi
\fi
\fi
\fi
\fi
\fi
\fi
\fi
\fi
\fi
\fi
\fi
\fi
\fi
\fi
\fi
\fi
\fi
\fi
\fi
\fi
\fi
\fi
\fi
\fi
\fi
\fi
\fi
\fi
}

\newcommand{\msgsymbnum}[2]{MSG #1.#2 (\msgsymb{#1}{#2})}

\begin{document}

\author{I\~{n}igo Robredo}
\affiliation{Donostia International Physics Center, P. Manuel de Lardizabal 4, 20018 Donostia-San Sebastian, Spain}
\affiliation{Max Planck Institute for Chemical Physics of Solids, 01187 Dresden, Germany}

\author{Yuanfeng Xu}
\affiliation{Center for Correlated Matter and School of Physics, Zhejiang University, Hangzhou 310058, China}

\author{Yi Jiang}
\affiliation{Donostia International Physics Center, P. Manuel de Lardizabal 4, 20018 Donostia-San Sebastian, Spain}

\author{Claudia Felser}
\affiliation{Max Planck Institute for Chemical Physics of Solids, 01187 Dresden, Germany}

\author{B. Andrei Bernevig}
\affiliation{Donostia International Physics Center, P. Manuel de Lardizabal 4, 20018 Donostia-San Sebastian, Spain}
\affiliation{Department of Physics, Princeton University, Princeton, New Jersey 08544, USA}
\affiliation{IKERBASQUE, Basque Foundation for Science, 48013 Bilbao, Spain}

\author{Luis Elcoro}
\affiliation{Department of Physics, University of the Basque Country UPV/EHU, Apartado 644, 48080 Bilbao, Spain}

\author{Nicolas Regnault}
\affiliation{Department of Physics, Princeton University, Princeton, New Jersey 08544, USA}
\affiliation{Laboratoire de Physique de l'Ecole normale sup\'{e}rieure, ENS, Universit\'{e} PSL, CNRS, Sorbonne Universit\'{e}, Universit\'{e} Paris-Diderot, Sorbonne Paris Cit\'{e}, 75005 Paris, France}

\author{Maia G. Vergniory}
\email{maia.vergniory@cpfs.mpg.de}
\affiliation{Donostia International Physics Center, P. Manuel de Lardizabal 4, 20018 Donostia-San Sebastian, Spain}
\affiliation{Max Planck Institute for Chemical Physics of Solids, 01187 Dresden, Germany}
\affiliation{D\'epartement de physique et Institut quantique, Universit\'e de Sherbrooke, Sherbrooke, Qu\'ebec, Canada J1K 2R1}

\title{New magnetic topological materials from high-throughput search}

\begin{abstract}
We conducted a high-throughput search for topological magnetic materials on {\MTQCDBRunTwoNbrBCSIDsWithPhaseDiagram} new, experimentally reported commensurate magnetic structures from \webBCSMAG, doubling the number of available materials on the {\webMTQC} database. This brings up to date the previous studies  \cite{ Magneticht,MTQC,completecatalogue,alltopobands} which had become incomplete due to  the discovery of new materials. 
For each material, we performed first-principle electronic calculations and diagnosed the topology as a function of the Hubbard $U$ parameter.
Our high-throughput calculation led us to the prediction of {\MTQCDBRunTwoNbrBCSIDsNonTrivialNonAtomic} experimentally relevant topologically non-trivial materials, which represent {\MTQCDBRunTwoNbrBCSIDsNonTrivialNonAtomicPercent} of the newly analyzed materials. 
We present five remarkable examples of these materials, each showcasing a different topological phase: \bcsidwebformula{1.508}{Mn${}_2$AlB${}_2$} (\bcsidmagndata{1.508}), which exhibits a nodal line semimetal to topological insulator transition as a function of SOC,
\bcsidwebformula{0.599}{CaMnSi} (\bcsidmagndata{0.599}), a narrow gap axion insulator,  \bcsidwebformula{0.594}{UAsS} (\bcsidmagndata{0.594}) a 5$f$-orbital Weyl semimetal, \bcsidwebformula{0.327}{CsMnF${}_4$} (\bcsidmagndata{0.327}), a material presenting a new type of quasi-symmetry protected closed nodal surface and \bcsidwebformula{0.613}{FeCr${}_2$S${}_4$} (\bcsidmagndata{0.613}), a symmetry-enforced semimetal with double Weyls and spin-polarised surface states.
\end{abstract}

\maketitle

\section{Introduction}\label{sec1}

For decades, electronic topological systems have been a research hotspot in the field of condensed matter physics. The development of topological quantum chemistry (TQC) \cite{TQC,MTQC} and symmetry-based indicators \cite{Fu-Kane,VishSI,SongSI,VishMSI,slager_space_2013} has facilitated the prediction of topological quantum materials in realistic systems. In particular, it allowed for high-throughput searches of topological quantum materials resulting in extensive databases \cite{completecatalogue,alltopobands,Chen_database,tang2019comprehensive} 
like the \webTQC. Recent developments have shown that the interplay of spin and charge degrees of freedom in magnetic systems can realize new topological phases such as axion insulators (AXI) \cite{li2010dynamical,mong2010antiferromagnetic,li2019intrinsic,zhang2019topological, xu2019higher},
or non-axionic magnetic higher-order topological insulators (MHOTI)
\cite{MTQC,Magneticht,doi:10.1126/sciadv.aat0346,Schindler2018, peng2022topological}.
These phases, unique to magnetic systems, combine the robustness of topological surface states with spin ordering, promising a wide array of applications, such as in materials for quantum computation \cite{Kitaev-QC}, spintronics \cite{Pesin2012} and catalysis \cite{Weyl_catalysis,3drsi2}. The current focus now lies on finding materials that display the aforementioned topological properties and can be readily grown. To enable the prediction and diagnosis of magnetic topological phases of matter, the formalism of TQC was extended to magnetic TQC (MTQC) \cite{MTQC}, resulting in the prediction of {\MTQCDBRunOneNbrBCSIDsNonTrivialNonAtomic} new magnetic topological materials \cite{Magneticht,3drsi}.

In this work, we use MTQC to conduct a high-throughput analysis on the new entries in {\webBCSMAG}, extending the {\webMTQC} \cite{Magneticht} database from {\MTQCDBRunOneNbrBCSIDsWithPhaseDiagram} to {\MTQCDBNbrBCSIDsWithPhaseDiagram} entries and adding {\MTQCDBRunTwoNbrBCSIDsWithPhaseDiagram} new experimentally reported stoichiometric magnetic materials with commensurate magnetic order. The main result of our work is the prediction of {\MTQCDBRunTwoNbrBCSIDsNonTrivialNonAtomic} experimentally relevant topologically non-trivial materials. We define a material as topologically non-trivial if we diagnose it as topological for at least one value of the interaction strength, parameterized by the Hubbard $U$ value \cite{Hubbard}. Interestingly, we found several examples of interaction-driven topology: systems in which the interaction can onset a topological phase transition.

In our search, we found {\MTQCDBRunTwoNbrBCSIDsTI} new topological insulators (TI), which can be broken down into different topological classes. We found {\MTQCDBRunTwoNbrBCSIDsAXI} axion insulators (AXI), {\MTQCDBRunTwoNbrBCSIDsMTCI} magnetic topological crystalline insulators (MTCI) and {\MTQCDBRunTwoNbrBCSIDsThreeDQAH} 3D quantum anomalous Hall (3DQAH) systems, as well as {\MTQCDBRunTwoNbrBCSIDsMOAI} magnetic obstructed atomic insulators (mOAI), a special case of topological insulators that can be explained in terms of localized orbitals but whose Wannier charge center is localized away from atomic positions \cite{TQC,Songrsi,3drsi,feoai,gao2022unconventional}.
The latter is relevant since their non-magnetic counterparts have been predicted to have good catalytic activity \cite{3drsi2,3drsi}; and magnetic systems are known to improve some catalytic reactions, such as oxygen evolution reaction (OER) \cite{OER_1, OER_2, OER_3}. We therefore speculate that mOAI may be promising materials displaying both features. In this category, we find a promising candidate, \bcsidwebformula{1.347}{CuFeO${}_2$} (\bcsidmagndata{1.347}), which holds the same topological diagnosis for all values of the Hubbard parameter $U$. We also predict {\MTQCDBRunTwoNbrBCSIDsSM} new enforced semimetals (ES) with symmetry-enforced degeneracies at high-symmetry points, lines or planes. One of the most relevant magnetic semimetals are Weyl semimetals, due to various proposed effects and applications, such as Veselago lenses, quantized circular photogalvanic effect or unconventional charge to spin conversion\cite{Weyl_app_1,Weyl_app_2,Weyl_app_3,Weyl_app_4}. Contrary to non-magnetic systems where predicting Weyl crossings is difficult in terms of symmetry indicators (SI) \cite{Nomag_Weyl_databse}, the MTQC formalism can predict Weyl crossings in magnetic systems under certain conditions \cite{MTQC}. Based on SI, MTQC can predict the difference in the Chern numbers between high-symmetry planes, which assures the existence of a quantum critical point along the momentum separating them. This degeneracy point carrying topological charge forces the system to be semi-metallic, thus indicating the presence of Weyl nodes. These systems are also known as Smith index semimetals or symmetry indicated Weyl semimetals \cite{SISM_chern,MTQC,Magneticht}.
In our high-throughput search, we found {\MTQCDBRunTwoNbrBCSIDsSISM} symmetry indicated Weyl semimetals, among which we found that the ferromagnetic Weyl semimetal \bcsidwebformula{0.860}{Co$_3$Sn$_2$S$_2$} (\bcsidmagndata{0.860}), which has been well-studied previously\cite{wang2018large,liu2018giant,liu2019magnetic}, can be reinterpreted by a topological index $\eta_{4I}=3$.

From our search results, we have identified five especially promising materials, two TIs and three ESs, which we selected for further analysis of their topological properties.
First, we present \bcsidwebformula{1.508}{Mn${}_2$AlB${}_2$} (\bcsidmagndata{1.508}), a layered antiferromagnetic system that displays a nodal line to TI transition when the spin-orbit coupling (SOC) is turned on. Then we consider materials within the CeFeSi family of small gap magnetic insulators \cite{CaMnSi-family}. We found that one material in the family, \bcsidwebformula{0.599}{CaMnSi} (\bcsidmagndata{0.599}), displays a band inversion that drives the topological transition to an AXI. In the ES class, we present three examples, namely \bcsidwebformula{0.594}{UAsS} (\bcsidmagndata{0.594}), \bcsidwebformula{0.327}{CsMnF${}_4$} (\bcsidmagndata{0.327}) and \bcsidwebformula{0.613}{FeCr${}_2$S${}_4$} (\bcsidmagndata{0.613}). The first is a ferromagnetic Weyl and nodal line semimetal with double Fermi arcs on the surface. The second is a layered square net ferromagnetic system that displays a symmetry-protected nodal line. While already reported in Ref.~\cite{CsMnF4_prev}, our analysis proves that there is a hidden quasi-symmetry \cite{Quasisymmetry1,Quasisymmetry2} that protects an almost closed 2D nodal surface, enhancing the semimetallic behavior at the Fermi level. Lastly, \bcsidwebformula{0.613}{FeCr$_2$S$_4$} (\bcsidmagndata{0.613}) is a ferrimagnetic system with symmetry-protected nodal lines and topologically protected Weyl nodes, as well as double Weyl nodes \cite{Multi-Weyl}. 

This work is divided as follows. In Section~\ref{sec:workflow}, we present the workflow and methods for the high-throughput search. In Section~\ref{sec:intmat}, we summarize the topological properties of the five selected materials. Section~\ref{sec:disc} provides the discussion of the main results of our search. In Appendix~\ref{app:mtqcreview} we give an overview of the formalism of MTQC and explain the origin of the symmetry indicators and real space invariants (RSI) \cite{Songrsi,3drsi} using magnetic space group (MSG) \msgsymb{2}{4} as an example.
Appendix~\ref{app:fullintmat} gives the full analysis of the topological properties of the selected example materials. Finally, Appendices~\ref{app:magmomcomp}, \ref{app:gaptopo}, \ref{app:ticlass} and \ref{app:fulltables} contain the full tables summarizing the results of our materials search: Appendix~\ref{app:magmomcomp} includes a comparison between experimental and calculated magnetic moments, Appendix~\ref{app:gaptopo} covers the topological classification and both direct and indirect gaps. Finally, in Appendix~\ref{app:ticlass} we give the physical interpretation of the systems with non-trivial symmetry indicators and Appendix~\ref{app:fulltables} contains detailed information on the crystal structure, magnetic space group, electronic band structure, and topological classification of all materials.

\section{Workflow}\label{sec:workflow}

In this section we present the main symmetry-diagnosable categories of topological materials and our computational methods.

\begin{figure}
    \centering
    \includegraphics[width=\linewidth]{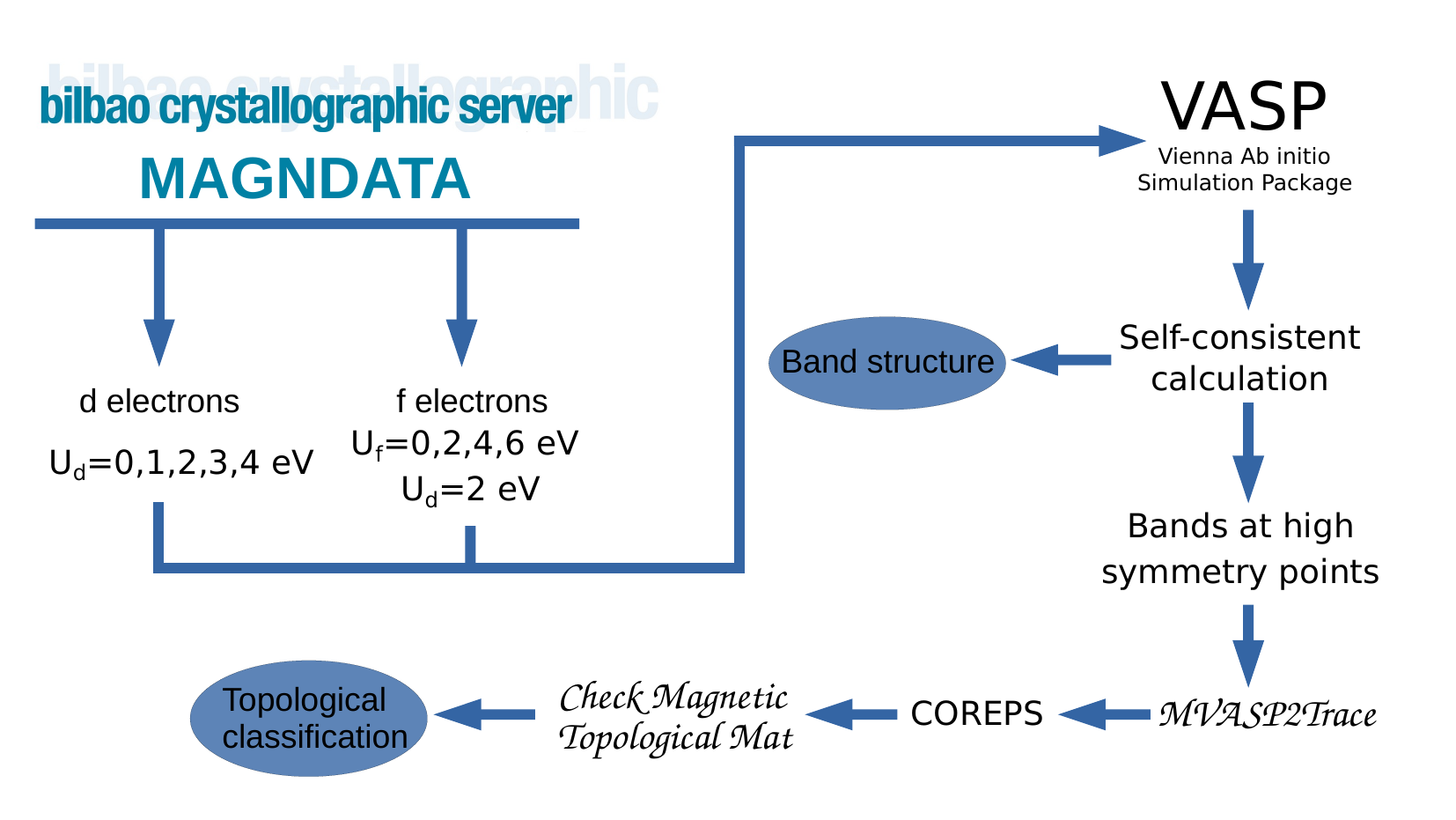}
    \caption{High throughput flowchart. We categorize the materials in MAGNDATA database into two groups: one containing \emph{d} electrons and no \emph{f} electrons, and another one containing \emph{f} electrons. We then conduct \emph{ab initio} calculations using VASP \cite{VASP1,VASP2,VASP3,VASP4} with multiple Hubbard $U$ parameters to obtain the band structure as well as the magnetic coreps \cite{bradley1972mathematical,MTQC}. Subsequently, the corresponding symmetry-data-vector can be uploaded to the BCS to obtain the topological classification.}
    \label{fig:flowchart}
\end{figure}

\begin{table*}[!htbp]
\centering
\resizebox{\textwidth}{!}{%
\begin{tabular}{|c|ccc|ccc|c|ccc|ccc|c|ccc|ccc|}
\toprule
MSG &  \multicolumn{3}{c|}{TI/OAI} & \multicolumn{3}{c|}{ES/ESFD} & MSG &  \multicolumn{3}{c|}{TI/OAI} & \multicolumn{3}{c|}{ES/ESFD} & MSG &  \multicolumn{3}{c|}{TI/OAI} & \multicolumn{3}{c|}{ES/ESFD}\\
\midrule
{}   & $U$=0   & $U$=2    & $U$=4 & $U$=0   & $U$=2    & $U$=4 & {} & $U$=0   & $U$=2    & $U$=4  & $U$=0   & $U$=2    & $U$=4 & {} & $U$=0   & $U$=2    & $U$=4 & $U$=0   & $U$=2    & $U$=4   \\\hline
 &  &  &  &  &  &  &  &  &  &  &  &  &  &  &  &  &  &  &  & \\ 
\msgsymb{2}{4} & 1 & 1 & 1 & 0 & 0 & 0 & \msgsymb{59}{410} & 0 & 0 & 0 & 2 & 2 & 1 & \msgsymb{122}{336} & 0 & 0 & 0 & 1 & 1 & 0\\ 
\msgsymb{2}{7} & 4 & 2 & 2 & 0 & 0 & 0 & \msgsymb{60}{431} & 1 & 0 & 1 & 0 & 1 & 0 & \msgsymb{123}{345} & 0 & 0 & 1 & 3 & 3 & 2\\ 
\msgsymb{11}{55} & 1 & 1 & 1 & 0 & 0 & 0 & \msgsymb{60}{432} & 2 & 2 & 0 & 0 & 1 & 3 & \msgsymb{124}{360} & 0 & 0 & 0 & 0 & 1 & 2\\ 
\msgsymb{11}{57} & 1 & 0 & 0 & 0 & 0 & 0 & \msgsymb{61}{439} & 0 & 0 & 1 & 1 & 1 & 0 & \msgsymb{126}{386} & 2 & 1 & 3 & 9 & 16 & 15\\ 
\msgsymb{12}{58} & 0 & 0 & 0 & 1 & 1 & 1 & \msgsymb{62}{441} & 4 & 1 & 0 & 0 & 0 & 0 & \msgsymb{127}{394} & 1 & 1 & 1 & 1 & 2 & 2\\ 
\msgsymb{12}{62} & 8 & 7 & 6 & 0 & 0 & 0 & \msgsymb{62}{446} & 2 & 1 & 1 & 2 & 1 & 1 & \msgsymb{128}{408} & 0 & 0 & 0 & 0 & 1 & 1\\ 
\msgsymb{12}{63} & 7 & 5 & 3 & 1 & 0 & 0 & \msgsymb{62}{448} & 1 & 0 & 0 & 2 & 2 & 1 & \msgsymb{128}{410} & 1 & 0 & 1 & 5 & 6 & 5\\ 
\msgsymb{13}{70} & 1 & 0 & 0 & 0 & 0 & 0 & \msgsymb{62}{449} & 0 & 0 & 0 & 2 & 0 & 0 & \msgsymb{129}{416} & 4 & 2 & 2 & 0 & 0 & 0\\ 
\msgsymb{14}{75} & 3 & 1 & 1 & 8 & 2 & 2 & \msgsymb{62}{450} & 2 & 2 & 2 & 1 & 0 & 0 & \msgsymb{129}{417} & 1 & 2 & 0 & 2 & 1 & 3\\ 
\msgsymb{14}{79} & 2 & 1 & 1 & 0 & 0 & 0 & \msgsymb{62}{451} & 0 & 0 & 0 & 1 & 0 & 1 & \msgsymb{130}{432} & 0 & 1 & 1 & 1 & 0 & 0\\ 
\msgsymb{14}{80} & 1 & 1 & 1 & 0 & 0 & 0 & \msgsymb{63}{459} & 0 & 0 & 0 & 2 & 2 & 1 & \msgsymb{139}{535} & 0 & 0 & 0 & 1 & 1 & 0\\ 
\msgsymb{14}{83} & 1 & 1 & 1 & 0 & 0 & 0 & \msgsymb{63}{462} & 0 & 0 & 2 & 3 & 4 & 2 & \msgsymb{139}{536} & 2 & 3 & 1 & 0 & 0 & 2\\ 
\msgsymb{14}{84} & 2 & 0 & 0 & 0 & 0 & 0 & \msgsymb{63}{463} & 1 & 0 & 0 & 1 & 2 & 3 & \msgsymb{139}{537} & 0 & 0 & 0 & 2 & 5 & 5\\ 
\msgsymb{15}{85} & 1 & 0 & 0 & 0 & 1 & 0 & \msgsymb{63}{464} & 1 & 0 & 0 & 10 & 10 & 9 & \msgsymb{140}{550} & 0 & 1 & 1 & 1 & 0 & 0\\ 
\msgsymb{15}{89} & 2 & 2 & 2 & 0 & 0 & 0 & \msgsymb{63}{466} & 2 & 0 & 2 & 0 & 2 & 0 & \msgsymb{141}{557} & 0 & 0 & 0 & 0 & 1 & 0\\ 
\msgsymb{15}{90} & 2 & 3 & 2 & 0 & 0 & 0 & \msgsymb{64}{480} & 5 & 2 & 1 & 1 & 1 & 2 & \msgsymb{166}{97} & 0 & 0 & 0 & 0 & 1 & 0\\ 
\msgsymb{15}{91} & 1 & 1 & 0 & 0 & 0 & 0 & \msgsymb{65}{486} & 0 & 0 & 0 & 5 & 5 & 3 & \msgsymb{166}{101} & 3 & 2 & 1 & 2 & 3 & 3\\ 
\msgsymb{36}{175} & 0 & 0 & 0 & 1 & 0 & 0 & \msgsymb{65}{489} & 1 & 0 & 1 & 0 & 0 & 0 & \msgsymb{167}{108} & 1 & 1 & 1 & 0 & 0 & 0\\ 
\msgsymb{38}{191} & 0 & 0 & 0 & 1 & 0 & 1 & \msgsymb{66}{498} & 0 & 0 & 1 & 1 & 1 & 0 & \msgsymb{189}{225} & 0 & 0 & 0 & 2 & 0 & 2\\ 
\msgsymb{39}{201} & 0 & 0 & 0 & 1 & 0 & 0 & \msgsymb{66}{500} & 1 & 1 & 0 & 0 & 0 & 0 & \msgsymb{191}{240} & 0 & 0 & 0 & 3 & 0 & 2\\ 
\msgsymb{44}{231} & 0 & 0 & 0 & 6 & 6 & 5 & \msgsymb{67}{509} & 3 & 3 & 3 & 0 & 0 & 0 & \msgsymb{192}{252} & 0 & 0 & 0 & 1 & 1 & 0\\ 
\msgsymb{46}{247} & 0 & 0 & 0 & 1 & 0 & 0 & \msgsymb{68}{520} & 1 & 1 & 1 & 0 & 0 & 0 & \msgsymb{194}{268} & 2 & 1 & 1 & 0 & 0 & 0\\ 
\msgsymb{47}{252} & 0 & 0 & 1 & 1 & 1 & 0 & \msgsymb{70}{530} & 0 & 0 & 0 & 1 & 0 & 0 & \msgsymb{194}{270} & 0 & 0 & 0 & 1 & 1 & 1\\ 
\msgsymb{52}{318} & 0 & 0 & 0 & 1 & 1 & 1 & \msgsymb{71}{535} & 0 & 0 & 0 & 1 & 2 & 2 & \msgsymb{198}{9} & 0 & 0 & 0 & 0 & 0 & 2\\ 
\msgsymb{54}{350} & 2 & 2 & 2 & 0 & 0 & 0 & \msgsymb{74}{559} & 0 & 0 & 0 & 1 & 1 & 1 & \msgsymb{227}{131} & 2 & 1 & 0 & 0 & 0 & 0\\ 
\msgsymb{58}{404} & 0 & 0 & 0 & 0 & 2 & 2 & \msgsymb{113}{273} & 0 & 0 & 1 & 0 & 0 & 0 & \msgsymb{229}{143} & 0 & 0 & 0 & 0 & 1 & 1\\ 
\msgsymb{59}{409} & 3 & 1 & 1 & 1 & 4 & 4 & \msgsymb{119}{319} & 0 & 0 & 0 & 0 & 1 & 1 &  &  &  &  &  &  & \\ 
 &  &  &  &  &  &  &  &  &  &  &  &  &  &  Total & 89 & 58 & 56 & 95 & 102 & 95\\ 
\bottomrule
\end{tabular}
}
\caption{Summary of topologically nontrivial materials per MSG as a function of Hubbard $U$ parameter. We group the topologically non-trivial systems into topological insulators (TI, which accounts for SEBR and NLC, and OAI) and enforced semimetals (both ES and ESFD). We display the $U$ values 0, 2 and 4 (in units of eV), which are the ones in common for all material types, both containing d and f electrons. The table contains the classification as a function of magnetic space group.}
\label{tab:topovsu}
\end{table*}

\subsection{MTQC brief overview}
The formalism of MTQC has been extensively described and reviewed in prior work \cite{MTQC,Magneticht,canomultifold,10.1209/0295-5075/ad6bbc}. Here we briefly review the fundamental building blocks and the resulting topological classification. The electronic band structures of solids can be classified according to the symmetry properties of the Bloch wavefunctions at high-symmetry momenta. This information is encoded in a vector containing the multiplicity of the irreducible co-representations (coreps) of the occupied set of bands called symmetry-data-vector. A symmetry-data-vector is identified as topologically trivial if it can be expressed as a sum of elementary band representations (EBR) with positive integer coefficients. The trivial symmetry-data-vectors, also called band representations (BR), conform the space of trivial insulators. Physically, the sum of EBRs can be understood as the addition of trivial atomic insulators. If a system can be expressed as a collection of trivial insulators, the studied band structure can be adiabatically connected to a trivial insulator, and it is thus, diagnosed as topologically trivial.

\begin{figure*}
    \centering
    \includegraphics[width=\linewidth]{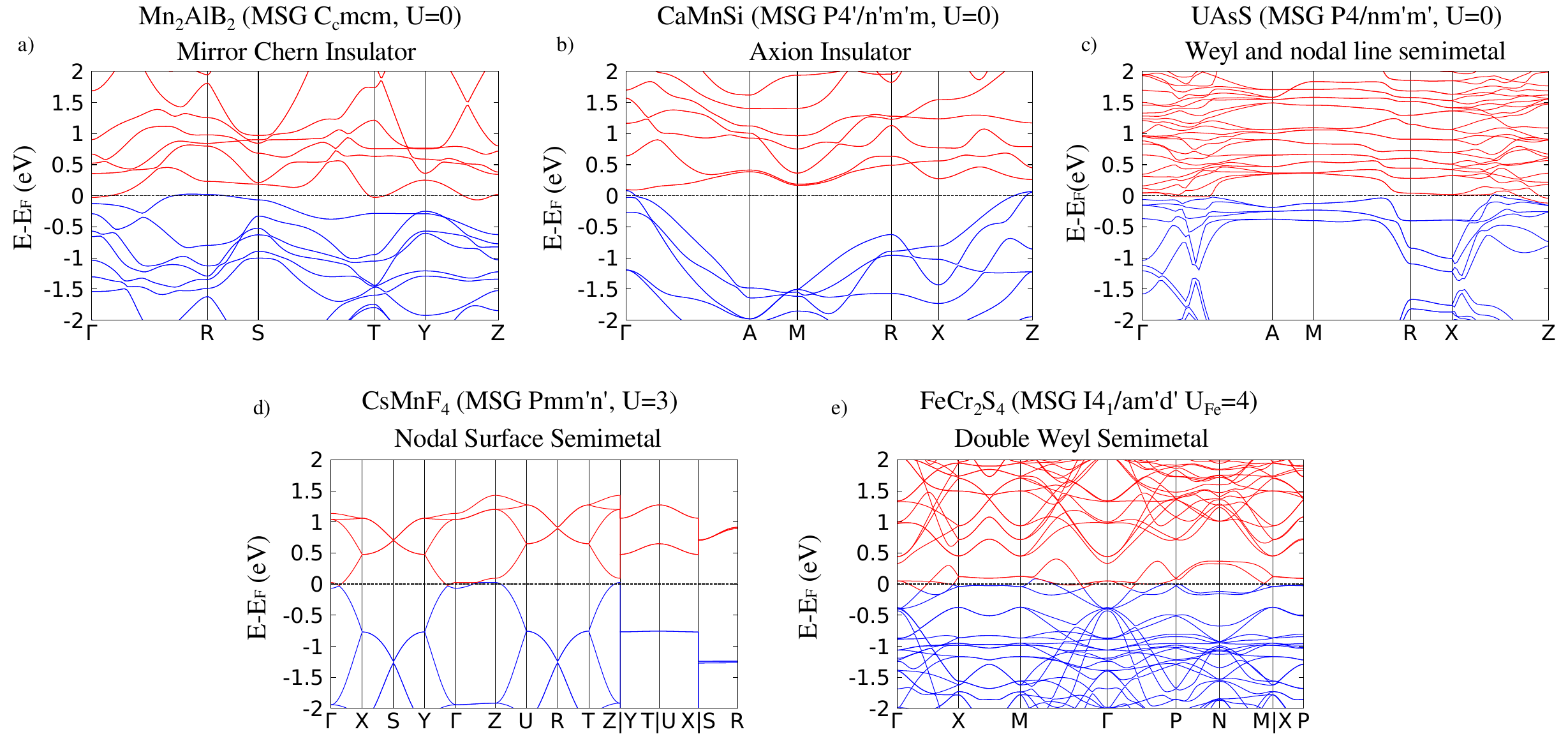}
    \caption{Electronic band structures for the five ideal topological systems: a) \bcsidwebformula{1.508}{Mn${}_2$AlB${}_2$} (\bcsidmagndata{1.508}), b) \bcsidwebformula{0.599}{CaMnSi} (\bcsidmagndata{0.599}), c) \bcsidwebformula{0.594}{UAsS} (\bcsidmagndata{0.594}), d) \bcsidwebformula{0.327}{CsMnF${}_4$} (\bcsidmagndata{0.327}) and e) \bcsidwebformula{0.613}{FeCr${}_2$S${}_4$} (\bcsidmagndata{0.613}). The blue (red) color stands for the valence (conduction) bands according to the band index. In the title above each figure, we display the chemical formula, MSG and Hubbard $U$ value used for the band structure and providing the best match for the magnetic moments. Under the title, we provide the topological classification.}
    \label{fig:Fig1}
\end{figure*}

There are 4 possible outcomes when we express a symmetry-data-vector in terms of EBRs:

\begin{itemize}
    \item Linear combination of EBRs (LCEBR). In this case, the symmetry-data-vector can be expressed as linear combination of EBRs with positive integer coefficients. Therefore, the system is compatible with a description in terms of localized magnetic orbitals in real space. In addition to atomic insulators (AI) that are topologically trivial, a topologically distinct phase exists, i.e., the mOAI. In the latter case, the occupied electronic bands are induced from localized magnetic orbitals away from the atomic positions and there exists an obstruction to reverting them to the original atomic location. These systems exhibit special surface states pinned by crystalline symmetries that do not connect valence and conduction bands and can be diagnosed via the RSI \cite{RSI,3drsi}.
    \item Linear combination of EBRs with, necessarily, at least one negative integer coefficient in the EBR decomposition (fragile). This solution is inconsistent with a trivial insulator, but the addition of extra orbitals to the system would render it topologically trivial \cite{TQC,MTQC,SongSI,VishSI,VishMSI}.
    \item Topological insulator. The solution of the decomposition into EBRs has fractional coefficients and indicates stable topology. For instance, AXI, MTCI or (M)HOTIs \cite{Axion1,Magneticht} belong to that category. They can be compactly diagnosed by symmetry indicators \cite{TQC,MTQC,SongSI,VishSI,VishMSI,Chen_database} (see Appendix \ref{app:mtqcreview} for further details).
    There are two subcategories, split EBR (SEBR), in which the solution can be expressed in terms of EBRs and parts of split EBRs, and `non linear combination' (NLC), in which case there is no decomposition into EBRs and parts of split EBRs.
    \item Enforced semimetal (ES). This case arises if the symmetry-data-vector \cite{TQC,MTQC,VishSI,VishMSI}
is incompatible with an insulator. There are two types of ESs: first, if by electron counting, a corep at a high symmetry point is only partially filled, we classify the material as enforced semimetal with Fermi degeneracy (ESFD). In case the Fermi level is clean from electron or hole pockets, the Fermi energy would exactly intersect the degenerate crossing enforced by symmetry. Second, even if the bands are gapped at every high-symmetry k-point, if the compatibility relations are not satisfied\cite{MTQC,Magneticht} the bands must be gapless at some point along a high-symmetry line or plane in the BZ. The formalism does not predict the exact point in which the crossing must occur, but it can predict certain high symmetry lines or planes in which the crossing happens.
   \item Smith-index SM (SISM). Unlike ESs in which the crossing is protected as well as predicted by symmetry, SISMs satisfy the compatibility relations \cite{TQC,MTQC}, i.e., the symmetry-data-vector is compatible with an insulator. However, a difference in the Chern number between high-symmetry planes assures the existence of a quantum critical point along the path \cite{SISM_chern}. This degeneracy point, carrying topological charge, forces the system to be semimetallic (e.g. Weyl semimetals).
    
\end{itemize}

The symmetry-based diagnosis poses certain limitations.
In general, symmetry-based topological diagnosis is a sufficient condition to classify a compound as topologically non-trivial but it is not a necessary condition.
Therefore, in cases with scarce symmetry, MTQC can diagnose nontrivial topology as trivial. This formalism, however, serves as a computationally inexpensive diagnosing method (as opposed to, e.g., Wilson loops) for high-throughput prediction of topological materials \cite{alltopobands,Magneticht,completecatalogue,Chen_database}.

\begin{figure*}[t]
    \centering
    \includegraphics[width=0.9\linewidth]{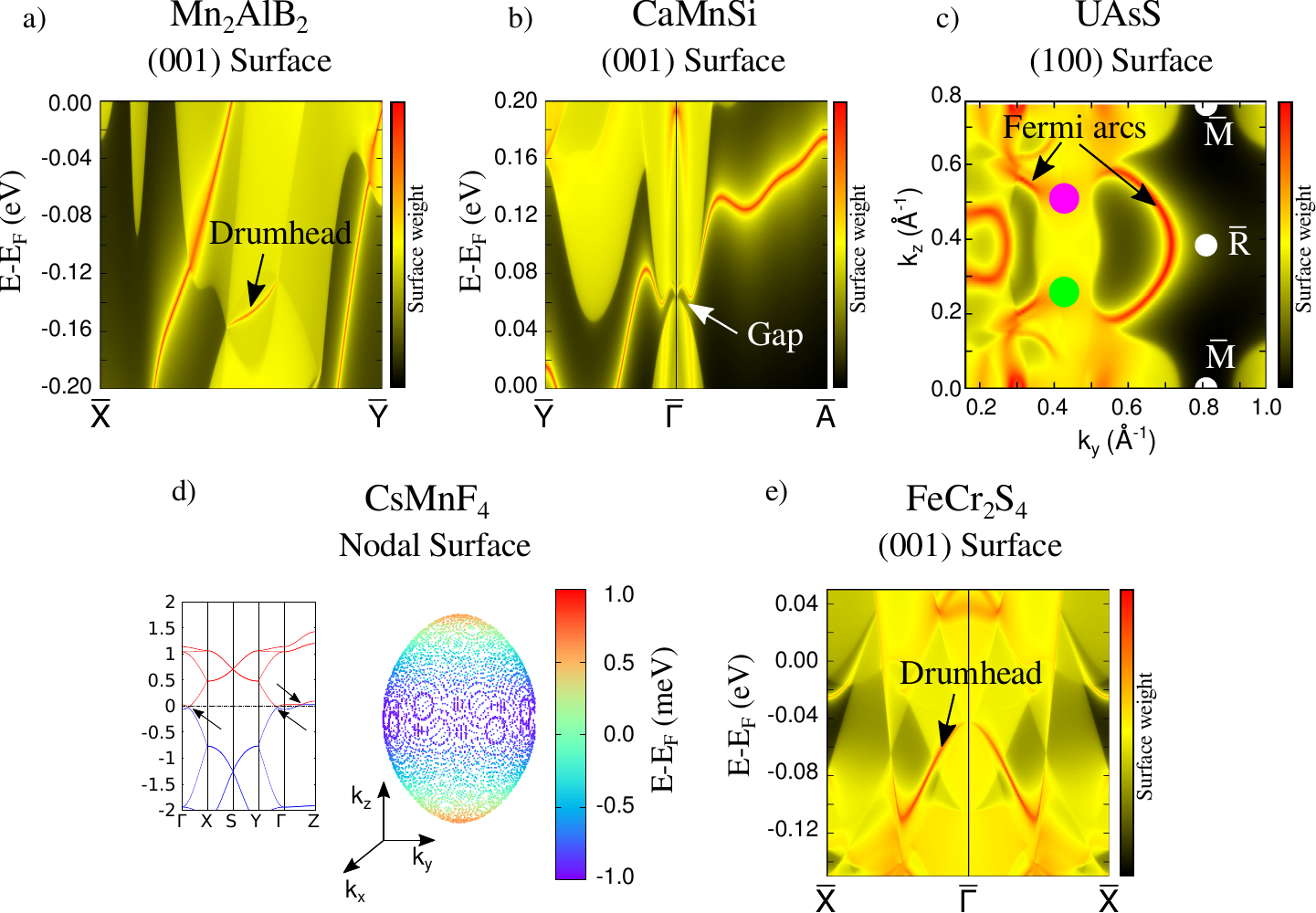}
    \caption{Surface spectrum and topological properties of the high-quality systems. For a), b), c) and e) we provide the surface direction in the title of the plot. a) Drumhead surface states of \bcsidwebformula{1.508}{Mn${}_2$AlB${}_2$} (\bcsidmagndata{1.508}) without SOC connecting the projections of the bulk nodal line in the surface. b) Gaped surface spectrum of \bcsidwebformula{0.599}{CaMnSi} (\bcsidmagndata{0.599}) axion insulator. c) Double Fermi arcs connecting the projection of Weyl nodes of \bcsidwebformula{0.594}{UAsS} (\bcsidmagndata{0.594}) at the Fermi level (Weyl projection of charge 2 in magenta, Weyl projection of charge -2 in green). d) 2D closed nodal surface enforced by quasi-symmetry of \bcsidwebformula{0.327}{CsMnF${}_4$} (\bcsidmagndata{0.327}). The arrows point at the nodal surface location in the band structure. e) Drumhead surface states of SOC \bcsidwebformula{0.613}{FeCr${}_2$S${}_4$} (\bcsidmagndata{0.613}). }
    \label{fig:main_surface}
\end{figure*}

\subsection{Methods}

We start by selecting {\MTQCDBRunTwoNbrBCSIDsWithPhaseDiagram} new material entries from the BCS magnetic material database {\webBCSMAG} \cite{MAGNDATA1,MAGNDATA2}. In this database, both the crystalline and magnetic structures have been derived from experiments. As the determination of the magnetic ground state is a computationally challenging task, we initialize our calculations with the experimental magnetization and ensure their self-consistent convergence. This is expected to yield much more accurate results than determining the magnetic ordering from first principles. The process will generally converge to distinct magnetic structures as a function of the Hubbard $U$ parameter, which accounts for the localization properties of magnetic elements' electrons in d- and f-shells. We summarize the resulting magnetization as a function of the Hubbard $U$ parameter in Appendix \ref{app:magmomcomp}. Based on these results, we can estimate the most probable topological classification for a given material, by selecting the one that deviates the least from the experimentally reported magnetic structure.

We performed DFT calculations as implemented in the Vienna ab initio simulation package (VASP) \cite{VASP1, VASP2, VASP3, VASP4}. The interaction between the ion cores and valence electrons was treated by the projector augmented-wave method \cite{PAW}, the generalized gradient approximation (GGA) was employed for the exchange-correlation potential with the Perdew–Burke–Ernzerhof functional for solid parameterization \cite{PBE}, and the spin-orbit coupling (SOC) was considered based on the second variation method \cite{DFT-SOC}. A $\Gamma$-centered Monkhorst-Pack k-point grid of ($9\times9\times9$) was used for reciprocal space integration and 500 eV energy cutoff of the plane-wave expansion.
The grid density of the largest materials was reduced up to ($5\times5\times5$) k-points. For materials with d-electron elements, we performed DFT+U calculations with Hubbard $U$ parameter values of $U$ = 0, 1, 2, 3, 4 eV. For f-electron elements, that tend to be more localized, we used larger Hubbard $U$ values, $U$ = 0, 2, 4, 6 eV. In the presence of both d- and f-electron elements, we set the Hubbard $U$ value of the d-electron element to $U$ = 2 eV and vary the f-electron element $U$ parameter, following the criteria of Ref.~\cite{Magneticht}. All materials presented herein were converged self-consistently to a magnetic ground state with an accuracy of $10^{-5}$ eV per unit cell.
We then run two non-self-consistent band calculations; one along high symmetry lines to get the electronic band dispersion and another one at the high-symmetry points to obtain the wavefunctions. We use \textit{MagVASP2trace} \cite{MTQC,Magneticht} on the latter to obtain the magnetic coreps and BR of the occupied set of bands, which we then upload to the BCS server \cite{BCS1,BCS2,BCS3} for symmetry-based topological prediction. The workflow is schematically depicted in Fig. \ref{fig:flowchart}.

\section{Ideal topological materials}\label{sec:intmat}

As representatives of the different topological classes found on this work, we selected five  examples, each showcasing a different topological phase: \bcsidwebformula{1.508}{Mn${}_2$AlB${}_2$} (\bcsidmagndata{1.508}), which exhibits a nodal line semimetal to topological insulator transition, \bcsidwebformula{0.599}{CaMnSi} (\bcsidmagndata{0.599}), a narrow gap axion insulator,
\bcsidwebformula{0.594}{UAsS} (\bcsidmagndata{0.594}), a 5f-shell ferromagnetic Weyl and nodal line semimetal, \bcsidwebformula{0.327}{CsMnF${}_4$} (\bcsidmagndata{0.327}), a material presenting a new type of quasi-symmetry protected closed nodal surface, and \bcsidwebformula{0.613}{FeCr${}_2$S${}_4$} (\bcsidmagndata{0.613}), a symmetry-enforced ferrimagnetic semimetal with double Weyls and spin-polarised surface states.

As an example of interaction strength-dependent topology, we selected \bcsidwebformula{1.508}{Mn${}_2$AlB${}_2$} (\bcsidmagndata{1.508}) in \msgsymbnum{63}{466} (see Fig. \ref{fig:Fig1}a). Interestingly, the case with $U = 0$ most accurately reproduced the experimentally measured magnetic momenta. This is also the only $U$ value for which the system is a topological insulator.
This points to \bcsidwebformula{1.508}{Mn${}_2$AlB${}_2$} being an itinerant ferromagnet (Stoner model of magnetism \cite{Stoner_model, Zeller_spin-polarizeddft,itin_mag_1,itin_mag_2}), in which magnetism does not originate from localized electrons. In particular, it is diagnosed as $\eta_{4I}=2,\delta_{2m}=1$. The $\eta_{4I}=2$ symmetry indicator implies that the system is in the non-trivial axionic phase, that is, the axion angle is quantized to $\theta=\pi$. Therefore, if the surface is fully gapped, the system will present half-quantized values of surface anomalous Hall conductivity\cite{yue2019symmetry,Axion1,xu2019higher,Axion2,Axion3}.  The other non-trivial symmetry indicator, $\delta_{2m}=1$, indicates a difference in the mirror Chern number of the $k_z=0,\pi$ planes. This forces the mirror Chern number to be nonzero in at least one of the planes, which implies the presence of symmetry-protected Dirac nodes on the surfaces preserving the mirror symmetry. Upon neglecting the SOC, we found symmetry-protected nodal lines, with their corresponding drumhead surface states (see Fig. \ref{fig:main_surface}a).

\bcsidwebformula{0.599}{CaMnSi} (\bcsidmagndata{0.599}), in \msgsymbnum{129}{416}, is an example of AXI which is indicated by index $z_2=1$ (see Fig. \ref{fig:Fig1}b). One of the major distinctions between AXI and 3DTIs is that the topological surface Dirac cones of a 3DTI are gapped in the AXI phase because of the TRS breaking (Fig. \ref{fig:main_surface}b). As stated previously in the case of \bcsidwebformula{1.508}{Mn${}_2$AlB${}_2$}, if the AXI surface is fully gapped, it presents a half-quantized surface anomalous Hall conductivity \cite{Axion1,Axion2,Axion3}. AXIs are also predicted to display large magnetoelectric response, as well as quantized Kerr rotation \cite{Axion4}. AXI materials are also a promising platform to study Majorana edge modes \cite{Axion5}. The computed gap of \bcsidwebformula{0.599}{CaMnSi} using PBE functional is small, and the topology is extremely sensitive to gap closings. Thus, we improved the accuracy of gap prediction by repeating the calculation with a meta-GGA functional (modified Becke–Johnson functional \cite{MBJ}). The smallest direct gap using this functional is 6.1 meV, and the symmetry-based topological diagnosis remains unchanged, which reassures the topological diagnosis (see Appendix~\ref {app:fullintmat} for further details).

We choose \bcsidwebformula{0.594}{UAsS} (\bcsidmagndata{0.594}) in \msgsymbnum{129}{417} as an example of 5f-electron heavy fermion ferromagnetic semimetal (see Fig. \ref{fig:Fig1}c). We selected this compound from the family due to its relative simplicity at the Fermi level. Having a clean Fermi surface allows an easier identification of topological signatures. Similar to \bcsidwebformula{1.508}{Mn${}_2$AlB${}_2$}, the Hubbard $U$ value that fits best is $U=0$eV. We find Weyl nodes and nodal lines, both with their characteristic surface states. In particular, even though there are only single-Weyls (topological charge 1), when projecting on the (100) surface termination, equal charge pairs of Weyl nodes are projected on the same point, thus giving rise to two Fermi arcs stemming from each of their surface projections (Fig. \ref{fig:main_surface}c).
Heavy fermion Weyl semimetal systems are of particular interest to study the interplay of electronic correlations and topological protection of Weyl nodes, such as Weyl-Kondo systems \cite{xu2017heavy,Heavy_Weyl}.

We also study a 3d-electron ferromagnetic enforced semimetal, \bcsidwebformula{0.327}{CsMnF${}_4$} (\bcsidmagndata{0.327}), in \msgsymbnum{59}{410} (see Fig. \ref{fig:Fig1}d). This material is an example of a ferromagnetic half metal, where one spin channel is fully gapped and the bands close to the Fermi level are spin polarized. Note that there are two different reported structures for this material \cite{CsMnF4_1,CsMnF4_2}. We choose the first structure for this work \cite{CsMnF4_1}, which is the one available in {\webBCSMAG} and has been studied with DFT before \cite{CsMnF4_prev}.
We find almost no change in magnetization with respect to the Hubbard $U$ parameter: the system remains half-metallic for all $U$ values but the band structure and topological properties are affected by $U$. Our calculations show that this system is an ES in the range $U \leq 3$eV and presents a topological transition from enforced semimetal to trivial insulator at $U$=4eV (see Appendix \ref{app:fulltables}). Further analysis identifies an approximate nodal surface in this phase (Fig.~\ref{fig:main_surface}d). The bands close to the Fermi level mainly stem from Mn atoms, which form a square lattice with a 2x2 supercell. As we detail in Appendix~\ref{app:fullintmat}, the extra translational symmetry associated with the 2x2 supercell is responsible for folding the bands and creating the approximate nodal surface.
The extra translational symmetry prevents the gap from opening (at first order), thereby demonstrating the recently proposed quasisymmetric-protected phase \cite{Quasisymmetry1,Quasisymmetry2}, which can enforce small, non-vanishing gaps that can be sources of large Berry curvature. Full details on the system can be found in Appendix~\ref{app:fullintmat}.

Finally, we analyze \bcsidwebformula{0.613}{FeCr${}_2$S${}_4$} (\bcsidmagndata{0.613}) in \msgsymbnum{141}{557}, which is an example of ferrimagnetic system with a clean Fermi level, where the low-energy physics is governed by a few connected bands (see Fig. \ref{fig:Fig1}e). In particular, there are 4 bands close to the Fermi level, and the relevant crossings occur between the last valence and the first conduction bands. We observed multi-Weyl nodes with a topological charge of 2 along the $\Gamma-$M line and a mirror symmetry-protected nodal line at $k_z=0$. The charge $C>1$ of these Weyl nodes is enforced by the 4-fold symmetry axis that leaves the $\Gamma-$M line invariant \cite{Multiweyl_1, Multiweyl_2, Multiweyl_3} and gives rise to 2 Fermi arcs per node. The nodal line gives rise to drumhead surface states in the (001) surface termination, whereas the other remnant gapped nodal lines' drumhead states are also gapped. The 4 bands crossing the Fermi level are purely spin-polarised: the resulting surface states are also spin-polarised.

\section{Discussion}\label{sec:disc}

In this study, we performed complete DFT+U calculations and topological diagnosis of {\MTQCDBRunTwoNbrBCSIDsWithPhaseDiagram} new experimentally reported materials from the {\webBCSMAG} database. The main result is the addition of {\MTQCDBRunTwoNbrBCSIDsNonTrivialNonAtomic} new experimentally relevant topological materials to the {\webMTQC} database, which now counts with {\MTQCDBNbrBCSIDsWithPhaseDiagram} entries, out of which {\MTQCDBNbrBCSIDsNonTrivialNonAtomicPercent} are reported as topologically nontrivial. We present a summary of topologically classified materials in Table \ref{tab:topovsu}. Among these, we report {\MTQCDBRunTwoNbrBCSIDsMOAI} new mOAI insulators, although most of them failed to satisfy our criteria of ideal topological material, in this case, having an indirect gap. The most promising candidate, \bcsidwebformula{1.347}{CuFeO${}_2$} (\bcsidmagndata{1.347}) is an mOAI for all Hubbard $U$ values.
During the preparation of this manuscript it was predicted that \bcsidwebformula{0.528}{CrSb} (\bcsidmagndata{0.528}) is an altermagnet Weyl semimetal \cite{Reimers2024,li2024topologicalweylaltermagnetismcrsb,lu2024observationsurfacefermiarcs}. Since it has an even number of Weyl nodes in each half of the BZ, its topology cannot be diagnosed using symmetry indicators. However, we discovered that the system is an mOAI at the Fermi level, providing new insights into the topological properties of this highly interesting material.

Since all additions to the {\webMTQC} database arise from experimentally reported systems, they are readily available to be grown and have their topological properties tested. Systems with stable topology as a function of the Hubbard $U$ parameter are the most promising ones to display topological properties, like \bcsidwebformula{0.599}{CaMnSi}. However, systems with variable topology can result in interesting platforms to study the interplay between topology and electron interactions, like the \bcsidwebformula{1.508}{Mn${}_2$AlB${}_2$} (\bcsidmagndata{1.508}) insulator or the \bcsidwebformula{0.613}{FeCr${}_2$S${}_4$} (\bcsidmagndata{0.613}) semimetal. The five materials we have selected exhibit the most distinct experimental signatures of their topological properties that can be measured experimentally.

\section{Acknowledgements}

M.G.V. and C.F. thank support from the Deutsche Forschungsgemeinschaft (DFG, German Research Foundation) GA3314/1-1 -FOR 5249 (QUAST).
M.G.V and I.R. thank support to the Spanish Ministerio de Ciencia e Innovacion grant PID2022-142008NB-I00 and the Ministry for Digital Transformation and of Civil Service of the Spanish Government through the QUANTUM ENIA project call - Quantum Spain project, and by the European Union through the Recovery, Transformation and Resilience Plan - NextGenerationEU within the framework of the Digital Spain 2026 Agenda. 
This project was partially supported by the European Research Council (ERC) under the European Union’s Horizon 2020 Research and Innovation Programme (Grant Agreement No. 101020833).
L.E. was supported by the Government of the Basque Country (Project IT1458-22) and the Spanish Ministry of Science and Innovation (PID2019-106644GB-I00).
B.A.B was supported by the Gordon and Betty Moore Foundation through Grant No.GBMF8685 towards the Princeton theory program, the Gordon and Betty Moore Foundation’s EPiQS Initiative (Grant No. GBMF11070), Office of Naval Research (ONR Grant No. N00014-20-1-2303), Global Collaborative Network Grant at Princeton University, BSF Israel US foundation No. 2018226, NSF-MERSEC (Grant No.MERSEC DMR 2011750), the Simons theory collaboration on frontiers of superconductivity, and Simons collaboration on mathematical sciences. 
Y.X. was supported by the National Natural Science Foundation of China (General Program no. 12374163) and the Fundamental Research Funds for the Central Universities (grant no. 226-2024-00200).
Y.J. is supported by the European Research Council (ERC) under the European Union’s Horizon 2020 research and innovation program (Grant Agreement No. 101020833) as well as by IKUR Strategy.

\bibliography{biblio.bib}

\onecolumngrid

\clearpage
\appendix

\setcounter{figure}{0}
\renewcommand{\figurename}{FIG.}
\renewcommand{\thefigure}{S\arabic{figure}}

\setcounter{table}{0}
\renewcommand{\tablename}{TAB.}
\renewcommand{\thetable}{S\arabic{table}}

\section{MTQC summary, symmetry indicators and real space invariants} \label{app:mtqcreview}

In this section, we will provide a concise overview of key topics, including the theory of magnetic topological quantum chemistry (TQC) \cite{TQC,MTQC},
topological phases determined by symmetry eigenvalues \cite{SongSI,elcorosmithSI,VishSI,VishMSI},
the definition of OAIs and RSI \cite{RSI,3drsi,3drsi2}.

\subsection{Magnetic Topological Quantum Chemistry}

Magnetic topological quantum chemistry (MTQC) integrates the principles of topological quantum chemistry (TQC) combined with magnetism and enables the investigation of topological phenomena emerging in magnetic materials. TQC \cite{TQC, MTQC} is a theory that characterizes the topology of electronic bands in crystalline solids and it is constructed on the basis of trivial atomic limits, which transform as elementary band representations (EBRs) \cite{Zak1,Zak2,Zak3,Bacry1988}. MTQC tabulates all the magnetic EBRs (MEBRs) in the 1651 magnetic space groups (MSGs). In particular, even within MSGs where symmetry eigenvalue labels cannot distinguish between trivial and topological states, the real-space indicators obtained from MEBRs describe stable and fragile topological insulating phases as well as magnetic OAIs. MTQC also tabulates  all band co-representations, described in momentum space, induced from magnetic atomic (Wannier) orbitals in position space (Wyckoff positions). MTQC provides a complete characterization of the symmetry properties of an isolated band through its decomposition into irreducible co-reps, where only the multiplicities ($m$) of the decomposition into (co)-irreps ($\chi^i_{k_j}$) at the high symmetry momenta (maximal K-points) are necessary. In general, we compute the multiplicities for the first $N_e$ bands (occupied bands, regardless of the Fermi level) and define the symmetry-data-vector $B$ of the occupied bands 
as follows:
\begin{equation}
B=\left(m(\chi^1_{K_1}),m(\chi^2_{K_1}),...,m(\chi^1_{K_2}),...,\right),
\end{equation}
where $m(\chi^i_{K_j}$) denotes the multiplicity of the $\chi^i$ irrep of the little cogroup of $K_j$. The symmetry properties of a set of occupied bands are fully described by the symmetry-data-vector $B$. The set of bands (and then the compound itself) is diagnosed as trivial whenever it is possible to describe the symmetry data vector $B$ as a linear integer (and positive) combination of the corresponding symmetry-data-vectors of the MEBRs. If such a combination does not exist, the set of bands is classified as topological.

Following the procedure that we will further detail in the next subsection, the solution of this problem can be encoded in a set of so-called symmetry indicators (SI) \cite{SongSI,VishSI,VishMSI,TQC,MTQC}, that are mapped to topological invariants. As discussed in the main text, various topological phases of matter can be diagnosed based on this analysis (see Sec. \ref{sec:workflow}). However, a band representation is well-defined only for insulators that satisfy the compatibility relations \cite{PhysRevE.96.023310,PhysRevLett.120.266401}. There are two main cases in which these relations are not satisfied: enforced semimetal with Fermi degeneracy (ESFD) and enforced semimetal (ES). In the former case, there is an $n$-dimensional corep at a high symmetry point which is occupied by $d$ number of electrons, with $d<n$. There, the band structure can be adiabatically deformed until the Fermi level precisely intersects the degeneracy point, and the Fermi surface becomes a single point. This is why they are referred to as "enforced semimetal with Fermi degeneracy". The situation with enforced semimetals is more intricate, as the bands exhibit local gaps at each high-symmetry k-point in the Brillouin Zone (BZ). Consider two maximal k-points $K_{a}$, $K_{b}$ and a path $k_p$ connecting them, which means that the co-irreps of the little cogroup associated with $k_p$ can be subduced from irreps at both $K_{a}$ and $K_{b}$. If the subsets of coreps of the $N_e$ occupied bands at both momentum maximal k-points  ($K_{a}$ and $K_{b}$) subduce into two different subsets of coreps at $k_p$, the $N_e$ bands cannot be  connected. At least one band (corep) at each maximal k-vector is connected to another band (corep) above the first $N_e$ bands in the other maximal k-vector. Therefore, at least two bands cross along the path $k_p$ and the Fermi level also crosses both bands. In this situation, the diagnosis is ES.

In the following sections we derive the symmetry indicators for stable topology in \msgsymbnum{2}{4} as well as the real space invariants, which are analogous to the symmetry indicators and can diagnose magnetic obstructed atomic insulators (see Sec. \ref{sec:workflow}).

\subsection{Symmetry indicators}
For illustration purposes, in this section we will derive the symmetry indicators and real space invariants of SG P$\bar{1}$. In three dimensions, the high-symmetry k-points in the BZ are labeled $\Gamma=(0,0,0)$, $X=(\pi,0,0)$, $Y=(0,\pi,0)$, $V=(\pi,\pi,0)$, $Z=(0,0,\pi)$, $U=(\pi,0,\pi)$, $T=(0,\pi,\pi)$ and $R=(\pi,\pi,\pi)$. The irreps of this MSG are particularly simple and can be labeled by the inversion eigenvalue, $\pm1\equiv\pm$. The symmetry information for any set of bands within this MSG can be encoded in an array that contains the multiplicities of the irreps, as previously described:
\begin{equation}\label{eq:brdef}
\begin{split}
    BR=&\big\{n(\Gamma^+),n(\Gamma^-),n(R^+),n(R^-),n(T^+),n(T^-),n(U^+),n(U^-),\\
    &n(V^+),n(V^-),n(X^+),n(X^-),n(Y^+),n(Y^-),n(Z^+),n(Z^-)\big\}.
\end{split}
\end{equation}

In \msgsymbnum{2}{4} all lines connecting high-symmetry points have the trivial little cogroup, so the compatibility relations reduce to having the same number of occupied bands (electrons) at each high-symmetry point. In terms of the coreps:
\begin{equation}\label{eq:comprel}
\begin{split}
    &n(\Gamma^+)+n(\Gamma^-)=n(X^+)+n(X^-)=n(Y^+)+n(Y^-)=n(V^+)+n(V^-)= \\
    &n(Z^+)+n(Z^-)=n(U^+)+n(U^-)=n(T^+)+n(T^-)=n(R^+)+n(R^-).
\end{split}
\end{equation}

There are $2^8=256$ possible insulating band structures with $n(\rho^+)+n(\rho^-)=1$ that satisfy the compatibility relations in \msgsymbnum{2}{4}. To classify the topological phases associated with all these band structures, we start with a basis of EBRs which span the space of trivial insulators within this specific MSG. This basis is not unique; we choose the convention of the BCS \cite{BCS1,BCS2,BCS3}, and we write the EBR matrix from BANDREP.  With the basis spanning the EBRs induced from alternating even/odd irreducible representations from the maximal 1a-1h Wyckoff positions:
\begin{equation}\label{eq:ebrmsg2}
    EBR=\left(
\begin{array}{cccccccccccccccc}
 1 & 0 & 1 & 0 & 1 & 0 & 1 & 0 & 1 & 0 & 1 & 0 & 1 & 0 & 1 & 0 \\
 0 & 1 & 0 & 1 & 0 & 1 & 0 & 1 & 0 & 1 & 0 & 1 & 0 & 1 & 0 & 1 \\
 1 & 0 & 0 & 1 & 0 & 1 & 0 & 1 & 1 & 0 & 1 & 0 & 1 & 0 & 0 & 1 \\
 0 & 1 & 1 & 0 & 1 & 0 & 1 & 0 & 0 & 1 & 0 & 1 & 0 & 1 & 1 & 0 \\
 1 & 0 & 0 & 1 & 0 & 1 & 1 & 0 & 0 & 1 & 0 & 1 & 1 & 0 & 1 & 0 \\
 0 & 1 & 1 & 0 & 1 & 0 & 0 & 1 & 1 & 0 & 1 & 0 & 0 & 1 & 0 & 1 \\
 1 & 0 & 0 & 1 & 1 & 0 & 0 & 1 & 0 & 1 & 1 & 0 & 0 & 1 & 1 & 0 \\
 0 & 1 & 1 & 0 & 0 & 1 & 1 & 0 & 1 & 0 & 0 & 1 & 1 & 0 & 0 & 1 \\
 1 & 0 & 1 & 0 & 0 & 1 & 0 & 1 & 1 & 0 & 0 & 1 & 0 & 1 & 1 & 0 \\
 0 & 1 & 0 & 1 & 1 & 0 & 1 & 0 & 0 & 1 & 1 & 0 & 1 & 0 & 0 & 1 \\
 1 & 0 & 1 & 0 & 1 & 0 & 0 & 1 & 0 & 1 & 0 & 1 & 1 & 0 & 0 & 1 \\
 0 & 1 & 0 & 1 & 0 & 1 & 1 & 0 & 1 & 0 & 1 & 0 & 0 & 1 & 1 & 0 \\
 1 & 0 & 1 & 0 & 0 & 1 & 1 & 0 & 0 & 1 & 1 & 0 & 0 & 1 & 0 & 1 \\
 0 & 1 & 0 & 1 & 1 & 0 & 0 & 1 & 1 & 0 & 0 & 1 & 1 & 0 & 1 & 0 \\
 1 & 0 & 0 & 1 & 1 & 0 & 1 & 0 & 1 & 0 & 0 & 1 & 0 & 1 & 0 & 1 \\
 0 & 1 & 1 & 0 & 0 & 1 & 0 & 1 & 0 & 1 & 1 & 0 & 1 & 0 & 1 & 0 \\
\end{array}
\right),
\end{equation}
where each column represents an EBR. The integers are multiplicities of the coreps in the same order as Eq. \ref{eq:brdef} of the corresponding EBR. These 16 EBRs represent the 16 different atomic limits in \msgsymbnum{2}{4} (two bands induced from the two irreps of the site-symmetry group of each of the 8 Wyckoff positions a-h). The rest (240) of the band structures are then topologically non-trivial. We can use this matrix to answer the following question: given a band structure characterised by a symmetry vector $B$, can it be expressed in terms of EBRs? This question is equivalent to solving the following system of equations \cite{TQC,MTQC,SongSI,VishSI,VishMSI}:
\begin{equation}\label{eq:ebrxb}
    EBR\cdot X=B
\end{equation}
with $X$ a vector of integer coefficients expressing the decomposition of $B$ in terms of EBRs. We can simplify Eq. \ref{eq:ebrxb} through the Smith decomposition. Since $EBR$ is an $m\times n$ integer matrix, there exists unimodular matrices $L$ ($m\times m$) and $R$ ($n\times n$) such that:
\begin{equation}
    \Delta=L\cdot EBR\cdot R,
\end{equation}
with $\Delta$ a matrix with non-zero entries only in the main diagonal, known as the Smith normal form of $EBR$. It is important to note that we can always choose $L$ and $R$ such that the first $r$ entries in the main diagonal of $\Delta$ are non-zero, with $r=\text{rank}(EBR)$. We can now rewrite Eq. \ref{eq:ebrxb} as:
\begin{equation}\label{eq:cisystem}
    L^{-1}\cdot\Delta\cdot R^{-1}\cdot X=B,\quad LB=\Delta R^{-1}X,\quad C=\Delta Y,
\end{equation}
where in the last step we redefined $C=LB$ and $Y=R^{-1}X$ for the sake of clarity. Because the matrix $\Delta$ only has non-zero entries in the main diagonal (denoted as $\Delta_{i,i}$), the system in Eq. \ref{eq:cisystem} only has an integer solution if (and only if) the following conditions are satisfied:
\begin{align}
    c_i=0 \quad &\text{for} \quad i>r \label{eq:solcond1}\\
    c_i/\Delta_{i,i}\in \mathbb{Z} \quad &\text{for} \quad i=1,\dots, r.\label{eq:solcond2}
\end{align}

If such a solution exists, the $X$ vector in Eq. \ref{eq:ebrxb} is:
\begin{equation}
    X=R\cdot Y \quad \text{with}\quad Y=(c_1/\Delta_{1,1},\dots,c_r/\Delta_{r,r},y_1,\dots,y_{l})
\end{equation}
with $l=m-r$ the number of 0 entries in the main diagonal of $\Delta$ and $y_i$ free variable parameters. The existence of a solution requires that $B$ satisfies both conditions defined in Eq. \ref{eq:solcond1} and Eq. \ref{eq:solcond2}. Eq. \ref{eq:solcond1} is equivalent to enforcing the compatibility relations we described earlier, while the condition in Eq. \ref{eq:solcond2} gives us the \emph{symmetry indicators}. We can rewrite the condition in Eq. \ref{eq:solcond2} as:
\begin{equation}
    c_i\equiv 0 \mod{\Delta_{i,i}}.
\end{equation}

We can apply the described procedure to the $EBR$ matrix in Eq. \ref{eq:ebrmsg2}. We compute the Smith normal decomposition and find:
\begin{equation}\label{eq:msg2delta}
    \Delta=\left(
\begin{array}{cccccccccccccccc}
 1 & 0 & 0 & 0 & 0 & 0 & 0 & 0 & 0 & 0 & 0 & 0 & 0 & 0 & 0 & 0 \\
 0 & 1 & 0 & 0 & 0 & 0 & 0 & 0 & 0 & 0 & 0 & 0 & 0 & 0 & 0 & 0 \\
 0 & 0 & 1 & 0 & 0 & 0 & 0 & 0 & 0 & 0 & 0 & 0 & 0 & 0 & 0 & 0 \\
 0 & 0 & 0 & 1 & 0 & 0 & 0 & 0 & 0 & 0 & 0 & 0 & 0 & 0 & 0 & 0 \\
 0 & 0 & 0 & 0 & 1 & 0 & 0 & 0 & 0 & 0 & 0 & 0 & 0 & 0 & 0 & 0 \\
 0 & 0 & 0 & 0 & 0 & 2 & 0 & 0 & 0 & 0 & 0 & 0 & 0 & 0 & 0 & 0 \\
 0 & 0 & 0 & 0 & 0 & 0 & 2 & 0 & 0 & 0 & 0 & 0 & 0 & 0 & 0 & 0 \\
 0 & 0 & 0 & 0 & 0 & 0 & 0 & 2 & 0 & 0 & 0 & 0 & 0 & 0 & 0 & 0 \\
 0 & 0 & 0 & 0 & 0 & 0 & 0 & 0 & 4 & 0 & 0 & 0 & 0 & 0 & 0 & 0 \\
 0 & 0 & 0 & 0 & 0 & 0 & 0 & 0 & 0 & 0 & 0 & 0 & 0 & 0 & 0 & 0 \\
 0 & 0 & 0 & 0 & 0 & 0 & 0 & 0 & 0 & 0 & 0 & 0 & 0 & 0 & 0 & 0 \\
 0 & 0 & 0 & 0 & 0 & 0 & 0 & 0 & 0 & 0 & 0 & 0 & 0 & 0 & 0 & 0 \\
 0 & 0 & 0 & 0 & 0 & 0 & 0 & 0 & 0 & 0 & 0 & 0 & 0 & 0 & 0 & 0 \\
 0 & 0 & 0 & 0 & 0 & 0 & 0 & 0 & 0 & 0 & 0 & 0 & 0 & 0 & 0 & 0 \\
 0 & 0 & 0 & 0 & 0 & 0 & 0 & 0 & 0 & 0 & 0 & 0 & 0 & 0 & 0 & 0 \\
 0 & 0 & 0 & 0 & 0 & 0 & 0 & 0 & 0 & 0 & 0 & 0 & 0 & 0 & 0 & 0 \\
\end{array}
\right)
\end{equation}

and
\begin{equation}\label{eq:msg2L}
   L=\left(
\begin{array}{cccccccccccccccc}
 0 & 0 & 1 & 0 & 0 & 0 & 0 & 0 & 0 & 0 & 0 & 0 & 0 & 0 & 0 & 0 \\
 0 & 1 & 0 & 0 & 0 & 0 & 0 & 0 & 0 & 0 & 0 & 0 & 0 & 0 & 0 & 0 \\
 1 & 0 & 0 & 0 & 0 & 0 & 0 & 0 & 0 & 0 & 0 & 0 & 0 & 0 & -1 & 0 \\
 1 & 0 & 0 & 0 & 0 & 0 & 0 & 0 & 0 & 0 & 0 & 0 & -1 & 0 & 0 & 0 \\
 1 & 0 & 0 & 0 & 0 & 0 & 0 & 0 & 0 & 0 & -1 & 0 & 0 & 0 & 0 & 0 \\
 1 & 0 & 0 & 0 & 0 & 0 & 0 & 0 & 1 & 0 & -1 & 0 & -1 & 0 & 0 & 0 \\
 1 & 0 & 0 & 0 & 0 & 0 & 1 & 0 & 0 & 0 & -1 & 0 & 0 & 0 & -1 & 0 \\
 1 & 0 & 0 & 0 & 1 & 0 & 0 & 0 & 0 & 0 & 0 & 0 & -1 & 0 & -1 & 0 \\
 1 & 0 & -1 & 0 & 1 & 0 & 1 & 0 & 1 & 0 & -1 & 0 & -1 & 0 & -1 & 0 \\
 -1 & -1 & 0 & 0 & 0 & 0 & 0 & 0 & 1 & 1 & 0 & 0 & 0 & 0 & 0 & 0 \\
 -1 & -1 & 0 & 0 & 1 & 1 & 0 & 0 & 0 & 0 & 0 & 0 & 0 & 0 & 0 & 0 \\
 -1 & -1 & 0 & 0 & 0 & 0 & 0 & 0 & 0 & 0 & 1 & 1 & 0 & 0 & 0 & 0 \\
 -1 & -1 & 1 & 1 & 0 & 0 & 0 & 0 & 0 & 0 & 0 & 0 & 0 & 0 & 0 & 0 \\
 -1 & -1 & 0 & 0 & 0 & 0 & 0 & 0 & 0 & 0 & 0 & 0 & 1 & 1 & 0 & 0 \\
 -1 & -1 & 0 & 0 & 0 & 0 & 1 & 1 & 0 & 0 & 0 & 0 & 0 & 0 & 0 & 0 \\
 -1 & -1 & 0 & 0 & 0 & 0 & 0 & 0 & 0 & 0 & 0 & 0 & 0 & 0 & 1 & 1 \\
\end{array}
\right).
\end{equation}

The $\Delta$ matrix for \msgsymbnum{2}{4} has three diagonal elements $\Delta_{i,i}=2$ (i=6,7,8) and one $\Delta_{9,9}=4$, so there are 4 SI in the group and the classification is $\mathbb{Z}_2\times\mathbb{Z}_2\times\mathbb{Z}_2\times\mathbb{Z}_4$, which is in agreement with previous work \cite{elcorosmithSI, SongSI, VishSI, MTQC}. From Eq. \ref{eq:cisystem} we extract that $c_i=L_{ij}\cdot B_j$, so that the rows of $L$ serve as symmetry indicators, which we denote as \emph{Smith} symmetry indicators (SSI):
\begin{equation}\label{eq:ssi}
    \begin{split}
        z_{2,a}&=n(\Gamma^+)+n(V^+)-n(X^+)-n(Y^+)\mod{2}\\
        z_{2,b}&=n(\Gamma^+)+n(U^+)-n(X^+)-n(Z^+)\mod{2}\\
        z_{2,c}&=n(\Gamma^+)+n(T^+)-n(Y^+)-n(Z^+)\mod{2}\\
        z_{4}&=n(\Gamma^+)-n(R^+)+n(T^+)+n(U^+)+n(V^+)-n(X^+)-n(Y^+)-n(Z^+)\mod{4}.
    \end{split}
\end{equation}

Notice that the SSI as given by Eq.~\ref{eq:ssi} do not have the same form the usual  \cite{SongSI,MTQC} indices found in the literature, which read:
\begin{equation}\label{eq:SG2SI}
        z_{2I,i}=\sum_{K\in k_i=\pi} \frac{n(K^-)-n(K^+)}{2} \mod{2} =  \sum_{K\in k_i=\pi} \frac{n^-_K-n^+_K}{2} \mod{2}\\
\end{equation}
\begin{equation}\label{eq:SG2SI_z4_FK}
        \eta_{4I} =\sum_{K\in \text{TRIM}}\frac{n(K^-)-n(K^+)}{2}\text{ mod 4} = \sum_{K\in \text{TRIM}}\frac{n^-_K-n^+_K}{2}\text{ mod 4},
\end{equation}
where the sum runs over the TRIM k-points in the $k_i=\pi$ planes (i=x,y,z) for Eq.~\ref{eq:SG2SI} and it runs over all TRIM k-points for Eq.~\ref{eq:SG2SI_z4_FK}. 
Note that we introduce the compact notation $n^\pm_K=n(K^\pm)$ used in the literature \cite{SongSI}. The Smith decomposition scheme fully determines the product group of the symmetry indicators, $\mathbb{Z}_2\times\mathbb{Z}_2\times\mathbb{Z}_2\times\mathbb{Z}_4$ in this case, and it gives a possible choice of symmetry indicators. As a consequence, distinct topological phases possess unique symmetry indicators, even though the translation from symmetry indicators to topological invariants is generally non-trivial \cite{SongSI,VishSI,MTQC}. 
To prove this idea, we select representatives of each of the $2\times2\times2\times4=32$ topological classes and compute the Fu-Kane-like symmetry indicators $(z_{2I,1},z_{2I,2},z_{2I,3},\eta_{4I})$ as well as our SSI, showing that, even though the individual values are different, there is a one-to-one mapping between the topological invariants and our SSI (see Table \ref{tab:zhidavssmith}).

\begin{table}
    \centering
    \begin{tabular}{c|c|c}
    Band Representation & Fu-Kane-like symmetry indicators & Smith symmetry indicators \\
      & $(z_{2I,1},z_{2I,2},z_{2I,3},\eta_{4I})$ & $(z_{2,a},z_{2,b},z_{2,c},z_{4})$ \\\hline
 \{1,0,1,0,1,0,1,0,1,0,1,0,1,0,1,0\} & \{0,0,0,0\} & \{0,0,0,0\} \\
 \{1,0,0,1,1,0,0,1,0,1,1,0,0,1,0,1\} & \{0,0,0,1\} & \{1,1,1,1\} \\
 \{2,0,0,2,2,0,1,1,1,1,2,0,1,1,1,1\} & \{0,0,0,2\} & \{0,0,0,2\} \\
 \{1,0,1,0,1,0,1,0,1,0,0,1,0,1,0,1\} & \{0,0,0,3\} & \{1,1,1,3\} \\
 \{0,1,0,1,0,1,1,0,0,1,0,1,0,1,1,0\} & \{0,0,1,0\} & \{1,0,0,2\} \\
 \{0,1,1,0,1,0,0,1,0,1,1,0,1,0,1,0\} & \{0,0,1,1\} & \{0,1,1,3\} \\
 \{0,1,1,0,0,1,1,0,0,1,1,0,1,0,0,1\} & \{0,0,1,2\} & \{1,0,0,0\} \\
 \{1,0,0,1,0,1,1,0,1,0,0,1,0,1,0,1\} & \{0,0,1,3\} & \{0,1,1,1\} \\
 \{0,1,0,1,0,1,0,1,1,0,1,0,0,1,0,1\} & \{0,1,0,0\} & \{0,1,0,2\} \\
 \{0,1,1,0,0,1,1,0,0,1,1,0,1,0,1,0\} & \{0,1,0,1\} & \{1,0,1,3\} \\
 \{0,1,1,0,1,0,0,1,0,1,1,0,1,0,0,1\} & \{0,1,0,2\} & \{0,1,0,0\} \\
 \{1,0,0,1,1,0,0,1,1,0,0,1,0,1,0,1\} & \{0,1,0,3\} & \{1,0,1,1\} \\
 \{0,1,0,1,0,1,1,0,1,0,1,0,0,1,1,0\} & \{0,1,1,0\} & \{1,1,0,0\} \\
 \{0,1,1,0,0,1,0,1,0,1,1,0,1,0,0,1\} & \{0,1,1,1\} & \{0,0,1,1\} \\
 \{0,1,1,0,0,1,0,1,0,1,0,1,0,1,1,0\} & \{0,1,1,2\} & \{1,1,0,2\} \\
 \{1,0,0,1,1,0,1,0,1,0,0,1,0,1,1,0\} & \{0,1,1,3\} & \{0,0,1,3\} \\
 \{1,0,0,1,1,0,1,0,0,1,1,0,1,0,1,0\} & \{1,0,0,0\} & \{0,0,1,2\} \\
 \{1,0,1,0,1,0,0,1,1,0,1,0,0,1,0,1\} & \{1,0,0,1\} & \{1,1,0,3\} \\
 \{1,0,0,1,1,0,0,1,1,0,1,0,0,1,0,1\} & \{1,0,0,2\} & \{0,0,1,0\} \\
 \{1,0,0,1,1,0,1,0,0,1,0,1,0,1,0,1\} & \{1,0,0,3\} & \{1,1,0,1\} \\
 \{0,1,1,0,1,0,1,0,0,1,1,0,0,1,0,1\} & \{1,0,1,0\} & \{1,0,1,0\} \\
 \{0,1,0,1,0,1,1,0,1,0,0,1,0,1,1,0\} & \{1,0,1,1\} & \{0,1,0,1\} \\
 \{0,1,1,0,0,1,0,1,0,1,0,1,1,0,0,1\} & \{1,0,1,2\} & \{1,0,1,2\} \\
 \{1,0,0,1,1,0,1,0,1,0,0,1,1,0,0,1\} & \{1,0,1,3\} & \{0,1,0,3\} \\
 \{0,1,1,0,1,0,1,0,0,1,0,1,1,0,0,1\} & \{1,1,0,0\} & \{0,1,1,0\} \\
 \{0,1,1,0,0,1,0,1,0,1,0,1,1,0,1,0\} & \{1,1,0,1\} & \{1,0,0,1\} \\
 \{0,1,1,0,0,1,0,1,0,1,1,0,0,1,0,1\} & \{1,1,0,2\} & \{0,1,1,2\} \\
 \{1,0,0,1,1,0,1,0,1,0,1,0,0,1,0,1\} & \{1,1,0,3\} & \{1,0,0,3\} \\
 \{0,1,0,1,0,1,0,1,1,0,0,1,0,1,1,0\} & \{1,1,1,0\} & \{1,1,1,2\} \\
 \{1,0,0,1,0,1,1,0,1,0,0,1,1,0,1,0\} & \{1,1,1,1\} & \{0,0,0,3\} \\
 \{0,1,1,0,1,0,0,1,0,1,0,1,1,0,1,0\} & \{1,1,1,2\} & \{1,1,1,0\} \\
 \{0,1,0,1,1,0,1,0,1,0,0,1,0,1,0,1\} & \{1,1,1,3\} & \{0,0,0,1\} \\\hline
    \end{tabular}
    \caption{Symmetry indicators for the 32 different topological phases in \msgsymbnum{2}{4}. In the first column we express the band representation as described in Eq. \ref{eq:brdef}. In the second and third columns, we show the resulting Fu-Kane-like indicators $(z_{2I,1},z_{2I,2},z_{2I,3},\eta_{4I})$ (Eq. \ref{eq:SG2SI},\ref{eq:SG2SI_z4_FK}) and Smith symmetry indicators $(z_{2,a},z_{2,b},z_{2,c},z_{4})$ (Eq. \ref{eq:ssi}). Even though they do not coincide, there is a one-to-one mapping between the indicators.}
    \label{tab:zhidavssmith}
\end{table}

In the context of MSG, the $\mathbb{Z}_4$ index of \msgsymbnum{2}{4} (in the absence of TRS) we described was renamed $\eta_{4I}$ \cite{MTQC,Magneticht}. Interestingly, odd values of this symmetry indicator ($\eta_{4I}=1,3$) indicate a difference in Chern number parity between the $k_z=0,\pi$ planes. This is only possible if there exists a quantum critical point at a finite $k_z=C$ plane at which the bands become degenerate. That is, an odd $\eta_{4I}$ indicator implies the presence of an odd number of Weyl nodes in each half of the BZ. Thus, the $\eta_{4I}$ indicator can diagnose Weyl semimetals from symmetry.

\subsection{Real space invariants}

In the previous section, we derived the symmetry indicators of \msgsymbnum{2}{4} and determined the conditions for a non-trivial stable topological insulator. When all the symmetry indicators are 0, the band structure can be deconstructed as a sum of atomic insulators, and the possible phases are fragile (negative coefficients in the EBR decomposition), trivial, or obstructed atomic insulator. In the last case, the analyzed band structure can be induced by a sum of atomic insulators but the decomposition \emph{necessarily} contains EBRs induced from unoccupied Wyckoff positions or positions that cannot be occupied by moving the atomic orbitals in a symmetry conserving fashion. There is thus a symmetry obstruction in which the charge is pinned on a non-occupied Wyckoff position, which represents a different atomic limit. This system can exhibit surface states inside the bulk gap that \emph{do not} connect valence and conduction bands but are nevertheless pinned by symmetry at the surfaces
\cite{3drsi,3drsi2,RSI,SSH}. These bands are not pinned in energy and generically can move up and down in the conduction or valence band. In order to diagnose these states, the theory of real space invariants (RSI) has been recently developed \cite{3drsi,3drsi2,RSI}. We will now derive the RSI for \msgsymbnum{2}{4} to illustrate the procedure.

Let us start by focusing on the 1a Wyckoff position $q_{\text{1a}}=(0,0,0)$ and its surroundings. The site-symmetry group of $q_{\text{1a}}$ is $\bar{1}$, which has two 1-dimensional irreducible representations, $A_g$ and $A_u$, which have opposite inversion eigenvalues ($A_g$ is even while $A_u$ is odd). If we place an orbital at a random position $q_r=(x,y,z)$, inversion symmetry forces us to place another orbital in the inversion-related position, $-q_r=(-x,-y,-z)$.
These orbitals transform into each other under inversion symmetry, so that the matrix representation of inversion symmetry is  $\rho(\mathcal{I})=\begin{pmatrix} 0 & 1 \\ 1 & 0 \end{pmatrix}$, which can be diagonalized to obtain $\rho(\mathcal{I})=\begin{pmatrix} 1 & 0 \\ 0 & -1 \end{pmatrix}=A_g\oplus A_u$. The two states (orbitals) at $\pm q_r$  transform under the representation $A_g\oplus A_u$  at $q_{\text{1a}}$. Since this process is reversible, whenever we have a pair of orbitals transforming in the $A_g\oplus A_u$ representation at $q_{\text{1a}}$ we can move them away form this position without breaking inversion symmetry.
In general, when the number of $A_g$ and $A_u$ orbitals is exactly the same, it is possible to move them away from the $1a$ Wyckoff position keeping the inversion symmetry, i.e., the orbitals are not pinned. Conversely, if the numbers of both types of orbitals are different, not all of them can be moved away keeping the inversion symmetry and there will be a finite subset of orbitals pinned at the $1a$ position. Thus, the quantity $\delta_{\text{1a}}=n(A_g)-n(A_u)$ serves as an real space invariant representing the amount of symmetry pinned orbitals at the $1a$ Wyckoff position. This procedure is based on the subduction/induction relations of Wyckoff positions and high-symmetry lines, as sketched in the previous example, which can be generalized to all Wyckoff positions in all MSGs. The formulas for computing the different RSIs are listed in Ref.~\cite{RSI}.

\section{Detailed discussion on ideal topological materials}\label{app:fullintmat}
In this section we will provide an in depth analysis of the ideal topological materials outlined in the main text.

\subsection{\texorpdfstring{$\text{Mn}_2$Al$\text{B}_2$}: nodal line semimetal to mirror Chern insulator transition}

\begin{figure}
    \centering
    \includegraphics[width=0.9\linewidth]{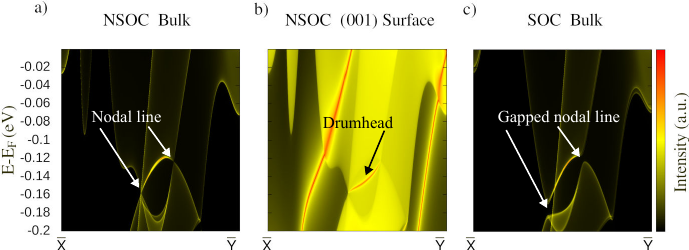}
    \caption{a) \bcsidwebformula{1.508}{Mn${}_2$AlB${}_2$} (\bcsidmagndata{1.508}) bulk bands projection on the (001) surface without SOC. We can clearly see the nodal line bandcrossing. b) Surface spectrum of a semi-infinite slab in the (001) direction. Drumhead states connect the projection of the nodal line crossings. c) Same calculation as in a) including SOC. The nodal line is gapped and the drumhead states are gapped as well.
    }
    \label{fig:1.508_nodal}
\end{figure}

\bcsidwebformula{1.508}{Mn${}_2$AlB${}_2$} (\bcsidmagndata{1.508}) crystallizes in inversion symmetric SG Cmmm (65). When the magnetic moments of the Mn atoms are ordered, the resulting \msgsymbnum{63}{466}. This MSG preserves inversion symmetry and $m_z$ mirror; these determine the symmetry indicators of this MSG. In the absence of SOC, the system exhibits a mirror symmetry-protected nodal line in the $k_z=0$ plane, formed by the crossing of 2 spin degenerate bands, so that the degeneracy at the nodal line is 4. The projection of the nodal line on the (001) surface and its corresponding drumhead states are displayed in Fig. \ref{fig:1.508_nodal}a and \ref{fig:1.508_nodal}b. 
If the SOC is included, the nodal line will open a gap as shown in Fig.~\ref{fig:1.508_nodal}c).
We can then characterize its topology from symmetry indicators. Since inversion symmetry protects the $\eta_{4I}$ symmetry indicator,
it can be computed from inversion eigenvalues of the occupied set of bands \cite{MTQC},
\begin{equation}
    \eta_{4I}=\sum_{K\in \text{TRIM}}\frac{n^-_K-n^+_K}{2}\text{ mod 4},
\end{equation}
where $K$ runs over the inversion symmetric momenta in the BZ and $n^\pm_K$ are the number of positive/negative inversion eigenvalues of the occupied Bloch wavefunctions at point $K$. Notice that this indicator is equivalent to the one in Eq. \ref{eq:ssi}. This index, when equal to 2, implies that the axion angle $\theta$ is non-trivial and it equals $\theta=\pi$. Thus, the system is an axion insulator. Moreover, this MSG has another symmetry indicator, $\delta_{2m}$, and can be computed as follows:
\begin{equation}
    \delta_{2m}=\sum_{K=Z,D,C,E}n^{\frac{1}{2},+i}_K - \sum_{K=\Gamma,A,B,Y}n^{\frac{1}{2},-i}_K \text{ mod 2,}
\end{equation}
where $n^{\frac{1}{2},\pm i}_K$ is the number of occupied states with $C_2$ eigenvalue $e^{-i\pi/2}$ and mirror eigenvalue $\pm i$. This indicator can be understood as the difference between the mirror Chern numbers in planes $k_z=0,\pi$ modulo 2, i.e., 
\begin{equation}
    \delta_{2m}=C_{k_z=\pi} - C_{k_z=0}\text{ mod 2}.
\end{equation} 

We calculated the system's symmetry indicator and found that $\delta_{2m}=1$, indicating that the system behaves as a mirror Chern insulator when spin-orbit coupling (SOC) is considered. The effect of SOC is, however, small. In particular, the nodal line gaps approximately $\sim 8$ meV (see Fig.~\ref{fig:1.508_nodal}), and makes it impossible to converge the Wilson loops. Moreover, there are many bulk states that project onto the surface, thus making it impossible to pinpoint the surface states associated with the mirror Chern number.

\subsection{CaMnSi Axion Insulator}

\begin{figure}
    \centering
    \includegraphics[width=0.5\linewidth]{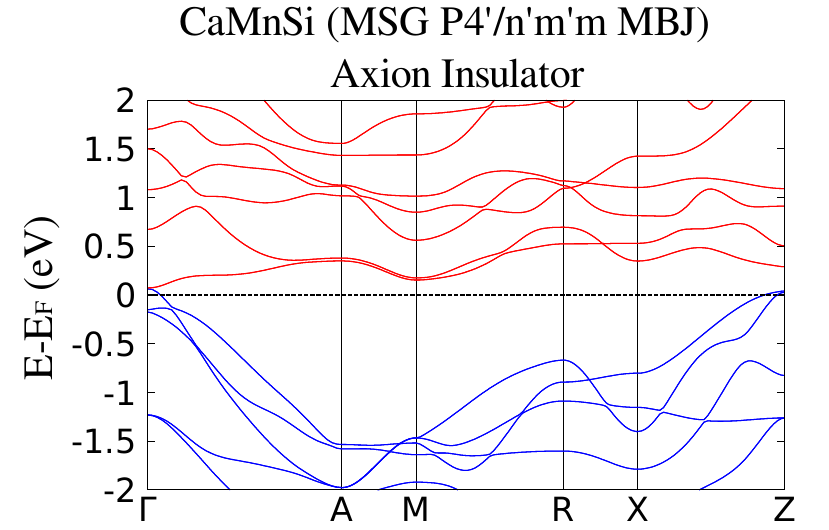}
    \caption{\bcsidwebformula{0.599}{CaMnSi} bulk band calculation with MBJ functional. In the title, we show the chemical formula, magnetic space group and exchange correlation used (MBJ). The topological classification remains the same as with Hubbard+U.}
    \label{fig:0.599_mbj}
\end{figure}

\bcsidwebformula{0.599}{CaMnSi} (\bcsidmagndata{0.599}) crystallizes on tetragonal SG P4/nmm (129). When the magnetic ordering is also considered, the resulting \msgsymbnum{129}{416} \cite{CaMnSi-family}.
Notice that both materials with \bcsidmagndata{0.599} and \bcsidmagndata{0.600} are almost isostructural ($\sim$0.5\% difference on lattice parameter.
The main difference between them is the experimentally measured magnetic moments and experimental temperature ($T_{0.599}=2K$, $T_{0.600}=300K$)
Based on the results of Table \ref{tab:magmoms_typ1} we observe that experimentally, $\mu_{\text{0.599}}=3.27$ and $\mu_{\text{0.600}}=1.96$. Our calculations show very good agreement for the 0.599 structure (0.8\% error) while they show rather bad agreement with 0.600. Thus, we selected the 0.599 structure to explore its topological properties in depth.

The insulators in \msgsymbnum{129}{416} exhibit the symmetry indicator $z_2$. This symmetry indicator is protected by $S_4$ rotoinversion symmetry, and can be computed as follows:
\begin{equation}
    z_2=\sum_{K=\Gamma,M,Z,A} \frac{n^{\frac{1}{2}}_K-n^{-\frac{1}{2}}_K}{2}\text{ mod 2,}
\end{equation}
where $n^{\pm\frac{1}{2}}_K$ denotes the number of bands with eigenvalue $e^{\mp i\frac{\pi}{4}}$ under $S_4$ symmetry and $\Gamma,M,Z,A$ represent the k-points invariants under $S_4$. We computed the index for CaMnSi  to be $z_2=1$.

Since the smallest direct gap computed with GGA functionals is small (it ranges from 15.3 meV for $U=0$ eV to 2 meV with $U=3$ eV),
we performed additional DFT calculations with the modified Becke-Johnson functional (MBJ) \cite{MBJ}. The MBJ predicted gap remains small (6.1 meV) and the topological diagnosis is the same (see Fig.~\ref{fig:0.599_mbj}). The surface state calculation shows gapped surface states, as expected for an axion insulator (see Fig. \ref{fig:main_surface}b).

\subsection{UAsS enforced semimetal}
\begin{figure*}[t]
    \centering
    \includegraphics[width=\linewidth]{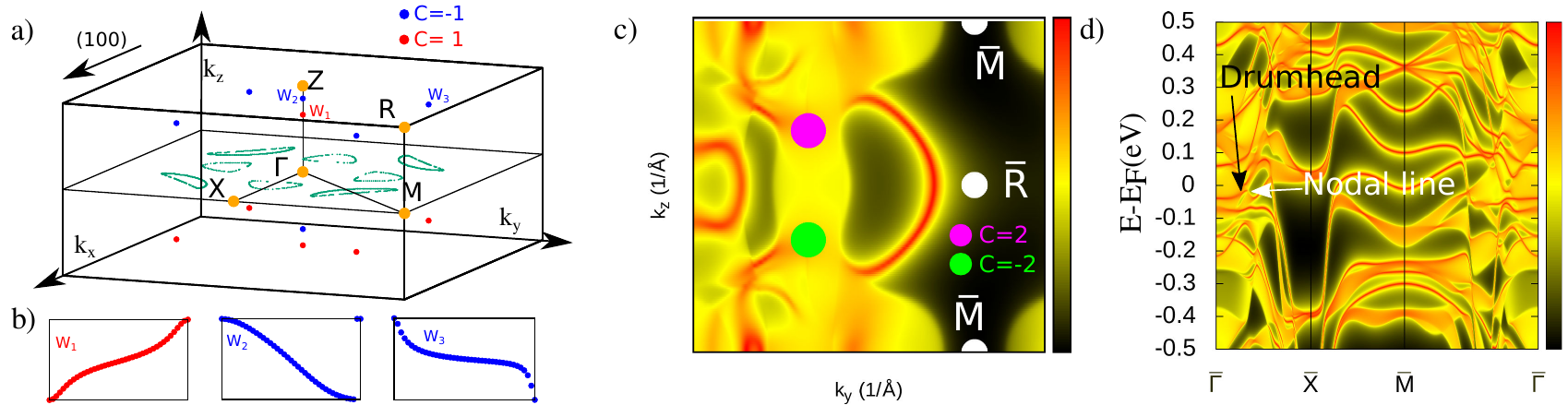}
    \caption{a) BZ and the spatial distribution of Weyl nodes and nodal lines in \bcsidwebformula{0.594}{UAsS} (\bcsidmagndata{0.594}). In red (blue) Weyl nodes with charge +1 (-1). b) Wilson loop on a sphere surrounding the location of Weyl nodes W$_1$, W$_2$ and W$_3$ indicated in a). c) (100) surface termination Fermi surface showing double Fermi arcs stemming from the projection of same-charge Weyl nodes. Magenta (green) color stands for two +1 (-1) Weyls projected in the same point, effectively doubling the charge. d) Drumhead states emerging from the projection of the nodal lines onto the (001) surface termination. 
    }
    \label{fig:0.594}
\end{figure*}

We studied a family of Uranium pnictide chalcogens (\bcsidwebformula{0.593}{UPSe} (\bcsidmagndata{0.593}), \bcsidwebformula{0.594}{UAsS} (\bcsidmagndata{0.594}), \bcsidwebformula{0.595}{UPTe} (\bcsidmagndata{0.595}) and \bcsidwebformula{0.596}{UAsTe} (\bcsidmagndata{0.596})). We  identified all of them as enforced semimetals. Among these, we selected \bcsidwebformula{0.594}{UAsS} (\bcsidmagndata{0.594}) for further investigation, because it displays the cleanest Fermi level (see Fig.~\ref{fig:Fig1}c).
\bcsidwebformula{0.594}{UAsS} crystallizes in SG P4/nmm (129), with the ferromagnetic ground state assigned to \msgsymbnum{129}{417}. Our findings indicate that $U$=0eV provides the best fit for the magnetic moment, with an average error of 24.9\%.

Similar to other ferromagnetic semimetals, we find Weyl nodes close to the Fermi level, which can be split into three families; 2 pairs of 2 Weyl nodes on the $\Gamma-Z$ line (W${}_1=(0,0,0.25)\text{\AA}^{-1}$ and W${}_2=(0,0,0.33)\text{\AA}^{-1}$) and a family of 8 symmetry-related Weyl nodes (W${}_3=(0.43,0.43,0.26)\text{\AA}^{-1}$). We show their location in Fig.~\ref{fig:0.594}a.
The topological charge for a member of each family is displayed in Fig.~\ref{fig:0.594}b.
The Weyl nodes along the $\Gamma-Z$ direction lie below the Fermi level ($E_{\text{W}_1}=-0.08\text{eV}$, $E_{\text{W}_2}=-0.11\text{eV}$), in an energy window densely populated by bulk states,
and are close to each other, making it challenging to identify their characteristic Fermi arcs. However, we did observe double Fermi arcs originating from the (100) surface projections of the other family of Weyl nodes ($E_{\text{W}_3}=-0.02\text{eV}$), which lie closer to the Fermi level. Double arcs arise because same-charge Weyl nodes' projections coincide on that particular surface, as illustrated in Fig. \ref{fig:0.594}c.

We also found a set of 8 nodal lines at the $k_z=0$ plane protected by the $\{m_z|\frac{1}{2}\frac{1}{2}0\}$ glide plane. The (001) surface spectrum reveals that the outer part of the nodal line corresponds to the topologically non-trivial region, where drumhead states are anticipated. In Fig \ref{fig:0.594}a we show the nodal lines in momentum space, located at $k_z=0$ and the path used for the surface. Along the surface path $\bar\Gamma-\bar X$, we observe the projection of the bulk nodal line and the emergence of drumhead states stemming from it (refer to Fig. \ref{fig:0.594}d).

\subsection{\texorpdfstring{CsMnF${}_4$}, nodal surface semimetal}

\begin{figure*}[t]
    \centering
    \includegraphics[width=0.8\linewidth]{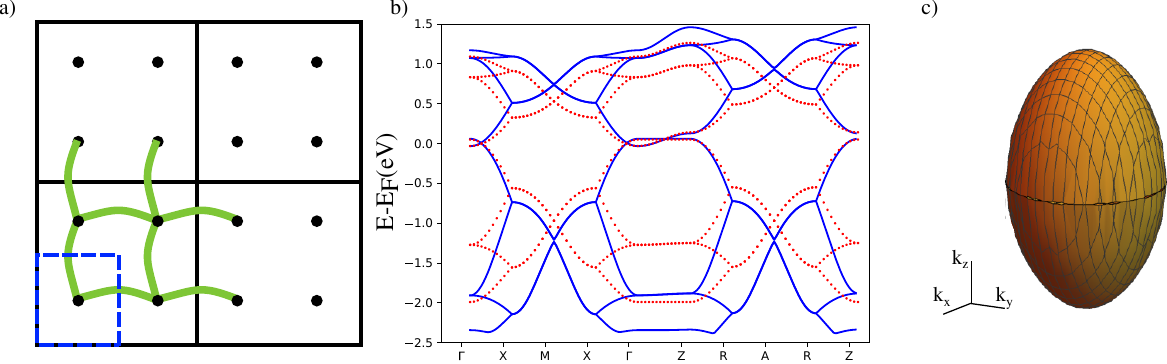}
    \caption{(a) TB with nearest neighbor model hoppings of \bcsidwebformula{0.327}{CsMnF${}_4$} (\bcsidmagndata{0.327}). In dashed blue lines, the `primitive' unit cell of the unfolded model. (b) Comparison between the TB effective model band dispersion and DFT bands. We observe a good agreement close to the Fermi level. c) Nodal surface of the effective tight binding model. The shape matches with the nodal surface reported in Fig.~\ref{fig:main_surface}d.}
    \label{fig:0.327}
\end{figure*}

\bcsidwebformula{0.327}{CsMnF${}_4$} (\bcsidmagndata{0.327}) crystallizes in the SG P4/nmm and its ferromagnetic phase is assigned to \msgsymbnum{59}{410}. Note that this is one of the experimentally reported structures for this compound \cite{CsMnF4_1}. It has been suggested that another structure could be the lowest energy state, potentially influenced by the Jahn-Teller effect \cite{CsMnF4_2}. The definitive structure, however, remains a subject of ongoing debate \cite{PhysRevB.76.094417,palacio1996}.
In this work we study the structure in \cite{CsMnF4_1}. We selected the system with $U$=3eV, as it exhibits the smallest error in magnetization while still being a metal.
We computed the orbital and atomic weight on the bands close to the Fermi level and found out they are formed exclusively from the d-orbitals originating from the Mn atoms at the 4c Wyckoff position. In total, four Mn atoms are present within the unit cell, each with 10 d-orbitals, thereby yielding a total of 40 bands. Owing to the magnetic order, the spin-up and spin-down bands are split. As the splitting is large in this case ($\sim 3-4$eV, depending on the value of Hubbard $U$),
we can limit the current scope of research to the spin-up sector. Moreover, as the SOC is very small, we can discard it as a first approximation. Therefore, we studied the remaining 20 spin-polarized bands crossing the Fermi level. Owing to the crystal field splitting, the T${}_{2g}(d_{xy}, d_{xz}, d_{yz})$ and E${}_{g}(d_{z^2},d_{x^2-y^2})$ orbitals are separated by a gap. The T${}_{2g}$ orbitals are fully occupied, while the E${}_{g}$ are only half-occupied and, thus, lie at the Fermi level. Subsequently, to analyze the crossings at the Fermi level,
we constructed a Wannier Hamiltonian based on the E${}_{g}(d_{z^2},d_{x^2+y^2})$ irrep.
The resulting Wannier TB at nearest neighbors can then be expressed in blocks, with the sites $q_0=(0,0,0)$, $q_1=(\frac{1}{2},0,0)$, $q_2=(\frac{1}{2},\frac{1}{2},0)$, $q_3=(0,\frac{1}{2},0)$ and the basis $\mathcal{B}=\{d_{z^2},d_{x^2-y^2}\}\otimes\{q_0,q_1,q_2,q_3\}$ (see Fig.~\ref{fig:0.327}a):
\begin{equation}\label{eq:hamNN}
    H_{\text{NN}}=\left(
    \begin{array}{cc}
    \epsilon_{d_{z^2}}+t_{d_{z^2}}H_{d_{z^2}}+t^v_{d_{z^2}}H^v_{d_{z^2}} & t_{dd}H_{dd} \\
    t_{dd}H^\dagger_{dd} & \epsilon_{d_{x^2-y^2}}+t_{d_{x^2-y^2}}H_{d_{x^2-y^2}}
    \end{array}
    \right),
\end{equation}
with $\epsilon_{d_{z^2}}=0.582\text{eV}$ and $\epsilon_{d_{x^2-y^2}}=-1.004\text{eV}$ the on-site energies, $t_{d_{z^2}}=-0.140\text{eV}$, $t^v_{d_{z^2}}=-0.042\text{eV}$, $t_{dd}=0.185\text{eV}$, $t_{d_{x^2-y^2}}=-0.255\text{eV}$ the nearest neighbors hoppings and the blocks:
\begin{equation}
H_{d_{z^2}}(k_x,k_y)=\left(
\begin{array}{cccc}
 0 & 2 \cos \left(\frac{k_x}{2}\right) & 0 & 2 \cos \left(\frac{k_y}{2}\right) \\
 2 \cos \left(\frac{k_x}{2}\right) & 0 & 2 \cos \left(\frac{k_y}{2}\right) & 0 \\
 0 & 2 \cos \left(\frac{k_y}{2}\right) & 0 & 2 \cos \left(\frac{k_x}{2}\right) \\
 2 \cos \left(\frac{k_y}{2}\right) & 0 & 2 \cos \left(\frac{k_x}{2}\right) & 0 \\
\end{array}
\right)=H_{d_{x^2-y^2}}(k_x,k_y),
\end{equation}
\begin{equation}
    H_{dd}(k_x,k_y)=\left(
\begin{array}{cccc}
 0 & 2 e^{i k_x} \cos \left(\frac{k_x}{2}\right) & 0 & -2 e^{i k_y} \cos \left(\frac{k_y}{2}\right) \\
 2 e^{-i k_x} \cos \left(\frac{k_x}{2}\right) & 0 & -2 e^{i k_y} \cos \left(\frac{k_y}{2}\right) & 0 \\
 0 & -2 e^{-i k_y} \cos \left(\frac{k_y}{2}\right) & 0 & 2 e^{-i k_x} \cos \left(\frac{k_x}{2}\right) \\
 -2 e^{-i k_y} \cos \left(\frac{k_y}{2}\right) & 0 & 2 e^{i k_x} \cos \left(\frac{k_x}{2}\right) & 0 \\
\end{array}
\right)
\end{equation}
and
\begin{equation}
    H^v_{d_{z^2}}=2\cos\left(k_z\right)\mathbb{I}_{4\times4}.
\end{equation}

These blocks represent the  nearest neighbor hoppings between $d_{z^2}$ ($H_{d_{z^2}}$), $d_{x^2-y^2}$ ($H_{d_{x^2-y^2}}$) and $d_{z^2}-d_{x^2-y^2}$ ($H_{dd}$) orbitals. Notice that we added the (001) direction hopping for the $d_{z^2}$ orbitals but neglected it for $d_{x^2-y^2}$, due to being very small ($t^v_{d_{x^2-y^2}}\approx 2\,10^{-4}$eV). We can see a comparison between the DFT (blue solid lines) and the Wannier TB model (red dots) in Fig.~\ref{fig:0.327}b. MTQC formalism predicts this material to be an ES with a nodal degeneracy in the $k_x=0$ plane. This is due to the interchange of mirror $\{m_x|0\}$ eigenvalues between the last valence and first conduction bands at $\Gamma=(0,0,0)$ and $Z=(0,0,\frac{1}{2})$. In order to check that the model presents the same band inversion, we compute the eigenstates at $\Gamma$ and $Z$ of the last valence and first conduction bands:
\begin{equation}\label{eq:eigenstates}
\begin{split}
    &\ket{\Gamma,v}=\frac{1}{2}(1,1,1,1,0,0,0,0) \\
    &\ket{\Gamma,c}=\frac{1}{2}(0,0,0,0,-1,1,-1,1) \\
    &\ket{Z,v}=\frac{1}{2}(0,0,0,0,-1,1,-1,1) \\
    &\ket{Z,c}=\frac{1}{2}(1,1,1,1,0,0,0,0),
\end{split}
\end{equation}
where $v,c$ stand for last valence and first conduction bands. The representation of the mirror $\{m_x|0\}$ in the basis $\mathcal{B}$ is:
\begin{equation}\label{eq:mxrep}
    \sigma_0\otimes
    \left(
\begin{array}{cccc}
 0 & 1 & 0 & 0 \\
 1 & 0 & 0 & 0 \\
 0 & 0 & 0 & 1 \\
 0 & 0 & 1 & 0 \\
\end{array}
\right)
\end{equation}
where the $\sigma_0$ matrix acts on the orbital degree of freedom. From Eq. \ref{eq:eigenstates} and Eq. \ref{eq:mxrep} we extract the mirror eigenvalues:
\begin{equation}\label{eq:eigengammaZ}
\begin{split}
    m_x&\ket{\Gamma,v}=\ket{\Gamma,v} \\
    m_x&\ket{\Gamma,c}=-\ket{\Gamma,c} \\
    m_x&\ket{Z,v}=-\ket{Z,v} \\
    m_x&\ket{Z,c}=\ket{Z,c}, \\
\end{split}
\end{equation}
which determines that the model reproduces the band inversion. As we explained in the main text, instead of a nodal line,  we find in this system an approximate nodal surface. We can reproduce this  analytically here. At first order in perturbation theory, we construct the low energy effective TB model close to $\Gamma$ as follows:
\begin{equation}
    H^\text{eff}_{i,j}=\bra{i}H_{\text{NN}}\ket{j}=\left(
\begin{array}{cc}
 \epsilon_{d_{z^2}}+2 t_{d_{z^2}} \left(\cos \left(\frac{k_x}{2}\right)+\cos \left(\frac{k_y}{2}\right)\right)+2 t^v_{d_{z^2}} \cos\left(k_z\right) & 0 \\
 0 & \epsilon_{d_{x^2-y^2}}-2 t_{d_{x^2-y^2}} \left(\cos \left(\frac{k_x}{2}\right)+\cos \left(\frac{k_y}{2}\right)\right) \\
\end{array}
\right),
\end{equation}
where $\ket{i},\ket{j}=\ket{\Gamma,v}$, $\ket{\Gamma,c}$ following the notation of Eq.~\ref{eq:eigengammaZ} and the constants are the same as in Eq.~\ref{eq:hamNN}. Notice that at this order in perturbation theory, the effective tight binding does not depend on $t_{dd}$, which measures the hopping that mixes orbitals. The nodal surface will then be the set of $(k_x,k_y,k_z)$ points that make this block Hamiltonian degenerate. After some manipulation, we conclude that the points forming the nodal surface satisfy the following equation:
\begin{equation}\label{eq:nodaleq}
    \cos{\frac{k_x}{2}}+\cos{\frac{k_y}{2}}=f(k_z),
\end{equation}
with
\begin{equation}
    f(k_z) = \frac{\epsilon_{d_{x^2-y^2}}-\epsilon_{d_{z^2}}}{2(t_{d_{x^2-y^2}}+t_{d_{z^2}})} - \frac{t^v_{d_{z^2}}}{t_{d_{x^2-y^2}}+t_{d_{z^2}}}\cos\left(k_z\right)=2.006 - 0.108 \cos\left(k_z\right)
\end{equation}
where in the last step we introduced the numerical values of the Wannier TB. We can see a numerical rendering of the shape in Fig.~\ref{fig:0.327}c. We can also extract the energies of the nodal surface crossing by plugging Eq.~\ref{eq:nodaleq} into any of the diagonal entries in $ H^\text{eff}$, for instance, the lower one:
\begin{equation}
    E_{\text{nodal}}=\epsilon_{d_{x^2-y^2}}-2t_{d_{x^2-y^2}}f(k_z).
\end{equation}
Interestingly, it only depends on $k_z$ and it is symmetric under $k_z\rightarrow -k_z$, both properties that we can see in the nodal surface from the full Wannier TB in Fig.~\ref{fig:main_surface}d.

Notice that the Wannier TB seems to have a unit cell that is $2\times2$ larger than it should, due to the intra-cell and inter-cell hoppings being equal. We can then construct a TB model in the `primitive' unit cell, highlighted in dashed blue lines in Fig.~\ref{fig:0.327}a. This new primitive model does not present any crossing; the crossings in the Wannier TB come from band folding. Since the bands at the Fermi level come exclusively from Mn atoms, the corrections to the hoppings from the interaction with Cs and F atoms will be very small. Indeed, there is no difference in the numerical Wannier TB. We can however push the difference between the intra-cell and inter-cell hoppings to gap the nodal surface while keeping the nodal line at $k_x=0$. In particular, we can add a term of the form:
\begin{equation}
H^{\text{b}}_{dd}=t^b_{dd}
\left(
\begin{array}{cccc}
 0 & -e^{i k_x} & 0 & 0 \\
 -e^{-i k_x} & 0 & 0 & 0 \\
 0 & 0 & 0 & e^{-i k_x} \\
 0 & 0 & e^{i k_x} & 0 \\
\end{array}
\right),
\end{equation}
which introduces a modulation in the hopping in the (100) direction between the intra-cell and inter-cell hoppings that mix different orbitals, with the modulation being different for orbitals in $q_i^y=0$ and $q_i^y=\frac{1}{2}$. Any finite value of $t^b_{dd}$ will gap the nodal surface except at the $k_x=0$ plane, thus giving rise to the predicted nodal line. The nodal surface degeneracy, which is exactly protected by band folding, is thus gapped away when taking into account all symmetry allowed terms. We can then say that, in the absence of interaction with the crystal environment, there is a `quasisymmetry' protected nodal surface, similar to the recent results found in other materials systems \cite{Quasisymmetry1,Quasisymmetry2}.

\subsection{\texorpdfstring{FeCr${}_2$S${}_4$}, enforced semimetal}

\begin{figure*}[t]
    \centering
    \includegraphics[width=\linewidth]{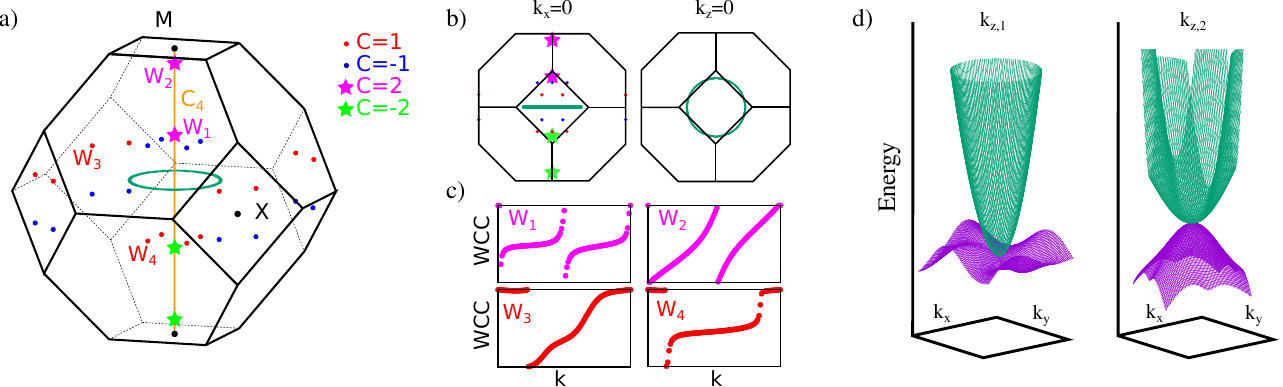}
    \caption{a) Location of topological crossings in the Brillouin Zone of \bcsidwebformula{0.613}{FeCr${}_2$S${}_4$} (\bcsidmagndata{0.613}). In red and magenta Weyls nodes with topological charge 1 and 2, respectively. In blue and green Weyl nodes with topological charge -1 and -2. Charge 2 Weyl nodes are represented by stars. In dark green, mirror-symmetry protected nodal line. b) Projection of crossing location. c) Wilson loop on a sphere surrounding the location of Weyl nodes W$_1$, W$_2$, W$_3$ and W$_4$ indicated in a). d) Fixed $k_z$ plane dispersion of double Weyls. As expected, the dispersion is quadratic in the $k_xk_y$ direction.
    }
    \label{fig:0.613}
\end{figure*}

\bcsidwebformula{0.613}{FeCr${}_2$S${}_4$} (\bcsidmagndata{0.613}) crystallizes in the cubic SG Fd$\bar{3}$m (227) and its anti-ferrimagnetic phase (with magnetic momenta aligned along  the `z' direction) is assigned to the tetragonal \msgsymbnum{141}{557}.
To examine the electronic properties of the system, we initially fit the Hubbard $U$ parameter values that best fit the experimental magnetic momentum, varying it independently for Fe and Cr, in a grid of values ranging from $U$=0eV-4eV. We observed that the best values are $U_{\text{Fe}}=4$eV and $U_{\text{Cr}}=0$eV, with an error of 2.2\% for Fe and 3.2\% for Cr, and leaving the magnetic momenta of S as close as possible to 0.

As typical both for ferro- and ferri-magnetic materials, this system has several Weyls nodes with topological charge $\pm 1$ ("single-Weyl" nodes), which are displayed in \ref{fig:0.613}a. Owing to the presence of the $C_4$ symmetry (orange axis in Fig. \ref{fig:0.613}a), this compound can also display double-Weyl nodes \cite{Multi-Weyl}, doubling the usual charge. This follows from the $C_4$ symmetry constraining the local Weyl Hamiltonian to have quadratic dispersion in the momentum orthogonal to the symmetry axis. In this material, we find two pairs of double-Weyls located at $k_{z,1}=\pm 0.24511\text{\AA}^{-1}$ and $k_{z,2}=\pm 0.56557\text{\AA}^{-1}$, which are displayed in Fig. \ref{fig:0.613}a, in magenta ($C=2$) and green ($C=-2$). The band dispersion of the two Weyls in the $k_z=k_{z,1}$ and $k_z=k_{z,2}$ planes is illustrated in Fig. \ref{fig:0.613}d showing the predicted quadratic dispersion. We further confirm their topological charge by computing the Wilson loops on a sphere surrounding the degeneracy points (see Fig.\ref{fig:0.613}c). The rest of the Weyls have topological charge 1 (see Fig.\ref{fig:0.613}c).
Moreover, in the $k_z=0$ plane there exists a symmetry protected nodal line, displayed in Fig. \ref{fig:0.613}a and Fig. \ref{fig:0.613}b.

\begin{figure*}[t]
    \centering
    \includegraphics[width=\linewidth]{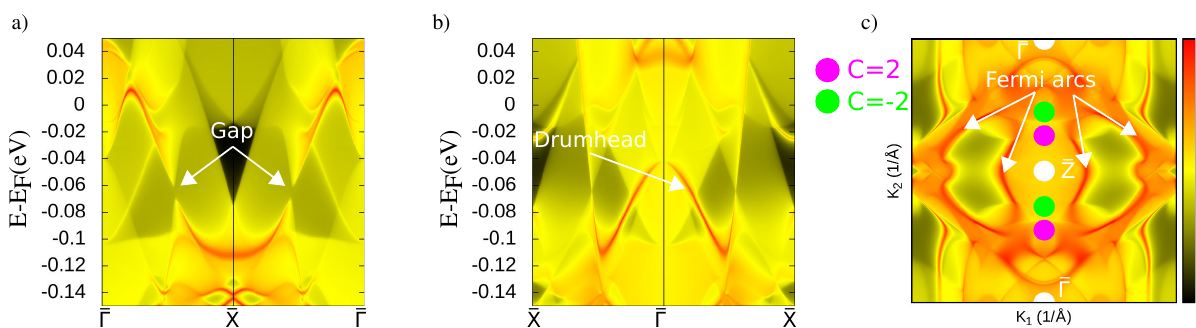}
    \caption{Surface spectrum of FeCr${}_2$S${}_4$ (\bcsidmagndata{0.613}) for two different surface terminations, a) (100) and b) (001). Notice that the pre-existing nodal line is gapped in the (100) termination, and the surfaces states are not protected. In the (001) termination the drumhead states stem from the nodal line and are topologically protected. c) (100) surface termination Fermi surface. We see two Fermi arcs stemming from the projection of each of the bulk double Weyls.}
    \label{fig:0.613_surf}
\end{figure*}

The surface state calculations in the (100) and (001) surface terminations are displayed in Fig. \ref{fig:0.613_surf}. In Fig. \ref{fig:0.613_surf}a we can observe the surface spectrum for the (100) termination where the band dispersion of a pre-existing nodal line that has been gapped owing to the magnetic symmetry subduction can be observed. Fig. \ref{fig:0.613_surf}b shows the drumhead states originate from the projection of the remaining nodal line. Finally, the Fermi surface calculations on the (100) surface are displayed in Fig. \ref{fig:0.613_surf}c. Two Fermi arcs stem from the projection of each of the bulk double Weyls, getting in total 4 Fermi arcs.

\section{Experimental magnetic momenta vs calculation}\label{app:magmomcomp}

The experimentally determined magnetic momenta of nonequivalent atoms for all computed systems and those within DFT+U for various values of the Hubbard $U$ parameter are presented herein in two tables. In Table \ref{tab:magmoms_typ1}, we consider the compounds with magnetic elements containing only d electrons. In Table \ref{tab:magmoms_typ2}, we consider compounds with only f electrons and in Table~\ref{tab:magmoms_typ3} we consider compounds with both d and f electrons. The first three columns contain the BCS ID of the material, the chemical formula and MSG. In the next columns, we display the experimental magnetic momenta vs the computed one, with the resulting error quantified as follows:
\begin{equation}\label{eq:error}
    \text{Error(\%)}=\sum_{j}\left(\sum_{i=x}^z\left|\frac{\mu^{\text{exp}}_i-\mu^{\text{comp}}_i}{\mu^{\text{exp}}_i}\right|\right)\frac{100}{N^{\text{at}}_j},
\end{equation}
where $i$ runs over the 3 components of the magnetic momenta and $j$ runs over the magnetic elements. Missing entries correspond to cases where the numerical calculation failed to converge.
\vspace{1cm}



\setlength{\tabcolsep}{1.1ex}

\begin{longtable}{|c|c|c|c|c|c|c|c|c|c|}\hline
\label{tab:magmoms_typ1}
BCS ID & Formula &  MSG & M.E. & $\mu_{exp}$ & U=0 & U=1 & U=2 & U=3 & U=4  \\\hline 
\multirow{4}{*}{\scriptsize{\bcsidmagndatashort{0.229}}} & \multirow{4}{*}{\scriptsize{\bcsidwebformula{0.229}{Ba2 Mn1 Si2 O7}}} & \multirow{4}{*}{\scriptsize{\msgsymbnum{113}{267}}}
& \scriptsize{Mn} & \scriptsize{4.1}& \scriptsize{4.38} & \scriptsize{4.44} & \scriptsize{4.5} & \scriptsize{4.56} & \scriptsize{4.6} \\
&&& \scriptsize{Error(\%)} & & \scriptsize{6.9} \% & \scriptsize{8.41} \% & \scriptsize{9.8} \% & \scriptsize{11.1} \% & \scriptsize{12.29} \% \\\cline{4-10}
&&& \scriptsize{Mn} & \scriptsize{4.1}& \scriptsize{4.38} & \scriptsize{4.44} & \scriptsize{4.5} & \scriptsize{4.56} & \scriptsize{4.6} \\
&&& \scriptsize{Error(\%)} & & \scriptsize{6.9} \% & \scriptsize{8.41} \% & \scriptsize{9.8} \% & \scriptsize{11.1} \% & \scriptsize{12.29} \% \\\cline{4-10}\hline
\multirow{4}{*}{\scriptsize{\bcsidmagndatashort{0.230}}} & \multirow{4}{*}{\scriptsize{\bcsidwebformula{0.230}{K2 Co1 P2 O7}}} & \multirow{4}{*}{\scriptsize{\msgsymbnum{58}{395}}}
& \scriptsize{Co} & \scriptsize{3.03}& \scriptsize{2.57} & \scriptsize{2.62} & \scriptsize{2.66} & \scriptsize{2.7} & \scriptsize{2.74} \\
&&& \scriptsize{Error(\%)} & & \scriptsize{15.35} \% & \scriptsize{13.7} \% & \scriptsize{12.18} \% & \scriptsize{10.76} \% & \scriptsize{9.44} \% \\\cline{4-10}
&&& \scriptsize{Co} & \scriptsize{3.03}& \scriptsize{2.57} & \scriptsize{2.62} & \scriptsize{2.66} & \scriptsize{2.7} & \scriptsize{2.74} \\
\end{longtable}
\newpage

\setlength{\tabcolsep}{1.1ex}
\begin{longtable}{|c|c|c|c|c|c|c|c|c|}\hline
\label{tab:magmoms_typ2}
        BCS ID & Formula &  MSG & M.E. & $\mu_{exp}$ & U=0 & U=2 & U=4 & U=6   \\\hline 
\multirow{4}{*}{\scriptsize{\bcsidmagndatashort{0.237}}} & \multirow{4}{*}{\scriptsize{\bcsidwebformula{0.237}{Er2 Sn2 O7}}} & \multirow{4}{*}{\scriptsize{\msgsymbnum{141}{555}}}
& \scriptsize{Er} & \scriptsize{4.38}& \scriptsize{2.76} & \scriptsize{2.81} & \scriptsize{2.84} & \scriptsize{2.91} \\
&&& \scriptsize{Error(\%)} & & \scriptsize{74.16} \% & \scriptsize{71.87} \% & \scriptsize{70.45} \% & \scriptsize{67.16} \% \\\cline{4-9}
&&& \scriptsize{Er} & \scriptsize{4.38}& \scriptsize{2.74} & \scriptsize{2.8} & \scriptsize{2.84} & \scriptsize{2.91} \\
&&& \scriptsize{Error(\%)} & & \scriptsize{24.96} \% & \scriptsize{24.09} \% & \scriptsize{23.55} \% & \scriptsize{22.44} \% \\\cline{4-9}\hline
\multirow{4}{*}{\scriptsize{\bcsidmagndatashort{0.238}}} & \multirow{4}{*}{\scriptsize{\bcsidwebformula{0.238}{Er2 Sn2 O7}}} & \multirow{4}{*}{\scriptsize{\msgsymbnum{141}{555}}}
& \scriptsize{Er} & \scriptsize{4.81}& \scriptsize{2.76} & \scriptsize{2.81} & \scriptsize{2.84} & \scriptsize{2.91} \\
&&& \scriptsize{Error(\%)} & & \scriptsize{85.03} \% & \scriptsize{83.09} \% & \scriptsize{81.82} \% & \scriptsize{78.82} \% \\\cline{4-9}
&&& \scriptsize{Er} & \scriptsize{4.81}& \scriptsize{2.75} & \scriptsize{2.8} & \scriptsize{2.84} & \scriptsize{2.91} \\
\end{longtable}

\newpage

\setlength{\tabcolsep}{1.1ex}
\begin{longtable}{|c|c|c|c|c|c|c|c|c|}\hline
\label{tab:magmoms_typ3}
        BCS ID & Formula &  MSG & M.E. & $\mu_{exp}$ & U=0 & U=2 & U=4 & U=6   \\\hline 
\multirow{12}{*}{\scriptsize{\bcsidmagndatashort{0.231}}} & \multirow{12}{*}{\scriptsize{\bcsidwebformula{0.231}{Tm1 Mn3 O6}}} & \multirow{12}{*}{\scriptsize{\msgsymbnum{59}{410}}}
& \scriptsize{Tm} & \scriptsize{0.77}& \scriptsize{1.56} & \scriptsize{1.69} & \scriptsize{-} & \scriptsize{-} \\
&&& \scriptsize{Error(\%)} & & \scriptsize{151.49} \% & \scriptsize{159.55} \% & \scriptsize{-} \% & \scriptsize{-} \% \\\cline{4-9}
&&& \scriptsize{Tm} & \scriptsize{3.7}& \scriptsize{1.59} & \scriptsize{1.72} & \scriptsize{-} & \scriptsize{-} \\
&&& \scriptsize{Error(\%)} & & \scriptsize{71.5} \% & \scriptsize{26.81} \% & \scriptsize{-} \% & \scriptsize{-} \% \\\cline{4-9}
&&& \scriptsize{Mn} & \scriptsize{2.0}& \scriptsize{3.8} & \scriptsize{3.81} & \scriptsize{-} & \scriptsize{-} \\
&&& \scriptsize{Error(\%)} & & \scriptsize{44.91} \% & \scriptsize{45.11} \% & \scriptsize{-} \% & \scriptsize{-} \% \\\cline{4-9}
&&& \scriptsize{Mn} & \scriptsize{3.17}& \scriptsize{4.42} & \scriptsize{4.44} & \scriptsize{-} & \scriptsize{-} \\
\end{longtable}

\section{Topological classification and gap}\label{app:gaptopo}

The topological phase diagram with gap information is illustrated herein. Similar to the previous table, in this table we list the BCS ID, chemical formula and MSG of the materials in the first three columns. Thereafter, we present the topological diagnosis of the system, as well as the electronic gaps in eV; first the indirect and second the smallest direct gaps. Notice that a negative sign in the indirect denotes a metallic system. We tabulate these results as a function of the Hubbard U parameter. In Table~\ref{tab:gaps1} we show the described information for compounds with only d electrons, in Table~\ref{tab:gaps2} we display the compounds with only f electrons and in Table~\ref{tab:gaps3} we show compounds with both d and f electrons. If the gap between the last valence and first conduction bands is small, we denote these systems as accidental Fermi degeneracies (AF). Since the topology is very sensitive to band inversions, we cannot confidently predict the topological phase of the system. 
\begin{adjustwidth}{-100pt}{-100pt}

\setlength{\tabcolsep}{0.05ex}

\begin{longtable}{|c|c|c|c|c|c|c|c|}\hline
\label{tab:gaps1}

        BCS ID & Formula &  MSG  & U=0 & U=1 & U=2 & U=3 & U=4  \\\hline 
 \bcsidmagndatashort{0.229} & \bcsidwebformula{0.229}{Ba2 Mn1 Si2 O7} & \msgsymbnum{113}{267} & \begin{tabular}{c} AI \\ 2.212 \\ 2.235 \end{tabular} & \begin{tabular}{c} AI \\ 2.566 \\ 2.599 \end{tabular} & \begin{tabular}{c} AI \\ 2.861 \\ 2.906 \end{tabular} & \begin{tabular}{c} AI \\ 3.12 \\ 3.179 \end{tabular} & \begin{tabular}{c} AI \\ 3.346 \\ 3.422 \end{tabular} \\\hline
 \bcsidmagndatashort{0.230} & \bcsidwebformula{0.230}{K2 Co1 P2 O7} & \msgsymbnum{58}{395} & \begin{tabular}{c} AI \\ 1.112 \\ 1.14 \end{tabular} & \begin{tabular}{c} AI \\ 1.813 \\ 1.838 \end{tabular} & \begin{tabular}{c} AI \\ 2.507 \\ 2.531 \end{tabular} & \begin{tabular}{c} AI \\ 3.176 \\ 3.198 \end{tabular} & \begin{tabular}{c} AI \\ 3.797 \\ 3.817 \end{tabular} \\\hline
 \bcsidmagndatashort{0.234} & \bcsidwebformula{0.234}{Mn2 La1 Sb1 O6} & \msgsymbnum{13}{69} & \begin{tabular}{c} AI \\ -0.028 \\ 0.004 \end{tabular} & \begin{tabular}{c} AI \\ 1.504 \\ 1.848 \end{tabular} & \begin{tabular}{c} AI \\ 1.068 \\ 1.088 \end{tabular} & \begin{tabular}{c} AI \\ 1.517 \\ 1.537 \end{tabular} & \begin{tabular}{c} AI \\ 2.068 \\ 2.578 \end{tabular} \\\hline
 \bcsidmagndatashort{0.236} & \bcsidwebformula{0.236}{Ca1 Fe4 Al8} & \msgsymbnum{139}{535} & \begin{tabular}{c} ES \\ -0.306 \\ 0.002 \end{tabular} & \begin{tabular}{c} ES \\ -0.578 \\ 0.002 \end{tabular} & \begin{tabular}{c} ES \\ -0.724 \\ 0.006 \end{tabular} & \begin{tabular}{c}  \\ -0.896 \\ 0.009 \end{tabular} & \begin{tabular}{c}  \\ -0.882 \\ 0.007 \end{tabular} \\
\end{longtable}

\newpage
\setlength{\tabcolsep}{0.1ex}

\begin{longtable}{|c|c|c|c|c|c|c|}\hline
\label{tab:gaps2}

        BCS ID & Formula &  MSG  & U=0 & U=2 & U=4 & U=6  \\\hline 
 \bcsidmagndatashort{0.237} & \bcsidwebformula{0.237}{Er2 Sn2 O7} & \msgsymbnum{141}{555} & \begin{tabular}{c} AI \\ 0.925 \\ 0.933 \end{tabular} & \begin{tabular}{c} AI \\ 1.883 \\ 1.892 \end{tabular} & \begin{tabular}{c} AI \\ 2.629 \\ 2.637 \end{tabular} & \begin{tabular}{c} AI \\ 2.608 \\ 2.614 \end{tabular} \\\hline
 \bcsidmagndatashort{0.238} & \bcsidwebformula{0.238}{Er2 Sn2 O7} & \msgsymbnum{141}{555} & \begin{tabular}{c} AI \\ 0.903 \\ 0.913 \end{tabular} & \begin{tabular}{c} AI \\ 2.083 \\ 2.092 \end{tabular} & \begin{tabular}{c} AI \\ 2.85 \\ 2.862 \end{tabular} & \begin{tabular}{c} AI \\ 2.823 \\ 2.832 \end{tabular} \\\hline
 \bcsidmagndatashort{0.320} & \bcsidwebformula{0.320}{U2 Pd2 In1} & \msgsymbnum{127}{394} & \begin{tabular}{c} ESFD \\ -0.303 \\ 0.0 \end{tabular} & \begin{tabular}{c} ESFD \\ -1.1 \\ 0.0 \end{tabular} & \begin{tabular}{c} ESFD \\ -0.961 \\ 0.0 \end{tabular} & \begin{tabular}{c} ESFD \\ -1.23 \\ 0.0 \end{tabular} \\\hline
 \bcsidmagndatashort{0.321} & \bcsidwebformula{0.321}{U2 Sn1 Pd2} & \msgsymbnum{127}{394} & \begin{tabular}{c} NLC  \\ -0.259 \\ 0.02 \end{tabular} & \begin{tabular}{c} NLC  \\ -1.073 \\ 0.036 \end{tabular} & \begin{tabular}{c} NLC  \\ -0.761 \\ 0.033 \end{tabular} & \begin{tabular}{c} NLC  \\ -0.875 \\ 0.081 \end{tabular} \\\hline
 \bcsidmagndatashort{0.324} & \bcsidwebformula{0.324}{Cd1 Yb2 S4} & \msgsymbnum{141}{551} & \begin{tabular}{c} AF \\ -0.484 \\ 0.0 \end{tabular} & \begin{tabular}{c}  \\ - \\ - \end{tabular} & \begin{tabular}{c}  \\ - \\ - \end{tabular} & \begin{tabular}{c} AI \\ 0.017 \\ 0.117 \end{tabular} \\
\end{longtable}

\newpage
\setlength{\tabcolsep}{0.1ex}

\begin{longtable}{|c|c|c|c|c|c|c|}\hline
\label{tab:gaps3}
        BCS ID & Formula &  MSG  & U=0 & U=2 & U=4 & U=6   \\\hline 
 \bcsidmagndatashort{0.231} & \bcsidwebformula{0.231}{Tm1 Mn3 O6} & \msgsymbnum{59}{410} & \begin{tabular}{c} ES \\ -0.057 \\ 0.0 \end{tabular} & \begin{tabular}{c} ES \\ -0.455 \\ 0.004 \end{tabular} & \begin{tabular}{c} ES \\ -0.603 \\ 0.004 \end{tabular} & \begin{tabular}{c}  \\ - \\ - \end{tabular} \\\hline
 \bcsidmagndatashort{0.232} & \bcsidwebformula{0.232}{Tm1 Mn3 O6} & \msgsymbnum{59}{409} & \begin{tabular}{c} AF \\ -0.04 \\ 0.0 \end{tabular} & \begin{tabular}{c} ES \\ -0.364 \\ 0.006 \end{tabular} & \begin{tabular}{c} ES \\ -0.514 \\ 0.006 \end{tabular} & \begin{tabular}{c} ES \\ -0.843 \\ 0.002 \end{tabular} \\\hline
 \bcsidmagndatashort{0.235} & \bcsidwebformula{0.235}{Mn2 Pr1 Sb1 O6} & \msgsymbnum{86}{67} & \begin{tabular}{c} AI \\ -0.015 \\ 0.005 \end{tabular} & \begin{tabular}{c} AI \\ 0.0 \\ 0.002 \end{tabular} & \begin{tabular}{c} AI \\ 0.829 \\ 0.848 \end{tabular} & \begin{tabular}{c} AI \\ 1.454 \\ 1.758 \end{tabular} \\\hline
 \bcsidmagndatashort{0.289} & \bcsidwebformula{0.289}{Nd1 Mn1 O3} & \msgsymbnum{62}{448} & \begin{tabular}{c} AF \\ -0.135 \\ 0.004 \end{tabular} & \begin{tabular}{c}  \\ -0.012 \\ 0.425 \end{tabular} & \begin{tabular}{c}  \\ 0.021 \\ 0.107 \end{tabular} & \begin{tabular}{c}  \\ 0.461 \\ 0.767 \end{tabular} \\\hline
 \bcsidmagndatashort{0.290} & \bcsidwebformula{0.290}{Ce1 Cu2} & \msgsymbnum{74}{560} & \begin{tabular}{c} AI \\ -0.341 \\ 0.021 \end{tabular} & \begin{tabular}{c} AI \\ -0.42 \\ 0.014 \end{tabular} & \begin{tabular}{c}  \\ -0.465 \\ 0.019 \end{tabular} & \begin{tabular}{c} AF \\ -0.532 \\ 0.004 \end{tabular} \\
\end{longtable}

\end{adjustwidth}

\section{Physical interpretation of systems with non-trivial symmetry indicators}\label{app:ticlass}

In this section we provide the physical interpretation of the topological phase predicted by symmetry indicators according to the definitions in Ref.~\cite{MTQC}. For each compound, for each Hubbard $U$ value, we compute the symmetry indicators. Then, by subducing to their minimal SSG (see Ref.~\cite{MTQC}) we provide the physical interpretation corresponding to their layer construction \cite{SongSI,MTQC}. As discussed in the main text (see Sec.~\ref{sec:workflow}), the symmetry diagnosable topological phases are AXI, 3DQAH, SISM and MTCI. If two topological phases cannot be distinguished by symmetry indicators, we write both possibilities. For example, a set of symmetry indicators $(\eta_{4I},z_{2I,1},z_{2I,2},z_{2I,3})=(2,0,0,0)$ indicates AXI. However, if the MSG does not forbid zero net Chern number, since the $z_{2I,i}$ indicators can only diagnose the Chern number mod(2), we cannot rule out that the system is in the 3DQAH phase.

\begin{longtable}{|c|c|c|c|c|c|c|}\hline
        BCS ID & Formula & U(eV) & MSG & minimal SSG & Indices & Interpretation  \\\hline 
\multirow{4}{*}{\bcsidmagndatashort{1.752}} & \multirow{4}{*}{\bcsidwebformula{1.752}{\scriptsize{Cs1 Co2 Mo2 O9 H1}}} & \multirow{4}{*}{2} & \multirow{4}{*}{\msgsymbnum{11}{55}} & \multirow{4}{*}{\msgsymbnum{2}{4}} & $\eta_{4I}=2$ & \multirow{4}{*}{AXI}\\
&&&&& $z_{2I,1}=0$ &\\
&&&&& $z_{2I,2}=0$ &\\
&&&&& $z_{2I,3}=0$ &\\\hline
\multirow{4}{*}{\bcsidmagndatashort{1.749}} & \multirow{4}{*}{\bcsidwebformula{1.749}{\scriptsize{Ho1 Sb1 Te1}}} & \multirow{4}{*}{0} & \multirow{4}{*}{\msgsymbnum{11}{55}} & \multirow{4}{*}{\msgsymbnum{2}{4}} & $\eta_{4I}=2$ & \multirow{4}{*}{AXI}\\
&&&&& $z_{2I,1}=0$ &\\
&&&&& $z_{2I,2}=0$ &\\
\end{longtable}

\section{Full tables with structure, bands and topological classification}\label{app:fulltables}

The complete tables with information on the crystal and magnetic structure, magnetic space group, topological classification, and electronic structure are presented herein.

\include{Tables/Topoplots/Type_1_r}
\begin{table}[t] 
    \centering
    %
    \caption{Topology phase diagram of Er2 Sn2 O7}
\end{table}

\begin{table}[t] 
    \centering
    %
    \caption{Topology phase diagram of Er2 Sn2 O7}
\end{table}

\begin{table}[t] 
    \centering
    %
    \caption{Topology phase diagram of U2 Pd2 In1}
\end{table}

\begin{table}[t] 
    \centering
    %
    \caption{Topology phase diagram of U2 Sn1 Pd2}
\end{table}

\begin{table}[t] 
    \centering
    %
    \caption{Topology phase diagram of Cd1 Yb2 S4}
\end{table}

\begin{table}[t] 
    \centering
    %
    \caption{Topology phase diagram of Nd2 Sn2 O7}
\end{table}

\begin{table}[t] 
    \centering
    %
    \caption{Topology phase diagram of Er1 Ge3}
\end{table}

\begin{table}[t] 
    \centering
    %
    \caption{Topology phase diagram of Nd2 Hf2 O7}
\end{table}

\begin{table}[t] 
    \centering
    %
    \caption{Topology phase diagram of Nd2 Zr2 O7}
\end{table}

\begin{table}[t] 
    \centering
    %
    \caption{Topology phase diagram of Tb1 Ge2}
\end{table}

\begin{table}[t] 
    \centering
    %
    \caption{Topology phase diagram of Tb1 Al1 O3}
\end{table}

\begin{table}[t] 
    \centering
    %
    \caption{Topology phase diagram of Gd1 Al1 O3}
\end{table}

\begin{table}[t] 
    \centering
    %
    \caption{Topology phase diagram of Ce1 Na1 O2}
\end{table}

\begin{table}[t] 
    \centering
    %
    \caption{Topology phase diagram of Er2 Si2 O7}
\end{table}

\begin{table}[t] 
    \centering
    %
    \caption{Topology phase diagram of U1 Se1 P1}
\end{table}

\begin{table}[t] 
    \centering
    %
    \caption{Topology phase diagram of U1 S1 As1}
\end{table}

\begin{table}[t] 
    \centering
    %
    \caption{Topology phase diagram of U1 Te1 P1}
\end{table}

\begin{table}[t] 
    \centering
    %
    \caption{Topology phase diagram of U1 Te1 As1}
\end{table}

\begin{table}[t] 
    \centering
    %
    \caption{Topology phase diagram of Ho1 B2}
\end{table}

\begin{table}[t] 
    \centering
    %
    \caption{Topology phase diagram of U2 In1 Pd2}
\end{table}

\begin{table}[t] 
    \centering
    %
    \caption{Topology phase diagram of Er2 Si2 O7}
\end{table}

\begin{table}[t] 
    \centering
    %
    \caption{Topology phase diagram of Sb3 Ce4}
\end{table}

\begin{table}[t] 
    \centering
    %
    \caption{Topology phase diagram of Pt1 Tb1}
\end{table}

\begin{table}[t] 
    \centering
    %
    \caption{Topology phase diagram of Er1 Pt1}
\end{table}

\begin{table}[t] 
    \centering
    %
    \caption{Topology phase diagram of Ho1 Pt1}
\end{table}

\begin{table}[t] 
    \centering
    %
    \caption{Topology phase diagram of Dy1 Pt1}
\end{table}

\begin{table}[t] 
    \centering
    %
    \caption{Topology phase diagram of Tm1 Pt1}
\end{table}

\begin{table}[t] 
    \centering
    %
    \caption{Topology phase diagram of Pt1 Pr1}
\end{table}

\begin{table}[t] 
    \centering
    %
    \caption{Topology phase diagram of Nd1 Pt1}
\end{table}

\begin{table}[t] 
    \centering
    %
    \caption{Topology phase diagram of Ho3 Al5 O12}
\end{table}

\begin{table}[t] 
    \centering
    %
    \caption{Topology phase diagram of Nd1 Sc1 O3}
\end{table}

\begin{table}[t] 
    \centering
    %
    \caption{Topology phase diagram of Nd1 In1 O3}
\end{table}

\begin{table}[t] 
    \centering
    %
    \caption{Topology phase diagram of Sr1 Gd2 O4}
\end{table}

\clearpage
\begin{table}[t] 
    \centering
    %
    \caption{Topology phase diagram of Ce1 Au1 Ge1}
\end{table}

\begin{table}[t] 
    \centering
    %
    \caption{Topology phase diagram of Dy1 Al1 O3}
\end{table}

\begin{table}[t] 
    \centering
    %
    \caption{Topology phase diagram of Gd2 Pt2 O7}
\end{table}

\begin{table}[t] 
    \centering
    %
    \caption{Topology phase diagram of Eu1 Cd2 As2}
\end{table}

\begin{table}[t] 
    \centering
    %
    \caption{Topology phase diagram of Er2 O3}
\end{table}

\begin{table}[t] 
    \centering
    %
    \caption{Topology phase diagram of Er1 La1 O3}
\end{table}

\begin{table}[t] 
    \centering
    %
    \caption{Topology phase diagram of Yb1 Pd1 Si1}
\end{table}

\begin{table}[t] 
    \centering
    %
    \caption{Topology phase diagram of Ho1 Pd1 In1}
\end{table}

\begin{table}[t] 
    \centering
    %
    \caption{Topology phase diagram of Ho1 Pd1 In1}
\end{table}

\begin{table}[t] 
    \centering
    %
    \caption{Topology phase diagram of Er1 Pd1 In1}
\end{table}

\begin{table}[t] 
    \centering
    %
    \caption{Topology phase diagram of Er1 Pd1 In1}
\end{table}

\begin{table}[t] 
    \centering
    %
    \caption{Topology phase diagram of Nd1 Pd1 In1}
\end{table}

\begin{table}[t] 
    \centering
    %
    \caption{Topology phase diagram of Nd1 Pd1 In1}
\end{table}

\begin{table}[t] 
    \centering
    %
    \caption{Topology phase diagram of Eu1 Pd3 Si2}
\end{table}

\begin{table}[t] 
    \centering
    %
    \caption{Topology phase diagram of Ce1 Pd2 Si2}
\end{table}

\begin{table}[t] 
    \centering
    %
    \caption{Topology phase diagram of Ce1 Rh2 Si2}
\end{table}

\begin{table}[t] 
    \centering
    %
    \caption{Topology phase diagram of Au2 Ce1 Si2}
\end{table}

\begin{table}[t] 
    \centering
    %
    \caption{Topology phase diagram of Pr2 Pd2 In1}
\end{table}

\begin{table}[t] 
    \centering
    %
    \caption{Topology phase diagram of Eu1 As3}
\end{table}

\begin{table}[t] 
    \centering
    %
    \caption{Topology phase diagram of Dy1 Ge1}
\end{table}

\begin{table}[t] 
    \centering
    %
    \caption{Topology phase diagram of Pu2 O3}
\end{table}

\begin{table}[t] 
    \centering
    %
    \caption{Topology phase diagram of Ce1 Rh1 Al4 Si2}
\end{table}

\begin{table}[t] 
    \centering
    %
    \caption{Topology phase diagram of Eu1 Bi2 Mg2}
\end{table}

\begin{table}[t] 
    \centering
    %
    \caption{Topology phase diagram of Gd1 Ag1 Sn1}
\end{table}

\begin{table}[t] 
    \centering
    %
    \caption{Topology phase diagram of Al2 Pd5 Nd1}
\end{table}

\begin{table}[t] 
    \centering
    %
    \caption{Topology phase diagram of Ce1 C2}
\end{table}

\begin{table}[t] 
    \centering
    %
    \caption{Topology phase diagram of Pr1 C2}
\end{table}

\begin{table}[t] 
    \centering
    %
    \caption{Topology phase diagram of Nd1 C2}
\end{table}

\begin{table}[t] 
    \centering
    %
    \caption{Topology phase diagram of U1 Pd2 Si2}
\end{table}

\begin{table}[t] 
    \centering
    %
    \caption{Topology phase diagram of Nd1 Bi1 Pt1}
\end{table}

\begin{table}[t] 
    \centering
    %
    \caption{Topology phase diagram of O2 S1 Yb2}
\end{table}

\begin{table}[t] 
    \centering
    %
    \caption{Topology phase diagram of K1 Er1 Se2}
\end{table}

\begin{table}[t] 
    \centering
    %
    \caption{Topology phase diagram of Bi2 Eu1 Mg2}
\end{table}

\clearpage
\begin{table}[t] 
    \centering
    %
    \caption{Topology phase diagram of Ce1 K1 S2}
\end{table}

\begin{table}[t] 
    \centering
    %
    \caption{Topology phase diagram of Cl1 Dy1 O1}
\end{table}

\begin{table}[t] 
    \centering
    %
    \caption{Topology phase diagram of Nd2 O3}
\end{table}

\begin{table}[t] 
    \centering
    %
    \caption{Topology phase diagram of Dy1 Ga3}
\end{table}

\begin{table}[t] 
    \centering
    %
    \caption{Topology phase diagram of U1 Pd1 Ga5}
\end{table}

\begin{table}[t] 
    \centering
    %
    \caption{Topology phase diagram of Eu1 Zn2 As2}
\end{table}

\begin{table}[t] 
    \centering
    %
    \caption{Topology phase diagram of Ce1 Bi2 Au1}
\end{table}

\begin{table}[t] 
    \centering
    %
    \caption{Topology phase diagram of Ce1 Au1 Sb2}
\end{table}

\begin{table}[t] 
    \centering
    %
    \caption{Topology phase diagram of Pr1 Pd1 Sn1}
\end{table}

\begin{table}[t] 
    \centering
    %
    \caption{Topology phase diagram of Er1 In1 Au1}
\end{table}

\begin{table}[t] 
    \centering
    %
    \caption{Topology phase diagram of Ho1 Sb1 Te1}
\end{table}

\begin{table}[t] 
    \centering
    %
    \caption{Topology phase diagram of Ho1 Bi1}
\end{table}

\begin{table}[t] 
    \centering
    %
    \caption{Topology phase diagram of K1 Er1 Se2}
\end{table}

\begin{table}[t] 
    \centering
    %
    \caption{Topology phase diagram of Ho1 P1}
\end{table}

\begin{table}[t] 
    \centering
    %
    \caption{Topology phase diagram of Eu3 Pb1 O1}
\end{table}

\begin{table}[t] 
    \centering
    %
    \caption{Topology phase diagram of Ho1 Rh1}
\end{table}

\begin{table}[t] 
    \centering
    %
    \caption{Topology phase diagram of Ga3 Tm1}
\end{table}

\begin{table}[t] 
    \centering
    %
    \caption{Topology phase diagram of Eu3 Pb1 O1}
\end{table}

\begin{table}[t] 
    \centering
    %
    \caption{Topology phase diagram of Tm1 Mn3 O6}
\end{table}

\begin{table}[t] 
    \centering
    %
    \caption{Topology phase diagram of Tm1 Mn3 O6}
\end{table}

\begin{table}[t] 
    \centering
    %
    \caption{Topology phase diagram of Mn2 Pr1 Sb1 O6}
\end{table}

\begin{table}[t] 
    \centering
    %
    \caption{Topology phase diagram of Nd1 Mn1 O3}
\end{table}

\begin{table}[t] 
    \centering
    %
    \caption{Topology phase diagram of Ce1 Cu2}
\end{table}

\begin{table}[t] 
    \centering
    %
    \caption{Topology phase diagram of Tm2 Co1 Mn1 O6}
\end{table}

\begin{table}[t] 
    \centering
    %
    \caption{Topology phase diagram of Tm2 Co1 Mn1 O6}
\end{table}

\begin{table}[t] 
    \centering
    %
    \caption{Topology phase diagram of Eu1 Cr2 As2}
\end{table}

\begin{table}[t] 
    \centering
    %
    \caption{Topology phase diagram of Dy1 Cr1 O4}
\end{table}

\begin{table}[t] 
    \centering
    %
    \caption{Topology phase diagram of Tb1 Ni4 Si1}
\end{table}

\begin{table}[t] 
    \centering
    %
    \caption{Topology phase diagram of Gd1 Ni1 Si3}
\end{table}

\begin{table}[t] 
    \centering
    %
    \caption{Topology phase diagram of Ho1 Cr2 Si2}
\end{table}

\begin{table}[t] 
    \centering
    %
    \caption{Topology phase diagram of Ni1 Nd1 Ge2}
\end{table}

\begin{table}[t] 
    \centering
    %
    \caption{Topology phase diagram of Ge2 Ni1 Tb1}
\end{table}

\begin{table}[t] 
    \centering
    %
    \caption{Topology phase diagram of Tm1 Cr1 O3}
\end{table}

\begin{table}[t] 
    \centering
    %
    \caption{Topology phase diagram of Er1 Cr1 O3}
\end{table}

\begin{table}[t] 
    \centering
    %
    \caption{Topology phase diagram of Mn1 Nd1 As1 O1}
\end{table}

\begin{table}[t] 
    \centering
    %
    \caption{Topology phase diagram of Mn1 Nd1 As1 O1}
\end{table}

\begin{table}[t] 
    \centering
    %
    \caption{Topology phase diagram of Mn1 Nd1 As1 O1}
\end{table}

\begin{table}[t] 
    \centering
    %
    \caption{Topology phase diagram of Nd1 Mn2 Ge2}
\end{table}

\begin{table}[t] 
    \centering
    %
    \caption{Topology phase diagram of Pr1 Mn2 Ge2}
\end{table}

\begin{table}[t] 
    \centering
    %
    \caption{Topology phase diagram of Ce1 Mn1 Sb1 O1}
\end{table}

\begin{table}[t] 
    \centering
    %
    \caption{Topology phase diagram of Pr1 Mn1 Sb1 O1}
\end{table}

\begin{table}[t] 
    \centering
    %
    \caption{Topology phase diagram of Sr2 Tm1 Ru1 O6}
\end{table}

\begin{table}[t] 
    \centering
    %
    \caption{Topology phase diagram of Tb1 Mn6 Sn6}
\end{table}

\begin{table}[t] 
    \centering
    %
    \caption{Topology phase diagram of Tb1 Mn6 Sn6}
\end{table}

\begin{table}[t] 
    \centering
    %
    \caption{Topology phase diagram of Ho1 Mn6 Sn6}
\end{table}

\begin{table}[t] 
    \centering
    %
    \caption{Topology phase diagram of Ho1 Mn6 Sn6}
\end{table}

\begin{table}[t] 
    \centering
    %
    \caption{Topology phase diagram of Ho1 Mn6 Sn6}
\end{table}

\begin{table}[t] 
    \centering
    %
    \caption{Topology phase diagram of W1 Ho1 Cr1 O6}
\end{table}

\begin{table}[t] 
    \centering
    %
    \caption{Topology phase diagram of Er1 Ni4 B1}
\end{table}

\begin{table}[t] 
    \centering
    %
    \caption{Topology phase diagram of Tb1 Ni4 B1}
\end{table}

\clearpage
\begin{table}[t] 
    \centering
    %
    \caption{Topology phase diagram of Ho1 Ni4 B1}
\end{table}

\begin{table}[t] 
    \centering
    %
    \caption{Topology phase diagram of Ce1 Fe1 O3}
\end{table}

\begin{table}[t] 
    \centering
    %
    \caption{Topology phase diagram of Ce1 Fe1 O3}
\end{table}

\begin{table}[t] 
    \centering
    %
    \caption{Topology phase diagram of Pr1 Mn1 Si2}
\end{table}

\begin{table}[t] 
    \centering
    %
    \caption{Topology phase diagram of Pr1 Mn1 Si2}
\end{table}

\begin{table}[t] 
    \centering
    %
    \caption{Topology phase diagram of Nd1 Mn1 Si2}
\end{table}

\begin{table}[t] 
    \centering
    %
    \caption{Topology phase diagram of Nd1 Mn1 Si2}
\end{table}

\begin{table}[t] 
    \centering
    %
    \caption{Topology phase diagram of Nd1 Mn1 Si2}
\end{table}

\begin{table}[t] 
    \centering
    %
    \caption{Topology phase diagram of Ce1 Mn1 Si2}
\end{table}

\begin{table}[t] 
    \centering
    %
    \caption{Topology phase diagram of Ce1 Mn1 Si2}
\end{table}

\begin{table}[t] 
    \centering
    %
    \caption{Topology phase diagram of Ce1 Mn1 Si2}
\end{table}

\begin{table}[t] 
    \centering
    %
    \caption{Topology phase diagram of Nd1 Co1 O3}
\end{table}

\begin{table}[t] 
    \centering
    %
    \caption{Topology phase diagram of Ce1 Cu1 Si1}
\end{table}

\begin{table}[t] 
    \centering
    %
    \caption{Topology phase diagram of Sr2 Tb1 Ru1 O6}
\end{table}

\begin{table}[t] 
    \centering
    %
    \caption{Topology phase diagram of Sr2 Ho1 Ru1 O6}
\end{table}

\begin{table}[t] 
    \centering
    %
    \caption{Topology phase diagram of Sr2 Ho1 Ru1 O6}
\end{table}

\begin{table}[t] 
    \centering
    %
    \caption{Topology phase diagram of Dy1 Fe1 O3}
\end{table}

\begin{table}[t] 
    \centering
    %
    \caption{Topology phase diagram of Dy1 Fe1 O3}
\end{table}

\begin{table}[t] 
    \centering
    %
    \caption{Topology phase diagram of Ir1 Nd2 Ni1 O6}
\end{table}

\begin{table}[t] 
    \centering
    %
    \caption{Topology phase diagram of Ni1 Ir1 Pr2 O6}
\end{table}

\begin{table}[t] 
    \centering
    %
    \caption{Topology phase diagram of Ni1 Ir1 Pr2 O6}
\end{table}

\begin{table}[t] 
    \centering
    %
    \caption{Topology phase diagram of Ni1 Ir1 Pr2 O6}
\end{table}

\begin{table}[t] 
    \centering
    %
    \caption{Topology phase diagram of Ni1 Ir1 Nd2 O6}
\end{table}

\begin{table}[t] 
    \centering
    %
    \caption{Topology phase diagram of Ni1 Ir1 Nd2 O6}
\end{table}

\begin{table}[t] 
    \centering
    %
    \caption{Topology phase diagram of Tb1 Mn2 Ge2}
\end{table}

\begin{table}[t] 
    \centering
    %
    \caption{Topology phase diagram of Dy1 Mn2 Ge2}
\end{table}

\begin{table}[t] 
    \centering
    %
    \caption{Topology phase diagram of Ni1 Si2 Tb1}
\end{table}

\begin{table}[t] 
    \centering
    %
    \caption{Topology phase diagram of Eu1 Mn1 Bi2}
\end{table}

\begin{table}[t] 
    \centering
    %
    \caption{Topology phase diagram of N1 P1 Mn1 Th1}
\end{table}

\begin{table}[t] 
    \centering
    %
    \caption{Topology phase diagram of N1 As1 Mn1 Th1}
\end{table}

\begin{table}[t] 
    \centering
    %
    \caption{Topology phase diagram of N1 As1 Mn1 Th1}
\end{table}

\clearpage
\begin{table}[t] 
    \centering
    %
    \caption{Topology phase diagram of Yb2 Ir2 O7}
\end{table}

\begin{table}[t] 
    \centering
    %
    \caption{Topology phase diagram of Yb2 Ir2 O7}
\end{table}

\begin{table}[t] 
    \centering
    %
    \caption{Topology phase diagram of Nd2 Ir2 O7}
\end{table}

\begin{table}[t] 
    \centering
    %
    \caption{Topology phase diagram of Pr1 Ru2 Si2}
\end{table}

\begin{table}[t] 
    \centering
    %
    \caption{Topology phase diagram of Er1 In1 Ni1}
\end{table}

\begin{table}[t] 
    \centering
    %
    \caption{Topology phase diagram of Tm1 V1 O3}
\end{table}

\begin{table}[t] 
    \centering
    %
    \caption{Topology phase diagram of Tm1 V1 O3}
\end{table}

\begin{table}[t] 
    \centering
    %
    \caption{Topology phase diagram of Tm1 V1 O3}
\end{table}

\begin{table}[t] 
    \centering
    %
    \caption{Topology phase diagram of Ce1 Ni1 As1 O1}
\end{table}

\begin{table}[t] 
    \centering
    %
    \caption{Topology phase diagram of Ho1 Ni2 B2 C1}
\end{table}

\begin{table}[t] 
    \centering
    %
    \caption{Topology phase diagram of Nd1 Ni2 B2 C1}
\end{table}

\begin{table}[t] 
    \centering
    %
    \caption{Topology phase diagram of Ho1 Ni2 B2 C1}
\end{table}

\begin{table}[t] 
    \centering
    %
    \caption{Topology phase diagram of Dy1 Ni2 B2 C1}
\end{table}

\begin{table}[t] 
    \centering
    %
    \caption{Topology phase diagram of Pr1 Ni2 B2 C1}
\end{table}

\begin{table}[t] 
    \centering
    %
    \caption{Topology phase diagram of Ho1 Ni2 B2 C1}
\end{table}

\begin{table}[t] 
    \centering
    %
    \caption{Topology phase diagram of Ba1 Nd2 Co1 O5}
\end{table}

\begin{table}[t] 
    \centering
    %
    \caption{Topology phase diagram of Tb1 Cu2 Si2}
\end{table}

\begin{table}[t] 
    \centering
    %
    \caption{Topology phase diagram of Ho1 Cu2 Si2}
\end{table}

\begin{table}[t] 
    \centering
    %
    \caption{Topology phase diagram of Cu2 Si2 Tb1}
\end{table}

\begin{table}[t] 
    \centering
    %
    \caption{Topology phase diagram of Cu2 Ho1 Si2}
\end{table}

\begin{table}[t] 
    \centering
    %
    \caption{Topology phase diagram of Ho1 Fe2 Ge2}
\end{table}

\begin{table}[t] 
    \centering
    %
    \caption{Topology phase diagram of Ce1 Ir1 Al4 Si2}
\end{table}

\begin{table}[t] 
    \centering
    %
    \caption{Topology phase diagram of Pr1 Mn2 Si2}
\end{table}

\begin{table}[t] 
    \centering
    %
    \caption{Topology phase diagram of Gd1 Cu1 Sn1}
\end{table}

\begin{table}[t] 
    \centering
    %
    \caption{Topology phase diagram of Ni2 Tb1 Si2}
\end{table}

\begin{table}[t] 
    \centering
    %
    \caption{Topology phase diagram of Er1 Co2 Si2}
\end{table}

\clearpage
\begin{table}[t] 
    \centering
    %
    \caption{Topology phase diagram of U2 In1 Ni2}
\end{table}

\begin{table}[t] 
    \centering
    %
    \caption{Topology phase diagram of Ba2 Yb1 Ru1 O6}
\end{table}

\begin{table}[t] 
    \centering
    %
    \caption{Topology phase diagram of Ba2 Tm1 Ru1 O6}
\end{table}

\begin{table}[t] 
    \centering
    %
    \caption{Topology phase diagram of Cu2 Gd1 Si2}
\end{table}

\begin{table}[t] 
    \centering
    %
    \caption{Topology phase diagram of Pr1 Fe1 As1 O1}
\end{table}

\begin{table}[t] 
    \centering
    %
    \caption{Topology phase diagram of Pr1 Fe1 As1 O1}
\end{table}

\begin{table}[t] 
    \centering
    %
    \caption{Topology phase diagram of Pr1 Fe1 As1 O1}
\end{table}

\begin{table}[t] 
    \centering
    %
    \caption{Topology phase diagram of Tb1 Cu1 Sb2}
\end{table}

\begin{table}[t] 
    \centering
    %
    \caption{Topology phase diagram of Er1 Fe2 Si2}
\end{table}

\begin{table}[t] 
    \centering
    %
    \caption{Topology phase diagram of Er1 Mn2 Si2}
\end{table}

\begin{table}[t] 
    \centering
    %
    \caption{Topology phase diagram of Er1 Mn2 Si2}
\end{table}

\begin{table}[t] 
    \centering
    %
    \caption{Topology phase diagram of Er1 Mn2 Ge2}
\end{table}

\begin{table}[t] 
    \centering
    %
    \caption{Topology phase diagram of Er1 Mn2 Ge2}
\end{table}

\begin{table}[t] 
    \centering
    %
    \caption{Topology phase diagram of Er1 Mn2 Ge2}
\end{table}

\begin{table}[t] 
    \centering
    %
    \caption{Topology phase diagram of Dy1 V1 O4}
\end{table}

\begin{table}[t] 
    \centering
    %
    \caption{Topology phase diagram of Tb1 Co1 Ga5}
\end{table}

\begin{table}[t] 
    \centering
    %
    \caption{Topology phase diagram of Ho1 Co1 Ga5}
\end{table}

\begin{table}[t] 
    \centering
    %
    \caption{Topology phase diagram of Np1 Fe1 Ga5}
\end{table}

\begin{table}[t] 
    \centering
    %
    \caption{Topology phase diagram of Ga5 Ni1 U1}
\end{table}

\begin{table}[t] 
    \centering
    %
    \caption{Topology phase diagram of Tm1 Mn2 Ge2}
\end{table}

\begin{table}[t] 
    \centering
    %
    \caption{Topology phase diagram of Mn2 Tb1 Ge2}
\end{table}

\begin{table}[t] 
    \centering
    %
 & \\ \hline 
    \end{tabular}
    \caption{Topology phase diagram of Mn2 Tb1 Ge2}
\end{table}

\end{document}